\documentclass[5p, times]{elsarticle}

\usepackage{lineno,hyperref}
\modulolinenumbers[5]

\journal{Future Generation Computer Systems}

\usepackage{paper}

\begin{document}

\begin{frontmatter}

\title{Benchmarking Parallelism in FaaS Platforms}



\cortext[mycorrespondingauthor]{Corresponding author}

\author{Daniel Barcelona-Pons\corref{mycorrespondingauthor}}
\ead{daniel.barcelona@urv.cat}

\author{Pedro Garc\'{i}a-L\'{o}pez}
\ead{pedro.garcia@urv.cat}
\address{Departament d'Enginyeria Inform\`{a}tica i Matem\`{a}tiques, Universitat Rovira i Virgili, Tarragona, Spain}

\begin{abstract}
Serverless computing has seen a myriad of work exploring its potential.
Some systems tackle Function-as-a-Service (FaaS) properties on automatic elasticity and scale to run highly-parallel computing jobs.
However, they focus on specific platforms and convey that their ideas can be extrapolated to any FaaS runtime.

An important question arises: do all FaaS platforms fit parallel computations?
In this paper, we argue that not all of them provide the necessary means to host highly-parallel applications.
To validate our hypothesis, we create a comparative framework and categorize the architectures of four cloud FaaS offerings, emphasizing parallel performance.
We attest and extend this description with an empirical experiment that consists in plotting in deep detail the evolution of a parallel computing job on each service.

The analysis of our results evinces that FaaS is not inherently good for parallel computations and architectural differences across platforms are decisive to categorize their performance.
A key insight is the importance of virtualization technologies and the scheduling approach of FaaS platforms.
Parallelism improves with lighter virtualization and proactive scheduling due to finer resource allocation and faster elasticity.
This causes some platforms like AWS and IBM to perform well for highly-parallel computations, while others such as Azure present difficulties to achieve the required parallelism degree.
Consequently, the information in this paper becomes of special interest to help users choose the most adequate infrastructure for their parallel applications.

\end{abstract}


\end{frontmatter}


\section{Introduction}
\label{sec:intro}

In the path to make the cloud more accessible, new services appear to abstract application developers from the underlying infrastructure.
Serverless architectures are the newest development in this line with the objective to eliminate servers from the developer concerns.
Serverless computing is mainly represented by Function-as-a-Service (FaaS) platforms.
This ``cloud functions'' model has clear benefits: users do not manage servers or resources and only provide their code to the cloud, where it is automatically handled and executed on demand.
To peak simplicity, user code is provided as code snippets or functions, with a focused purpose, that can run anywhere.
Even better, users only pay for individual invocations of those functions, with execution time accounted for at millisecond level.

FaaS has picked the interest of many applications due to its simplicity.
One of such applications is highly-parallel computing jobs.
Elastic scale and on-demand resource availability look like a good substrate to run embarrassingly parallel tasks at scale.
Consequently, it motivated the appearance of several research and industry projects that adopt FaaS to run highly-parallel jobs.
On a first take, the ``Occupy the cloud''~\cite{PyWren2017} and ExCamera papers~\cite{excamera} demonstrated inspiring results from using FaaS for data analytics applications.
On their basis, several works~\cite{crucial2019,gg,numpywren,locus,GIMENEZALVENTOSA2019259} evolved on the idea of running compute-intensive parallel workloads on cloud functions and showed interesting results against traditional IaaS cluster computation.
Some papers~\cite{servermix,berkeleyServerless,hellersteinServerless} analyze these efforts and focus on the challenges and viability to run data analytics workloads on FaaS platforms.
Their conclusions show enticing results despite some issues (e.g., they discuss open challenges such as cost efficiency and statefulness).
In sum, they convey that FaaS platforms are a good fit for data-processing parallel applications~\cite{PyWren2017}.

In parallel computing, many compute-intensive tasks or processes are executed simultaneously.
Simultaneity is important since these tasks usually collaborate.
Data analytics jobs, linear algebra, and iterative machine learning training algorithms are some examples.
This requires a set of very specific properties in terms of resources, scale, and latency that allow to run all tasks at the same time without interleaving for compelling performance.
Indeed, the information presented on the above papers shows that parallel applications on FaaS only make sense when the platform provides the necessary properties to enable their parallelism.
However, they do not investigate them.

Simple function concurrency is not enough if each function invocation does not get its full isolated resources (we refer to each unit of resources as function instance).
Otherwise, computation faces throttling and resource interference and becomes too slow and expensive compared to traditional clusters.
In fact, most works on parallel computing atop FaaS presuppose that function invocations will run simultaneously, each on isolated resources~\cite{PyWren2017,excamera,locus,crucial2019}.
Consequently, the FaaS service must be able to provide enough resources at low enough latency to run all invocations in parallel.

However, all the aforementioned works base their arguments exclusively on the performance of AWS Lambda.
While AWS seems to provide compelling values for the discussed properties~\cite{PyWren2017,benchmarkingBerlin}, they are not included in the simple FaaS definition of cloud functions that we presented above (e.g.~\cite{cloudflaredef}).
The properties are, in fact, particular of each implementation of the FaaS model and usually detailed on each platform's documentation.
Still, cloud-offered FaaS platforms do not guarantee any of them: there are no service-level agreements (SLA) for these properties.
More so, while function resources, timeouts, or even concurrency are clearly described by every platform, parallelism is not carefully addressed by any of them.

This raises an important question: \emph{do current FaaS platforms fit parallel computations?}
And also: what makes some FaaS platform a better fit for parallel applications than the others?
Several benchmarking papers~\cite{wang2018peeking,benchmarkingBerlin,parallelOrchWOSC,faasdom} compare different FaaS platforms.
These papers indicate that indeed not all services provide the same properties.
Unfortunately, existing literature approaches FaaS platforms from a high-level user perspective.
They tackle use cases that resemble the IO-bound, reactive applications FaaS is prepared for, and focus on properties such as latency, cold start, cost, and configuration capabilities.
Some go beyond and explore elasticity with extra detail.
However, they tackle elasticity as the ability to quickly handle dynamic workloads and disregard actual parallelism and the implications of the FaaS platform architecture.
While this methodology is logical due to the black-box nature of the platforms, it does not allow to evaluate their suitability for highly-parallel, compute-intensive applications.
Indeed, understanding why each platform behaves as it does when dealing with parallelism requires a deeper knowledge on their architecture.
And existing literature does not provide a detailed view of the architectures and management approaches of each platform, neither none of them tackle parallel computations in detail.

To address this, in this paper we carefully investigate the parallel performance of the four major cloud FaaS platforms.
Namely, we analyze the architecture and performance of AWS Lambda (AWS), Azure Functions (Azure), Google Cloud Functions (GCP), and IBM Cloud Functions (IBM).
We especially focus on details that would affect the ability of the services to provide a good substrate for highly-parallel computations.
First, we describe and analyze the design of each service based on available information.
We are interested in how functions are managed, the virtualization technology used, how invocations are scheduled and their approach to scale, the management of resources, and other components that directly affect parallelism.
To organize all these traits, we build a comparative framework that helps the description and posterior discussion on the differences between platforms.
Second, we perform an experiment that allows to clearly visualize the parallelism of executions in a FaaS platform.%
\footnote{
  The experiment code and results are accessible at \url{https://github.com/danielBCN/faas-parallelism-benchmark}, including extra plots.
}
The experiment runs a job split into several function invocations (tasks) and produces plots with their execution timeline, drawing a complete view of the parallelism achieved.
Combined with the information from their architectures, this visualization allows us to understand when new resources (function instances) are allocated to process function invocations, and whether resources are used simultaneously to handle different invocations in parallel.
We can also see if this scheduling and resource management affects the performance of parallel tasks, such as by throttling invocations or by sharing resources across invocations (interleaving them).

Our objective is hence to understand their performance, and be able to spot bottlenecks, limitations, and other issues that can severely influence applications.
In sum, we want to categorize characteristics of each service that must be considered and may help users understand the different platforms to choose the one that better fits their needs.

This paper makes the following contributions:

\begin{itemize}
  \item
    We present a detailed architectural analysis of the four major FaaS platforms: AWS Lambda, Azure Functions, Google Cloud Functions, and IBM Cloud Functions.
    We categorize their design through a comparative framework with special focus on parallelism.
    Two traits importantly influence parallelism of the platforms: virtualization technology and scheduling approach.
  \item
    We perform a detailed experiment to reveal invocation scheduling and parallelism on each platform.
    The experiment consists in running several function invocations concurrently and gather as much information as possible to draw a comprehensive timeline of the execution.
    This visualizes the parallelism achieved and reveals issues.
  \item
    We analyze the information gathered for the different platforms and their affinity to parallel computations.
    Generally, lighter virtualization technologies and proactive scheduling improve parallelism thanks to faster elasticity and finer resource allocation.
    Thus, platforms like AWS and IBM resolve parallel computations more satisfactorily than Azure, where our experiment only reaches a parallel degree of $11\%$.
\end{itemize}

\paragraph{Outline}
Section~\ref{sec:related} summarizes the related work on benchmarking FaaS platforms.
Section~\ref{sec:arch} details the architectures of the analyzed FaaS platforms.
Section~\ref{sec:experiment} describes our experimentation and Sections~\ref{sec:experiment:aws} through \ref{sec:experiment:ibm} present its results.
Finally, we discuss the overall results in Section~\ref{sec:fit} and close in Section~\ref{sec:conclusions}.

\section{Related work}
\label{sec:related}

The recent popularity of serverless computing has triggered the appearance of several benchmarks tackling FaaS platforms.
Since most services are offered as proprietary cloud platforms, these works explore them from a black-box perspective, and mostly from a high-level user point of view.

Some papers and websites analyze the cold start of functions across platforms~\cite{faasdom,serverlessbench}.
Invocation latency and CPU performance are also extensively explored in literature~\cite{SCHEUNER2020EVAL,wang2018peeking}.
Recently, new benchmarks~\cite{faasdom} include the invocation throughput that platforms are able to provide and an evaluation of the invocation costs.
Concurrency is also explored on different works~\cite{blog:Shilkov,pawlik2019performance,cloudbuttonbench,wang2018peeking}, but they only perform large-scale benchmarks from a high-level point of view.
The measurements on a recent paper~\cite{sequoia} regarding the QoS of different platforms also show special emphasis on their concurrency and explore different issues with resource allocation and function scheduling.

A very interesting topic in FaaS benchmarks is service elasticity~\cite{benchmarkingBerlin,lee2018evaluation,pawlik2019performance,allbutone}.
However, their experiments do not evaluate computation parallelism and performance.
They generate a dynamic workload of many invocations to observe how the platform behaves when there are changes in the demand scale.
Then, they analyze the capacity of each system to accept incoming requests rapidly.
Also, the workloads are usually IO-bound, like reactive web applications.
In this benchmark, we focus on compute-intensive workloads in parallel computing.
These applications need stronger guarantees on execution parallelism and resource isolation to achieve good performance, but this is not evaluated in literature benchmarks.
Indeed, they do not differentiate between resolving invocations concurrently and actually handling them in parallel to provide the necessary performance.
We explore the behavior of each platform with deeper detail to determine these characteristics.

Also, while the results of all these benchmark papers evidence performance variation across platforms, they usually disregard its causes and do not explore in detail the architecture of each platform for properties that affect performance.
Some papers partially dig into the internals of platforms~\cite{wang2018peeking}, but do not study its effects on parallelism.
We find extensive analysis of open-source FaaS platforms~\cite{opensource:bench,shahrad2019architectural}, but such evaluations are not possible for the major platforms offered in public clouds.

A few papers explore more complex applications.
An implementation of MapReduce~\cite{GIMENEZALVENTOSA2019259} is evaluated for large computations on AWS Lambda following the observations of PyWren~\cite{PyWren2017} and ExCamera~\cite{excamera}.
\textit{gg}~\cite{gg} and Pocket~\cite{pocketpre, pocket} also perform several analysis of large computations atop this particular platform.
The platform proves a good fit for batch computations, but other platforms are not studied nor their design regarding parallelism.
Other benchmarks perform their evaluation from an even higher level, and focus on orchestration tools atop FaaS~\cite{comparisonWOSC,parallelOrchWOSC}.
They show that some platforms do not achieve good parallelism, but do not explore why in detail.
A recent paper from Azure~\cite{microsoft2020serverless} offers some insights on how Azure Functions works and an analysis of the platform usage from the cloud provider perspective.
However, it disregards function parallelism and focuses its exploration on optimizing latency in cold starts and reduce resource waste.

None of the existing works investigates the architecture design of each FaaS platform, and how it affects their performance.
Especially for highly-parallel computations.

\section{Architecture analysis}
\label{sec:arch}

In this section we describe the architecture of each FaaS platform.
For an easy comprehension of the differences between services, we first create a comparative framework.
We use it to outline the general organization, configuration possibilities, and documented limitations, and we put them in context with a description of their deployment model.
Our interest is specially focused on \emph{resource provisioning and scalability} to meet on-demand requests.
Thus, we make emphasis on work distribution in terms of concurrency and parallelism.
The descriptions on this section are all based on official information available online, unless indicated otherwise.

\begin{figure}
  \centering
  \includegraphics[width=0.7\linewidth]{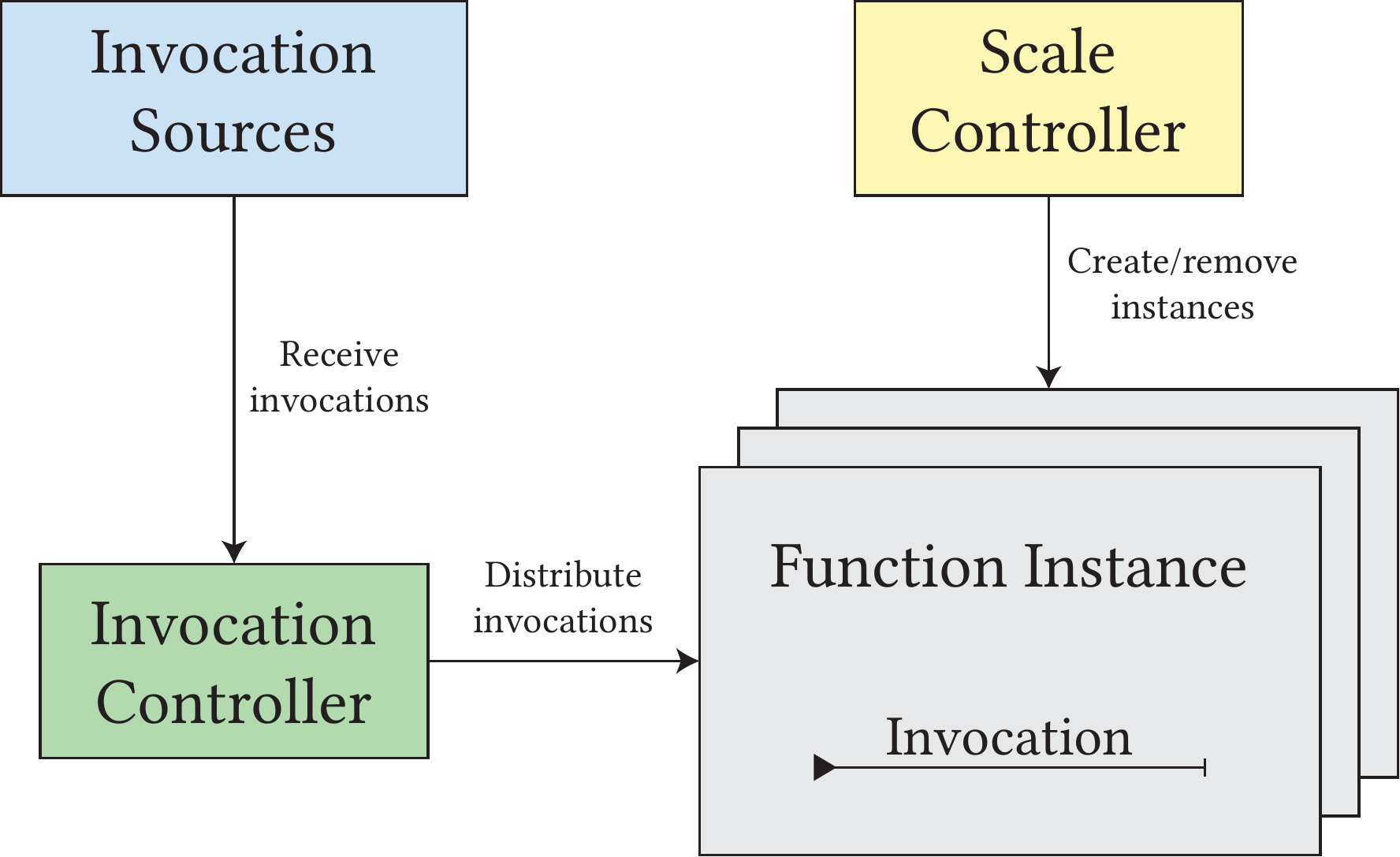}
  \caption{
    Abstract FaaS architecture.
  }
  \label{fig:arch:abstract}
\end{figure}

Figure~\ref{fig:arch:abstract} shows an abstract FaaS architecture with the main components we analyze in this section: function instances and invocations, the scale controller, the invocation controller, and invocation sources.
We draw this schema based on open-source platforms and the literature~\cite{openwhisk,opensource:bench,refArchFaas,shahrad2019architectural}.

There is an important distinction in a FaaS platform: \emph{function invocations} and \emph{function instances}.
Invocations are each one of the function executions in response to a request.
Instances refer to the resource units that are provided to run invocations.
If two function invocations are run on the same resource entity, we consider they run on the same function instance.
This can happen by reusing a container, or by running several invocations in the same VM.
While an invocation is easily identified on all platforms, each service manages instances differently.
As we will see, function instances are usually determined by the virtualization used on each architecture.

The scale controller represents the logic that decides when to create or remove instances.
The invocation controller is the logic that decides where to run each invocation that comes from invocation sources.
In practice, this components may be merged into a single one; or be part of another component.

\subsection{Comparative framework}   
\label{sec:arch:frame}

For a handy comparison between FaaS platforms, we design a comparative framework to collect the most relevant characteristics of each one.
It explores two items:
\begin{inparaenum}[(1)]
  \item the general model of function deployment and management, and  
  \item the architectural approach to scale and resource management.  
\end{inparaenum}
Our focus is specially on the second one, since it conditions scalability and parallelism for each service, while the first provides important context.
In this sense, we expand the second item by reviewing the following six traits:

\paragraph{Technology}
In this category we discuss the virtualization technology used to build \emph{function instances}.
Instances need to be isolated resource units to provide multi-tenant properties.
This is usually achieved with virtualization, but the chosen technology is very important for the design of a FaaS platform.
Traditional VMs are heavier than containers, what makes the latter better for the irregular, low-latency FaaS scaling.
But we also have microVMs, light as containers but with kernel-based virtualization.
Some providers may combine technologies to efficiently handle isolation and performance.

\paragraph{Approach}
This category analyzes the job of the invocation controller logic;
i.e. the scheduling approach used to distribute work (invocations) across resources (instances).
In particular, we categorize two kinds: push-based and pull-based.
We refer as push-based to architectures that follow a proactive policy where a control plane takes the role of the invocation controller: the controller pushes invocations to instances.
A pull-based architecture is more loose and reactive; the invocation controller logic is delegated to instances, which obtain work from the event sources: instances pull invocations from queues.

\paragraph{Scaling}
This describes the scale controller.
The scheduling approach heavily influences this component:
push-based architectures usually merge the scale and invocation controller logic to balance load on demand, while pull-based ones use a dedicated scale controller to manage the instance pool.
Here we also focus on the decisions of this component.
For example, when does the controller create or remove instances?

\paragraph{Resources}
Most platforms let users configure the resources that each function gets.
We determine the \emph{minimum guaranteed resources} for a single invocation with a particular function configuration.
This is a product of the platform architecture and the tuning set by the provider.
On one hand, how an architecture manages resources may introduce interferences across invocations.
On the other hand, the service provider may set up some limits on the system that affect this category.
For instance, resources could be restricted to ensure the proper functioning of the system or the economic viability of the service.

\paragraph{Parallelism}
This category analyzes all information relative to function concurrency and parallelism.
Particularly, we want to quantify the \texttt{maximum amount of parallelism} that a platform can achieve.
It is important to remember that this is just an imposed limit and the service does not guarantee (through an SLA) to reach such parallelism.

\paragraph{Rate limits}
Providers protect their systems with use rates that block excessive request bursts and can limit parallelism.
We illustrate it with the number of invocations per second the system accepts, but also discuss other limits related to parallelism.

\subsection{AWS Lambda architecture}   
\label{sec:arch:aws}

\begin{figure}
  \centering
  \includegraphics[width=0.7\linewidth]{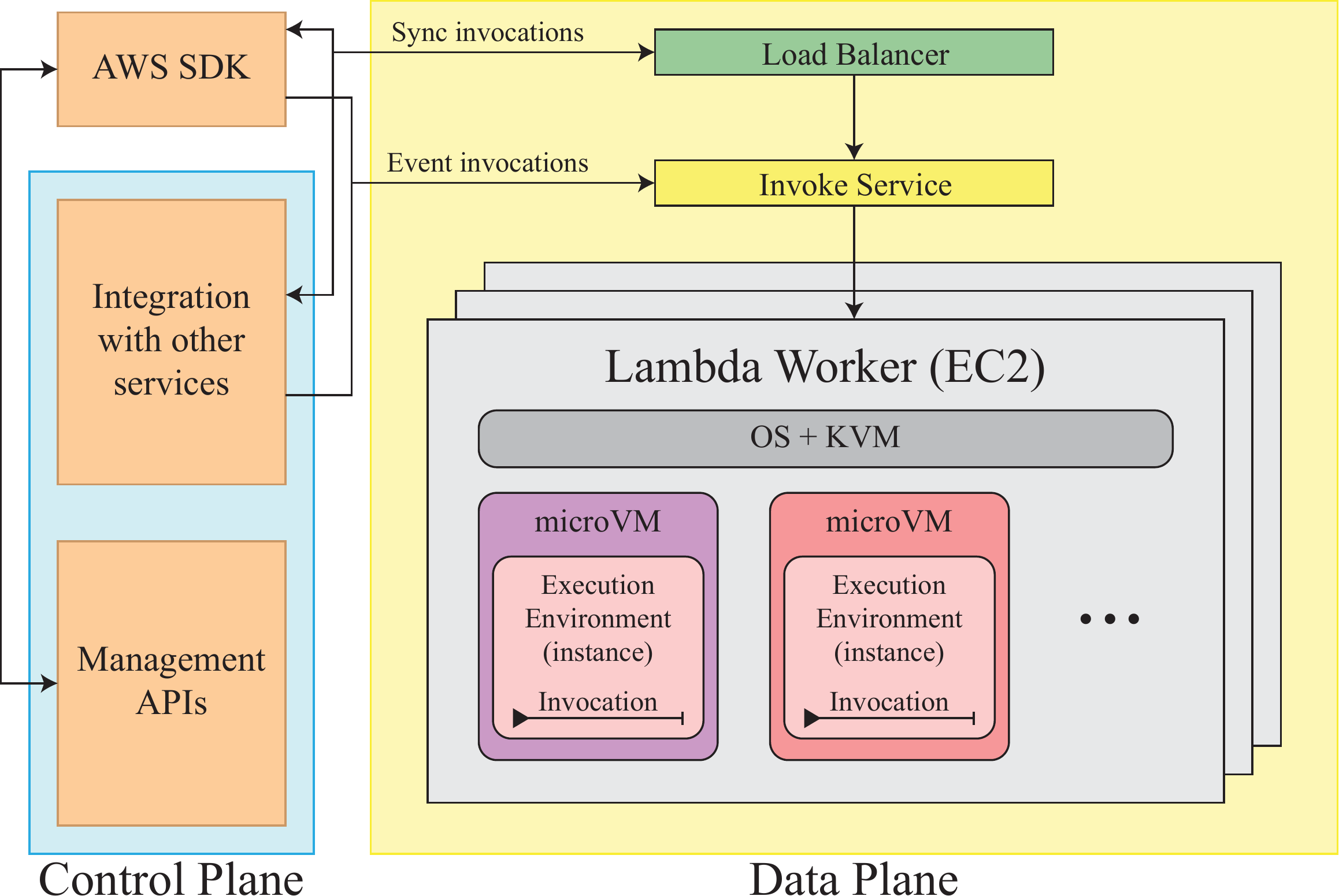}
  \caption{
    AWS Lambda architecture.
  }
  \label{fig:arch:aws}
\end{figure}

All AWS Lambda specification, configuration, and limitations are described in its documentation~\cite{lambda:config,lambda:limits}.
Additionally, a recent AWS whitepaper~\cite{aws:securitypaper} sketches its internals with more detail.
An architecture overview is depicted in Figure~\ref{fig:arch:aws}.
The service is split into the Control and Data planes.
The Control plane handles the management API, such as creating or updating functions, and also includes integrations with other cloud services (e.g., forwarding S3 events or polling SQS queues).
The Data plane manages resources and function invocations.
The Invoke service is the main control component taking the logic of the invocation and scale controllers.
Event-triggered invocations go directly to the Invoke, where they may be queued; synchronous invocations, which need extra management to respond to callers, are handled by a load balancer.

\subsubsection{Fuction deployment}
In AWS Lambda, the user deploys functions individually. 
The Management API, enables function creation and configuration (e.g., runtime and memory).
The function code is uploaded to the service in compressed packages and the configuration is updated with HTTP requests.
Functions may be invoked with an HTTP request, but the usual approach is to bind them to function triggers.
Triggers set up links with other cloud services that produce events, and allow enabling invocations in response to those events.
Configuration includes other features, such as limiting function concurrency and pre-provisioning resources.

\subsubsection{Resources and scale}

\paragraph{Technology}
Lambda uses several virtualization levels in its architecture~\cite{aws:securitypaper} (see Figure~\ref{fig:arch:aws}).
The general structure changed recently with Firecracker~\cite{aws:firecracker}, which enhances performance and management.
We focus on the new model.
The first level contains Lambda Workers, which are metal EC2 machines running a Firecracker hypervisor.
This technology allows to populate Lambda Workers with microVMs that are quick to spawn and provide strong isolation.
MicroVMs draw tenant boundaries, being each of them exclusive to a user.
Within a microVM, the service creates \emph{execution environments} to run the invocations.
Execution environments are the function instances, created with the help of \texttt{cgroups} and other container technologies.
Each of them is created especially for a function deployment, containing the appropriate runtime and function code, and can be reused for subsequent invocations.
MicroVMs are not tied to a single function deployment and may hold several execution environments of the same user.
With Firecracker, each microVM only contains a single execution environment at a time.

\paragraph{Approach}
An official AWS whitepaper~\cite{aws:securitypaper} depicts Lambda following a push-based scheduler.
The Invoke service proactively designates the instance for each invocation.
Upon a request, this component creates an execution environment (instance) inside a microVM or chooses an existing idle one.
To perform such decision, this component must monitor all system resources.
Then, the service pushes the invocation payload to the instance, where it is run.

\paragraph{Scaling}
The Invoke service controls the scale at a multi-tenant level.
It identifies the instance for each invocation among the cluster of Lambda Workers, which is common for all users.
Since multi-tenancy is achieved at microVM level, the service can easily fill Lambda Workers.
If the user performing a new invocation has no microVM available, the Invoke service finds the resources for a new one in the cluster.
If there is already a microVM running, it can be reused for two cases:
the existing execution environment is for the same function that is being invoked (and it is simply unfrozen and run with the new payload), or it is for another function (and a new container is created).

\paragraph{Resources}
User configure function memory from $128$~MiB to $10$~GiB.
Then, instances will grant exactly that much memory for each invocation.
To achieve so, a function instance only processes one invocation concurrently.
With memory, Lambda scales other resources proportionally.
In particular, $1792$~MiB corresponds to the equivalent to one vCPU~\cite{lambda:config}.

\paragraph{Parallelism}
The service imposes a limit of $1000$ concurrent executions per user---which can be increased under request~\cite{lambda:limits}.
Since there is no per-instance concurrency, the achievable parallelism shares this limit.

\paragraph{Rate limits}
The request per second rate is very ample: $10$ times the concurrent executions limit for synchronous and unlimited for asynchronous invocations.
However, instance creation is controlled by a burst limit~\cite{lambda:scale}.
Depending on the region, the service creates from $500$ to $3000$ instances without any limitation in a burst phase.
Reached that point, the number of instances created is limited to $500$ each minute.


\subsection{Azure Functions architecture}   
\label{sec:arch:azure}

\begin{figure}
  \centering
  \includegraphics[width=0.7\linewidth]{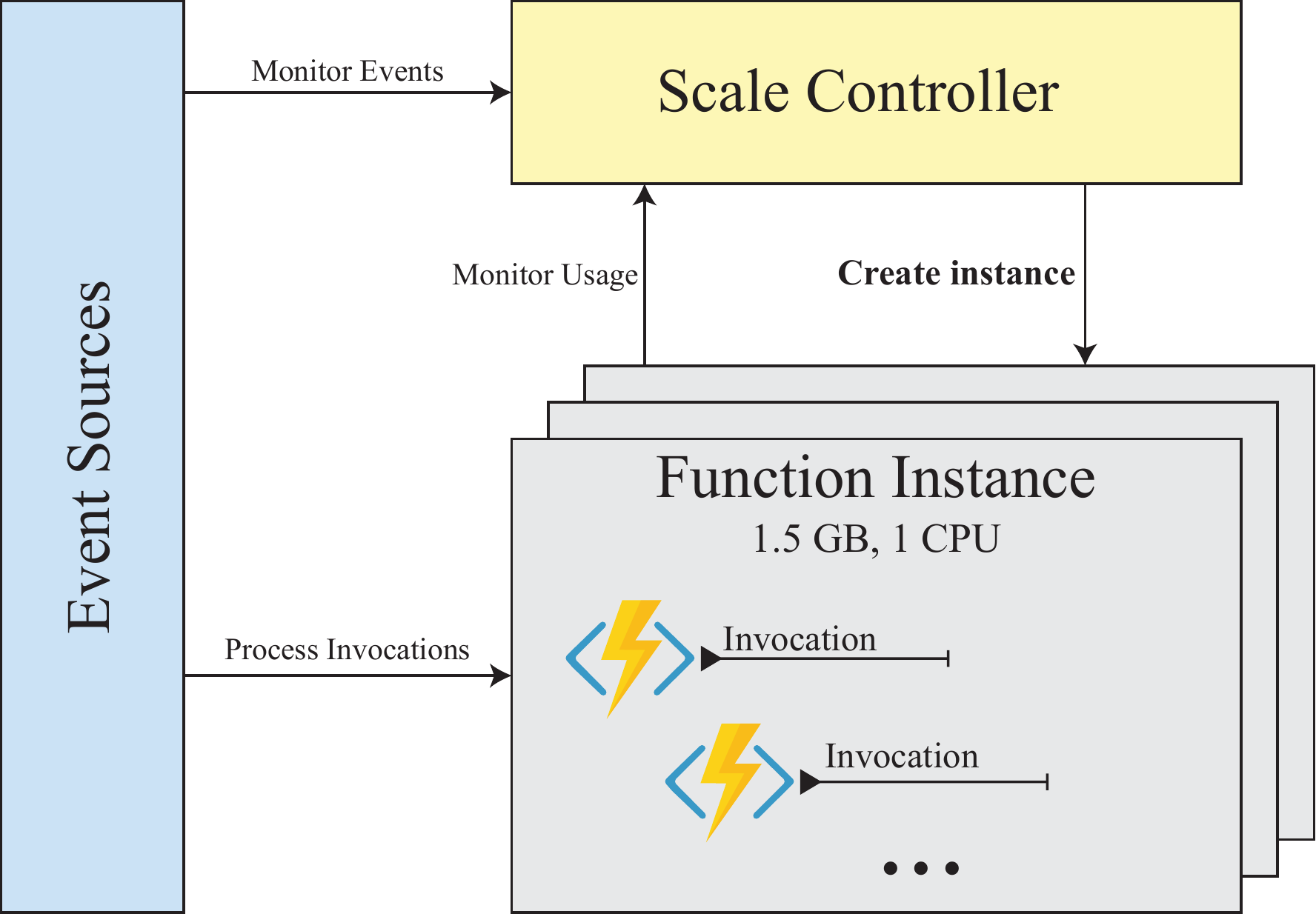}
  \caption{
    Azure Functions architecture.
  }
  \label{fig:arch:azure}
\end{figure}

An architecture overview of Azure Functions is shown in Figure~\ref{fig:arch:azure}.
A description of it is available in its documentation~\cite{azure:scale}.
In the service, a set of function instances run invocations in response to events from different sources.
The scale of this set is regulated by a long-running component that monitors the state of the service: the Scale Controller.

\subsubsection{Fuction deployment}
In Azure Functions, a Function App is the general unit of management and deployment.
Each Function App works as a bundle that may contain many function definitions and manages a pool of resources (function instances).
The application package the user uploads includes their code, dependencies, and configuration. 
Each function definition is a piece of code correctly annotated as an Azure Function.
The next part of configuration is function triggers and bindings, which define the events that will result in function invocations and enable functions to operate input and output streams.
Advanced configuration parameters allow to tune some extra features.

\subsubsection{Resources and scale}

\paragraph{Technology}
Azure Functions is built atop Azure WebJobs, a Web App PaaS service that auto-scales a VM cluster based on load.
Function instances are therefore VM hosts with fixed resources and the whole Function App deployment package installed.
The set of instances is managed by Azure WebJobs within each Function App.
Differently from all other platforms, Azure uses Windows hosts by default, instead of Linux.

\paragraph{Approach}
The documentation of Azure Functions depicts a pull-based scheduling approach~\cite{azure:scale}.
Function instances poll event sources to process function invocations.
When an instance finds an unprocessed request in one of its bound triggers, it runs it.
An instance can run any function definition in the Function App and several invocations may be taken by the same instance concurrently.
This means that different invocations (same Function App) may share resources.

\paragraph{Scaling}
The Scale Controller manages the number of Function Instances in a Function App.
This component monitors the event rates and instance usage to determine when to create or remove VMs.
The actions of this manager are dictated by a set of internal policies.
For example, it only creates one instance per second if invocations are by HTTP request~\cite{azure:scale}.

\paragraph{Resources}
The available resources on each instance depend on the plan the Function App is deployed on: Consumption plan or Premium plan~\cite{azure:scale}.
We focus on the Consumption plan, since it is the serverless one.
It presents the typical FaaS properties of fine-grained pay-per-use and scale to zero.
But differently from other platforms, resources are not configurable and all instances have $1.5$~GiB of memory and one CPU.
This means that an invocation may get \emph{at most} these resources.
Remember that each instance may take several invocations concurrently, so there is no guaranteed resources for each invocation.
The Premium plan allows increased performance by pre-provisioning resources.
The user defines a lower and upper limit to the number of instances, that do not scale to zero.

\paragraph{Parallelism}
The documentation of Azure Functions depicts the service clearly not focused on parallelism.
The number of instances per Function App is limited to $200$ and cannot be increased~\cite{azure:scale}.
However, it seems to be built for small IO-bound tasks that benefit from concurrency.
A single instance may chose to fetch several invocations from the event sources at the same time, allowing unlimited invocation concurrency by sharing instance resources, a good fit for IO operations.
The actual parallelism is thus limited to the number of instances since they only have one CPU each.
To avoid resource interference, a necessity for compute-heavy tasks, concurrency can be configured by the user by setting a per instance limit~\cite{azure:concurrency}.
There is a different limit for each trigger type and they are managed by the instances autonomously.
For example, HTTP requests have a default limit of $100$ concurrent invocations per instance, which after scaling to the maximum $200$ instances could offer $20$K concurrent invocations.
This does not improve parallelism.

\paragraph{Rate limits}
There is no service limit on the number of invocations processed per second.
It depends on the functions themselves (user code) and how many of them the available instances can process following the service polices (at which rate they pull from queues).
Note that there is a limit on the instance creation rate: one per second based on HTTP trigger load, and one every $30$ seconds for other triggers~\cite{azure:scale}.


\subsection{Google Cloud Functions architecture}   
\label{sec:arch:gcp}

The general concepts of the architecture of Google Cloud Functions are detailed in its documentation~\cite{gcp:concepts}.
However, it does not specify its internal components with clarity; such as which component runs the invocation and scale control logic.
Consequently, we do not present an overview scheme for this platform.
This only affects the scheduling approach and scaling categories of our comparative framework.
The documentation gives enough information for the other categories.

\subsubsection{Fuction deployment}
In Google's FaaS, the unit of deployment is a single function.
The system manages each function separately, even if deployed on the same package, and scales them individually.
To deploy a function, the CLI uploads the code directory and detects functions based on project structure conventions.
The configuration is updated through HTTP calls to the service API.
Functions may always be invoked with HTTP requests, but the user may also associate them with triggers to generate invocations in response to events from other services.

\subsubsection{Resources and scale}

\paragraph{Technology}
To isolate executions across tenants, Google Cloud Functions uses gVisor microVMs~\cite{gcp:usegvisor}.
gVisor~\cite{gcp:gvisor} is a kernel-based virtualization tool used to securely sandbox containers.
These containers are the function instances that run user code, taking only one invocation at a time~\cite{gcp:concepts}.
MicroVMs allow to strongly isolate real resources between tenants;
however, there is no information about how many containers can be packed in the same microVM or how the service ensures each of them has the resources configured for the function.

\paragraph{Approach}
There is no information available about the internals of the service that enables us to make any detailed evaluation of its scheduling policy.
Documentation points to a push-based approach~\cite{gcp:concepts}, where a controlling component manages invocations and scale.

\paragraph{Scaling}
Following the previous category, we sketch the existence of a controller component in the system that collects system information and decides when to scale in or out.
The reasons behind scaling decisions are listed in the documentation~\cite{gcp:limits} and include the usual running time (short functions scale more), the cold start time, the rate limits of the service, function error rates, and the load of the servers at the time.

\paragraph{Resources}
Users configure memory for their functions.
The service offers $5$ possible sizes from $128$~MiB to $2$~GiB, and assigns CPU therefrom~\cite{gcp:pricing'n'types}.
Note that this relation is not proportional: for instance, $256$~MiB functions are given $400$~MHz of CPU, while $2$~GiB, $2.4$~GHz.
The documentation states this numbers as approximations and not guaranteed resources.
Thus, we expand this information with a simple exploratory work.
Inspecting system information (\texttt{/proc}), we see that all containers run in VMs with $2$~GiB of memory and $2$ CPUs at $2.3$ or $2.7$~GHz.
This happens irrespective of the function configuration, which tells us that all microVMs are equally sized.
Again, it is unknown if different containers are packed in the same VM.

\paragraph{Parallelism}
There is no limit on invocation concurrency for functions called with HTTP requests~\cite{gcp:limits}.
Event-triggered invocations are limited to $1000$ concurrent executions per function (not increasable).
Advanced configuration also allows users to limit the number of concurrent instances.
Since each instance only allows one invocation at a time, parallelism is also bound by these limits.
Thus, maximum parallelism is unbound for HTTP-triggered functions.

\paragraph{Rate limits}
Google Cloud Functions sets a per region limit of $100$M function invocations per $100$ seconds~\cite{gcp:limits}.
Additionally, the CPU usage is limited by other rates.
These quotas are fairly generous for the majority of applications.


\subsection{IBM Cloud Functions architecture}   
\label{sec:arch:ibm}

\begin{figure}
  \centering
  \includegraphics[width=0.7\linewidth]{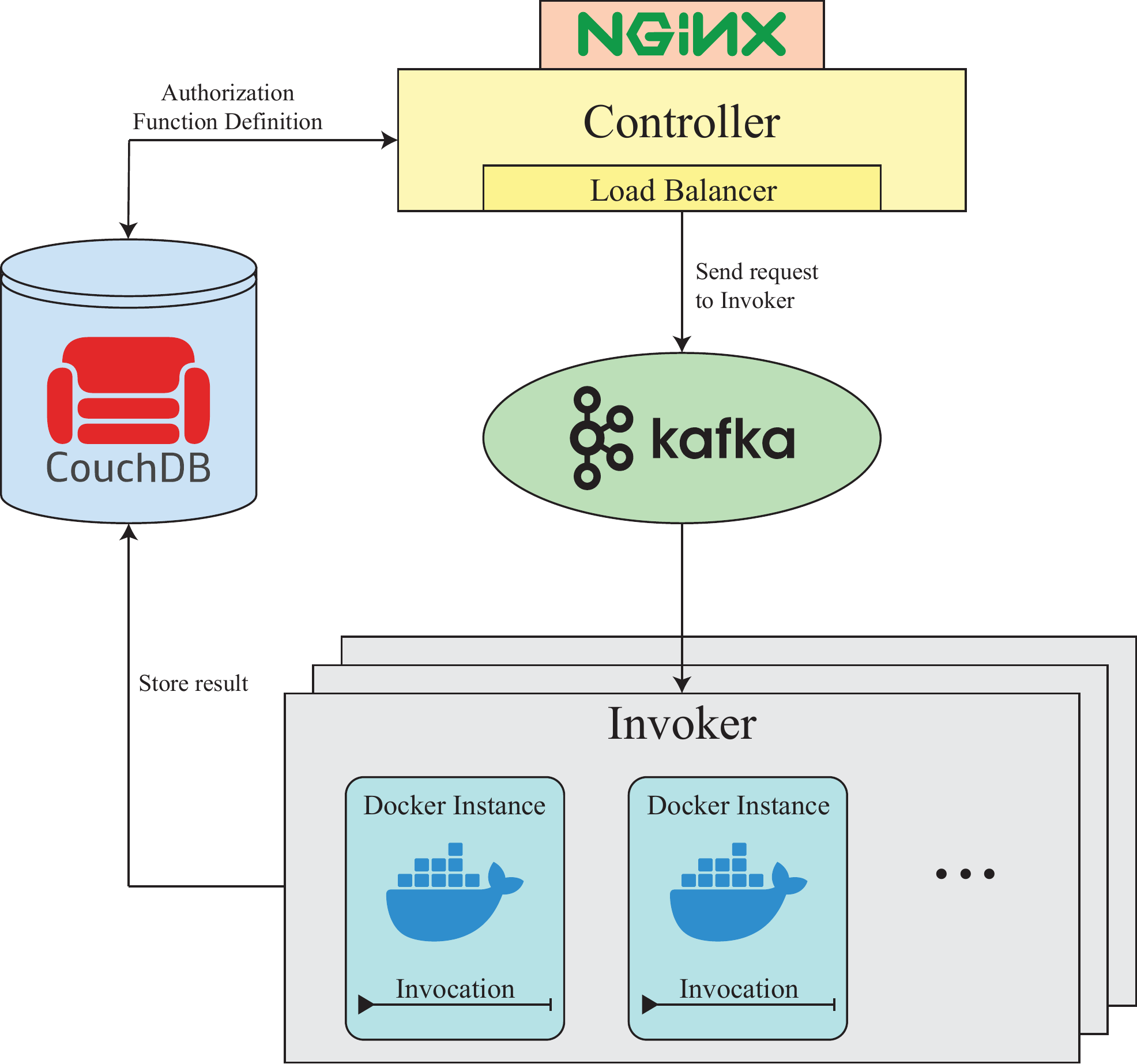}
  \caption{
    IBM Cloud Functions (OpenWhisk) architecture.
  }
  \label{fig:arch:ibm}
\end{figure}

IBM Cloud Functions is a cloud-managed Apache OpenWhisk~\cite{openwhisk} deployment, an open-source FaaS started by IBM and donated to the Apache Software Foundation.
Its design is expounded in both documentations~\cite{ibm:how-works}.
Figure~\ref{fig:arch:ibm} overviews the platform, with $4$ main components:
the Controller acts as a load balancer and manages instance resources;
the Invoker machines are VMs that run several containers (the function instances, usually Docker);
a Kafka deployment communicates them at scale;
and a database (CouchDB) stores function information, request data (payload and results), and logs.

\subsubsection{Fuction deployment}
In IBM Cloud, functions are called ``Actions" and deployed individually to the service.
Actions must be contained in a namespace, which belongs to a resource group, and may be organized in packages.
Action definitions (code and configuration) are registered in the database through the Controller, that exposes an HTTP API.
Like Actions, the user defines Triggers and Rules. 
Triggers identify event sources to monitor, while Rules are event filters to map Triggers to Actions.
Actions can always be invoked directly with HTTP requests.

\subsubsection{Resources and scale}

\paragraph{Technology}
Function instances are containers that are run on a cluster of Invoker machines, which are VMs.
Each Invoker manages its local pool of containers, while the Controller is responsible for the pool of Invoker machines.
Thus, IBM's FaaS platform has two levels of virtualization that we can analyze.

\paragraph{Approach}
OpenWhisk follows a push-based scheduling approach~\cite{ibm:how-works}.
The invocation control logic is split between two components.
The Controller, upon a request, forwards it to a designated Invoker machine.
The Invoker then creates or reuses an idle container (instance) to run it.
This one-to-one communication is performed asynchronously through Kafka.
The Controller acts as a load balancer while monitoring the state of all Invoker VMs.
Thus, the Controller proactively pushes function invocations to the instances that run them.

\paragraph{Scaling}
The scale control logic is also split.
Each Invoker machine locally manages its containers.
With a fixed set of memory assignable to containers on the machine and the functions' memory configuration, the Invoker responds to requests by creating containers with the right resources and informs the Controller of its usage levels.
The Controller manages the general pool of Invokers, and sends requests to them prioritizing the ones that already have warm, but idle containers.
If none is available, it choses one Invoker with enough free resources to create a new instance.
There is no information on when or how Invokers are created or removed, or if the set is fixed.

\paragraph{Resources}
Users configure function memory, and each instance provides those resources to each invocation.
The service does not ensure any CPU resources for a given memory, but it claims to scale resources proportionally.
To collect more specific information, we empirically study the platform (more details in Section~\ref{sec:experiment:ibm:disc}).
Inspecting system information (\texttt{/proc}) we see that all explored machines (Invokers) run $4$-core CPUs and $16$~GiB of RAM.
However, in our experiments, a single Invoker seems to dedicate only up to $8$~GiB for container hosting.
If resources scale proportionally, this CPU-memory relation tells us that we could ensure a full CPU core with $2$~GiB functions.

\paragraph{Parallelism}
The service has a limit of $1000$ executing or queued concurrent invocations per namespace---increasable under request~\cite{ibm:limits}.
Each instance only takes one invocation at a time, meaning that the maximum parallelism of the platform is the same as this imposed limit.
In fact, OpenWhisk offers a configuration parameter to manage per-instance concurrency~\cite{openwhisk:concurrency}, with which a single instance could take several invocations at the same time (unavailable on the IBM Cloud).
While this increases concurrency, it does not improve parallelism.

\paragraph{Rate limits}
No more than $5000$ invocations can be submitted per namespace per minute---also increasable~\cite{ibm:limits}.
It does not directly affect parallelism, since the concurrency limit is smaller.
However, for applications that run many small tasks, it can be easy to reach.
Examples are parallel computations with dynamic load balancing and consecutive batches of tiny tasks.


\subsection{Discussion}
\label{sec:arch:disc}


\begin{table}
  \caption{Traits of FaaS platforms (October 2020). See Section~\ref{sec:arch:frame}.}
  \label{tab:archs}
  \centering
  \footnotesize
  \begin{tabular}{llll}
    & \textbf{Technology} & \textbf{Approach} & \textbf{Scaling}  \\
    \toprule
    \textbf{AWS} & microVM & push-based & Control plane \\
    \textbf{Azure} & VM & pull-based & Scale Controller \\
    \textbf{GCP} & microVM & \textit{push-based} & \textit{Controller} \\
    \textbf{IBM} & VM + Container & push-based & Controller \\
    \bottomrule \\
    & \textbf{Resources} & \textbf{Parallelism} & \textbf{Rate limit} \\
    \toprule
    \textbf{AWS} & $1792$~MiB = $1$ CPU & \multicolumn{1}{r}{(ext.) $1000$} & $3000 + 500$/min \\
    \textbf{Azure} & $1$ CPU%
    \tablefootnote{
      With function instance concurrency limited (Section~\ref{sec:arch:azure}).
    }
    & \multicolumn{1}{r}{$200$} & unbound \\
    \textbf{GCP} & $2$~GiB = $2.4$~GHz & \multicolumn{1}{r}{unbound} & $100$M/$100$ s \\
    \textbf{IBM} & $2$~GiB = $1$ CPU%
    & \multicolumn{1}{r}{(ext.) $1000$} & $5000$/min \\
    \bottomrule
  \end{tabular}
\end{table}

Table~\ref{tab:archs} summarizes all traits collected above for the four FaaS platforms.
These are obtained directly from their documentation and official publications as of October 2020.
The exceptions are Google's scheduling approach and scaling, which are not clearly described; and the guaranteed resources at IBM, which we empirically assess.
We discuss all traits next:

\paragraph{Technology}
  Each provider uses a different virtualization technology.
  AWS and Google use several virtualization levels and include microVMs.
  This allows finer resource management with small start-up times and increased security.
  IBM also has several virtualization levels, but does not use microVMs.
  Consequently, packs of containers run on each VM, requiring a different approach to security.
  Azure only has one level of virtualization, simplifying resource management at the cost of elasticity.
  In sum, the schema of virtualization technologies is really important for the architecture, as it influences several factors that must be considered for scheduling and managing the service, e.g., security and the time it takes to start an instance.

\paragraph{Approach}
  Only Azure clearly uses a pull-based approach to scheduling work.
  The other providers build push-based architectures that create instances more eagerly.
  This benefits parallelism, as they are faster to create instances.
  From the table, the scheduling approach seems tightly related to the virtualization technologies used.
  Azure manages a single VMs level and takes a conservative approach to scale.
  Meanwhile, the others use lighter technologies and spawn instances with more liberty.

\paragraph{Scaling}
  There is always a controlling component that manages scale in the system.
  In push-based platforms, scale and invocation distribution logics are dealt by the same control component.
  In the pull-based, the controller manages scale based on the state of the system, but does not deal with invocations.

\paragraph{Resources}
  Instances usually have fixed resources, based on function configuration.
  Most providers let users configure function memory and scale other resources, like CPU, proportionally.
  Azure does not allow configuring resources, but monitors usage during execution to adjust billing~\cite{azure:pricing}.
  Even with their different configuration options, all providers offer at least $1$ vCPU with around $2$~GiB of memory.
  They allow users to ensure certain amounts of resources.

\paragraph{Parallelism}
  The achievable parallelism is quite good for AWS, GCP and IBM, with generous limits on concurrency.
  Azure, however, has restricted parallelism due to its scheduling approach, strict limits, and system tuning.

\paragraph{Rate limit}
  Invocation rate limits do not generally restrict parallelism in any platform.
  $5000$ invocations per minute at IBM is the most restrictive; but it can be increased under request.

\medskip
With this, we have a summary of the four platforms and several aspects that heavily affect the parallelism they offer.
Combined with some reasoning, we can start to shape our expectations for the different services.
However, none of them guarantee these properties through a Service Level Agreement.
For example, the instance resources described in the documentation should be taken as approximations and the maximum parallelism as just an upper limit.
For this reason, we empirically assess these properties in the next sections.

\section{Experiment methodology}
\label{sec:experiment}

We explore the different architectures by empirically evaluating a real parallel workload.
For this, we design an experiment to show how multiple simultaneous function invocations are distributed across instances.
This validates the performance of parallel tasks on each FaaS platform.

The general methodology consists in running many concurrent requests to a function while gathering execution information.
We use this information to draw an execution \textit{chart}.
In particular, we plot parallelism clearly by depicting the invocations on a timeline that identifies function instances.

This section starts by setting up a set of questions that motivate the experiment.
Followed by a description of the test function and its different configurations, including a big scale setup.
We discuss several considerations regarding the execution of this evaluation and define a common notation for the experiment parameters.
Then, we present a set of metrics that characterize each platform for parallel computations.
The section ends with the description of the plot resulting from the experiment.

\subsection{Questions to answer}
\label{sec:experiment:questions}
The benchmark is designed with the following questions in mind, which define our goals for evaluation.
We analyze them on a per-platform basis through Sections~\ref{sec:experiment:aws} to~\ref{sec:experiment:ibm}.

\paragraph{Q1}
\emph{Does the service scale function instances elastically to fit parallel tasks?}
Related to the technology and scheduling approach of each platform, this question validates if the design is useful to reach parallelism in practice.
In essence, do concurrent invocations actually get different instances?
Coincidentally, we also identify the maximum parallelism achieved in practice, in contrast to the upper bound described before.

\paragraph{Q2}
\emph{Does the service ensure instance resources so that there is no interference across function invocations?}
This question validates the actual resources gotten on each platform and if there are any issues, such as resource interference, when running a parallel workload.
The objective is to verify the information about instance resources from the documentation (Section~\ref{sec:arch}).

\paragraph{Q3}
\emph{What can we deduce from the scheduling of the system and its general performance?}
This last question embraces general information that can be learned from the experiments.
Including:
invocation latency and how it changes with scale;
possible performance issues;
tendencies in cold starts;
insights on internal policies and tuning for resource management and scheduling;
and any other useful information.

\subsection{Function definition}
\label{sec:experiment:funcdef}
The questions above determine the information that we need to collect in our experiments.
Next, we detail the definition of the function that will run our benchmark to collect that data on the different FaaS platforms.
The function has two jobs: gathering information and performing work.
We obtain as much execution information as possible for each platform, which means different code and resulting plots.
Nonetheless, there are three basic items that we require:
\begin{inparaenum}[(1)]
  \item client-side execution times for each invocation,
  \item intra-function execution times (actual invocation duration), and
  \item function instance identification.
\end{inparaenum}
Client-side times can be acquired irrespective of the platform.
However, the other two items may be obtained differently on each service.
Function instance identifiers are never exposed by the services and we use different techniques to obtain them (detailed on their respective sections).
We complement the data by inspecting \texttt{/proc} when available, since it can offer valuable information about the virtualization level and system configuration.
Extended discussion on how to obtain execution information can be found in the literature~\cite{wang2018peeking,lloyd:investigation}.

As for workload, we experiment with two kinds of tasks:
a simple sleep and a compute-intensive job.
The sleep is a baseline to explore the scheduling pattern of the service.
We use a $1$-second sleep, which is enough to plot a comprehensive timeline, while longer tasks could complicate the information due to concurrency.
The compute task is intended to mimic a real embarrassingly parallel workload and reveal issues with resource availability and interference.
For easy reasoning, this task has a clearly-defined time duration.
In particular, we run a Monte Carlo simulation where an invocation performs $x$ iterations to approximate $\pi$.
$x$ is configured and evaluated to represent a consistent amount of time, close to $1$ second.

In detail, the function does the following:
\begin{inparaenum}[(1)]
  \item get the current time,
  \item identify invocation and instance,
  \item perform the workload,
  \item get the current time, and
  \item return the collected data.
\end{inparaenum}
We obtain the initial time right from the start to represent when user code starts to run in the cloud.
We checked that the overhead of the second step is consistent across invocations and not significant against the actual workload under test.

The invocations are run with a Python script that performs synchronous HTTP requests concurrently with \texttt{asyncio}.
We use the \texttt{httpx} client with the authentication methods required by each platform.
For AWS, we use the \texttt{aiobotocore} client: a simple wrapper for signed HTTP calls.
The information collected is complemented with client-side data and appended to a file, that is later used to draw the execution plot.

\subsection{Function configuration}
\label{sec:experiment:config}

\begin{table}
  \caption{Function configuration for the different platforms.}
  \label{tab:config}
  \centering
  \footnotesize
  \begin{tabular}{lccl}
     & Memory (MiB) & CPU & Region \\
    \toprule
    AWS   & $256$, $2048$ & $1/7$, $8/7$ CPU & \texttt{us-east-1}   \\
    Azure & $1536$        & $1$ CPU          & France Central       \\
    GCP   & $256$, $2048$ & $0.4$, $2.4$ GHz & \texttt{us-central1} \\
    IBM   & $256$, $2048$ & $1/8$, $1$ CPU   & Washington DC        \\
    \bottomrule
  \end{tabular}
\end{table}

Table~\ref{tab:config} summarizes the function configuration parameters for each platform.
The default timeouts on all platforms are enough for our one-second functions (5 min on Azure and 1 on the others).
We test two memory configurations to assess performance and resource management for different function sizes.
One (\textit{big} -- $2048$~MiB) intends to reach a full CPU on all platforms;
the other (\textit{small} -- $256$~MiB) is small enough to reveal the scheduling of the system.%
\footnote{
  This does not apply for Azure, since resources are not configurable.
}
In Table~\ref{tab:config}, we include the presumed CPU for each platform and memory configuration;
refer to Section~\ref{sec:arch} for details on memory and CPU mapping.
Regions are chosen based on what they offer (availability zones, better network, more services, etc.) to ensure best function performance and parallelism.
Different regions may affect request latency, but not the service parallelism we analyze.
The function is written in Python for all platforms but Azure (C\#), whose support for the language was in preview during the experiments.
This does not affect the benchmark since we execute $1$ second of computation on all platforms either way.
To that end, the compute-bound task performs $5$M iterations on Python and $20$M on C\#.
See Table~\ref{tab:compute} for a complete relation of task duration on each platform and configuration.
While different languages may affect cold start time, configuration is consistent for each execution and the parallelism in the plots is unaffected.
We take this into consideration when comparing across platforms.
All functions are triggered by HTTP requests and have logs and monitoring services active.

\begin{table}
  \caption{Compute-intensive task on each platform.}
  \label{tab:compute}
  \centering
  \footnotesize
  \begin{tabular}{lcccc}
     & Runtime & Iterations & 2048 MiB & 256 MiB \\
    \toprule
    AWS    & Python & 5M  & $1.1$~s & $7.7$~s    \\
    Azure  & C\#    & 20M & \multicolumn{2}{c}{$1.2$~s ($1.5$ GiB)%
    \tablefootnote{Since Azure does not allow resource configuration, we only show one time.}} \\
    GCP    & Python & 5M  & $1.3$~s & $3.5$~s%
    \tablefootnote{
      Results from cold starts.
      Warm containers are slower.
      See Section~\ref{sec:experiment:gcp}.
    } \\
    IBM    & Python & 5M  & $1.3$~s & $1.3$~s%
    \tablefootnote{We discuss IBM's equal performance for both configurations in Section~\ref{sec:experiment:ibm}.} \\
    \bottomrule
  \end{tabular}
\end{table}

\subsection{On a bigger scale}
\label{sec:experiment:big}
We want to confirm our conclusions by assessing large scale executions of the benchmark.
Our detailed plot (Section~\ref{sec:experiment:plot}) becomes too noisy for analysis when targeting such configurations.
For this reason, we complement our results with an extra execution of $1000$ invocations that uses a simplified plot.
This plot includes the function execution time bars in a timeline together with a curve representing the number of function instances running at each instant, showing the evolution of the experiment concurrency.
In addition, we add a complementary histogram of the invocation execution time that helps identify resource interference between invocations.

With that many invocations, synchronous HTTP requests are inconvenient for parallel executions, so we opt for asynchronous invocations instead.
This difference may result in different strategies for the platforms to scale resources and we will keep this in mind when analyzing these executions.
In any case, the results are in line with the tendencies observed in the more detailed experiments, which tells us that the invocation method may not affect parallel performance.

This experiment runs a CPU-intensive task.
Specifically, each task computes several matrix multiplication calculations that last around a minute in total.
Function memory is fixed to $1024$~MiB for each FaaS service.
This implies that the portion of CPU assigned to a function varies between cloud providers.
The scale is $1000$ invocations of this task.
This means that the same workload performed on a single core would take approximately $16$ hours.
In this case, the use of asynchronous triggers requires the result to be sent to the object storage available in the cloud provider (e.g., S3 for AWS Lambda) and retrieved from the client after completion.
Note that the times displayed in the plot only represent the function execution time and not the overhead produced in uploading the result to object storage.

\subsection{Experiment execution}
\label{sec:experiment:exec}
The benchmark was run during May 2020 from a single client machine (a laptop with an $8$-core CPU @$2.6$~GHz) invoking functions to the different platforms.
The consequent invocation latency reflects only in the time between the client timestamp and the function start, and thus does not influence the display of parallelism in the plots.

Executions were run during different days and hours.
All configurations were tested several times and all showed similar results.
The complex nature of the plots (detailing a single execution to show its work distribution) makes it difficult to show all the data in an aggregated format that is readable and informative.
Therefore, we selected some of the executions to give the reader the general idea of the behavior of each platform.

The executions may find arbitrary numbers of warm and cold instances as they are executed in succession.
This is because warm starts depend on the platform and its particular policies for recycling instances.
Consequently, it is not possible to ensure a consistent number of warm instances across executions.
However, we can collect this information afterwards and compare it with the number of instances available in the previous execution.
Since we consider it important for evaluating parallelism, we include data on the number of warm starts experienced on each execution.
For example, when running an execution first with $10$ invocations and then with $50$, if the platform creates $10$ instances for the first run, the second one is expected to usually find $10$ instances warm.

To account for all the different configurations and system state, we use a simple notation system throughout the evaluation to describe the complete setup of each experiment.
The notation is: $I/W/T/M$.
Where $I$ is the number of invocations in that experiment, $W$ is the expected number of warm instances staying from a previous execution, $T$ is the workload type for the function ($S$--sleep or $C$--compute), and $M$ is the memory size for the functions ($s$--\textit{small} or $b$--\textit{big}, as introduced above).
For example, the notation $200/50/C/b$ indicates an execution with $200$ invocations, expecting $50$ warm instances, and performed the compute-intensive task on \textit{big} ($2048$~MiB) functions.

\subsection{Metrics}
\label{sec:experiment:metrics}
To summarize the results of our benchmark, we establish the following metrics that characterize the capabilities of the different FaaS platforms to host parallel computations:

\paragraph{Cold start}
Instance creation overhead is a direct result of the virtualization technology and the scheduling approach.
Other benchmarks~\cite{wang2018peeking,serverlessbench,faasdom} show that the cold start depends on the function runtime configuration and analyze it in detail.
We do not consider our values for cross-platform comparison due to different latencies to each cloud.
Hence we only point out general tendencies and its effects to the system in its behavior.

\paragraph{Completion time} 
This is a good indicator of the achieved parallelism, and specially of the simultaneity of invocations.
With this metric we quantify approximately how long it takes to run $200$ \textit{big} compute tasks on each platform.
Each task individually takes one second.
Hence a perfect system would run any number of this task within that second.
However, platforms add overhead to the execution, such as invocation delay.

\paragraph{Parallel degree} 
We define the \emph{parallel degree of a platform} in an experiment as the maximum number of instances used at the same time throughout the experiment.
We also include the percentage that this represents out of the total number of invocations.
We account this for the same setup as the previous metric, so $100\%$ parallelism means the use of $200$ instances at the same time.

\paragraph{Failed requests}
These are a hassle for parallelism, as they become stragglers, need retrying, and heavily impact total computation time.
With synchronous invocations, like our case, the platform delegates retries to the caller, making the process slower and increasing complexity for the user.

\subsection{Plot description}
\label{sec:experiment:plot}
The information gathered by all invocations in an experiment is represented in a Gantt-like plot showing the execution period of each function invocation in a global timeline of the run.
Using different colors, the plot shows on which function instance each invocation has run.
This allows to see the real parallelism achieved and spot concurrency problems (like per-instance concurrency or invocation throttling).

In the plots, the horizontal axis is the timeline.
Our time zero is the minimum timestamp playing in the experiment: the first client invocation (red \texttt{X}).
All other times are deltas to this one.
The vertical axis stacks the function invocations.
Each invocation is drawn as a horizontal bar indicating its time-span, i.e. the time it has been running in the cloud.
The yellow \texttt{X}s indicate the client-side invocation timestamp and the black ones, the request return.
Bar colors differentiate the instance where each invocation has been run.
Although colors are limited to four, since the plot groups invocations by instance, instances sharing a color are always separated by instances with other colors, making the distinction clear.

\begin{figure}
  \centering
  \begin{subfigure}[b]{0.48\linewidth}
    \includegraphics[width=\linewidth]{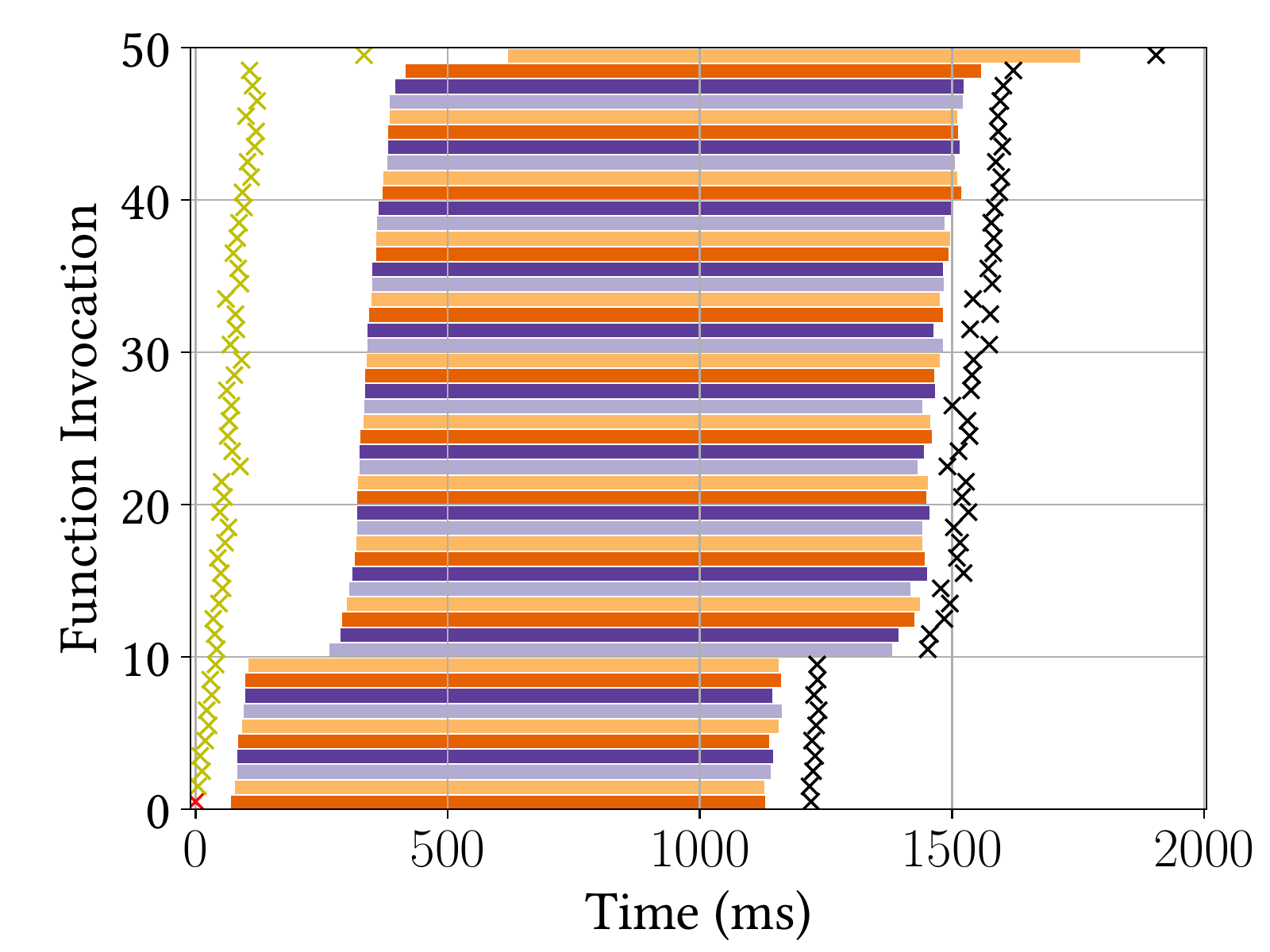}
    \caption{$50/10/S/s$}
    \label{fig:aws:sleep:50}
  \end{subfigure}
  ~
  \begin{subfigure}[b]{0.48\linewidth}
    \includegraphics[width=\linewidth]{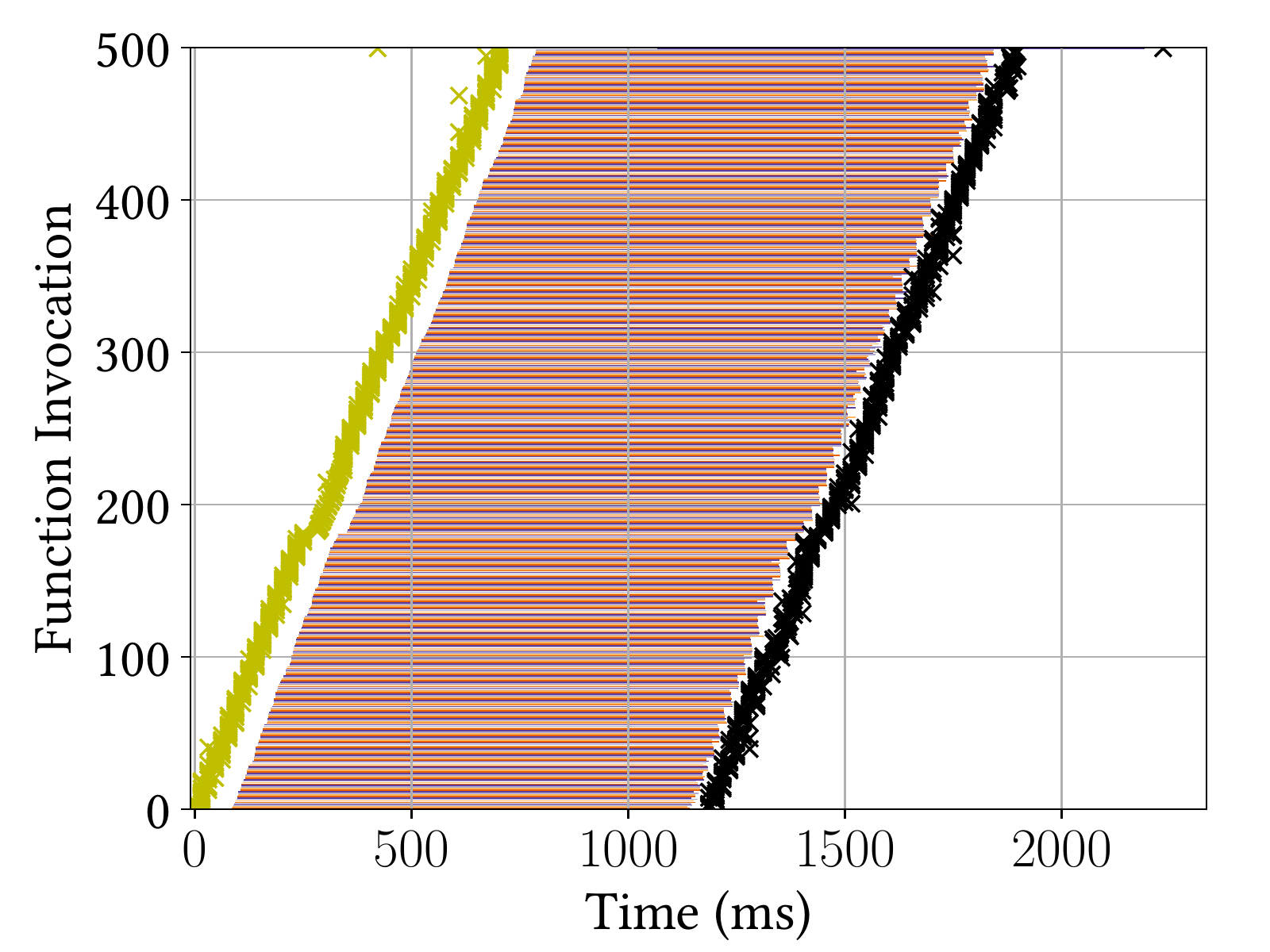}
    \caption{$500/500/S/s$}
    \label{fig:aws:sleep:500}
  \end{subfigure}

  \begin{subfigure}[b]{0.48\linewidth}
    \includegraphics[width=\linewidth]{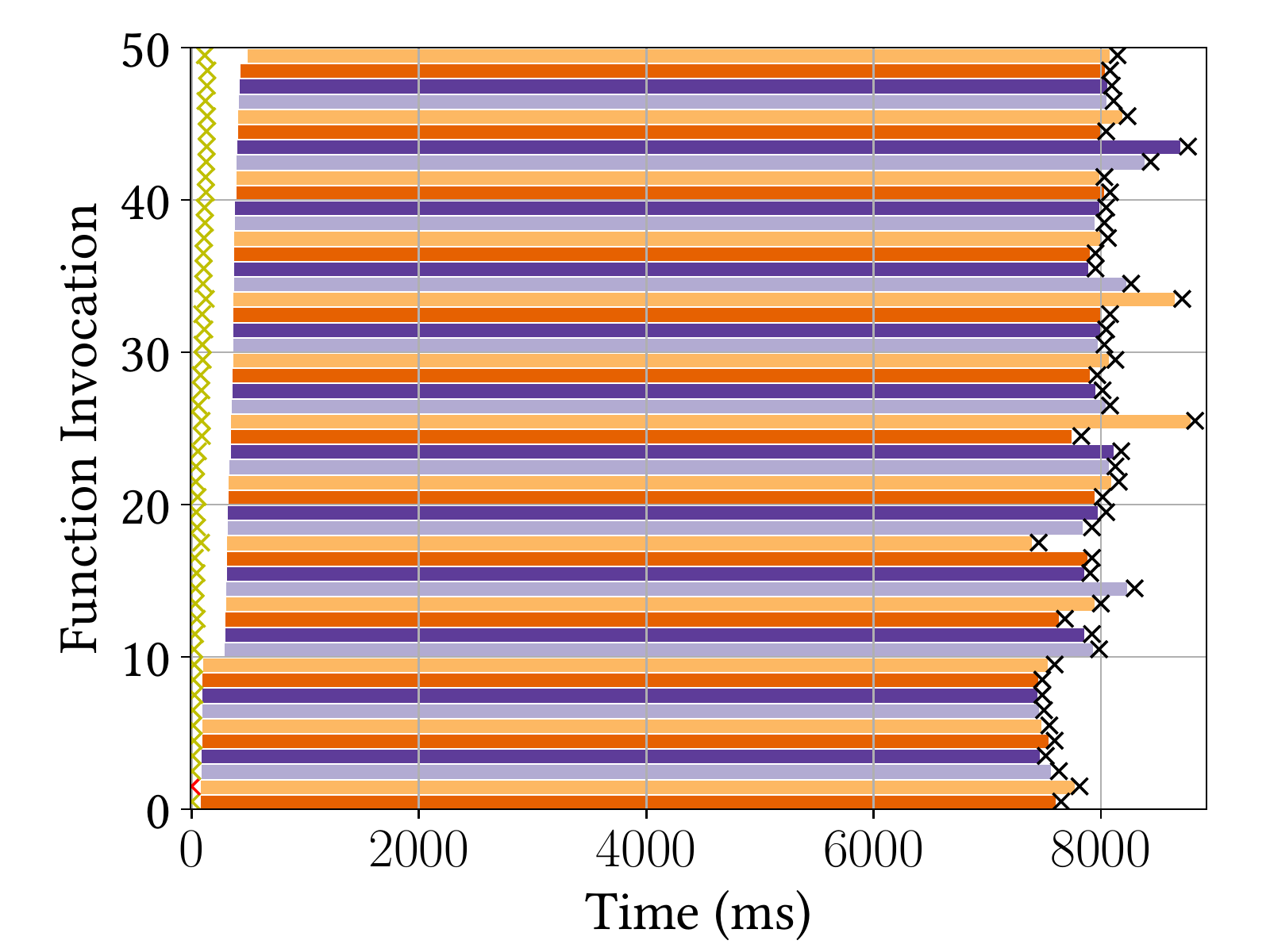}
    \caption{$50/10/C/s$}
    \label{fig:aws:work:50}
  \end{subfigure}
  ~
  \begin{subfigure}[b]{0.48\linewidth}
    \includegraphics[width=\linewidth]{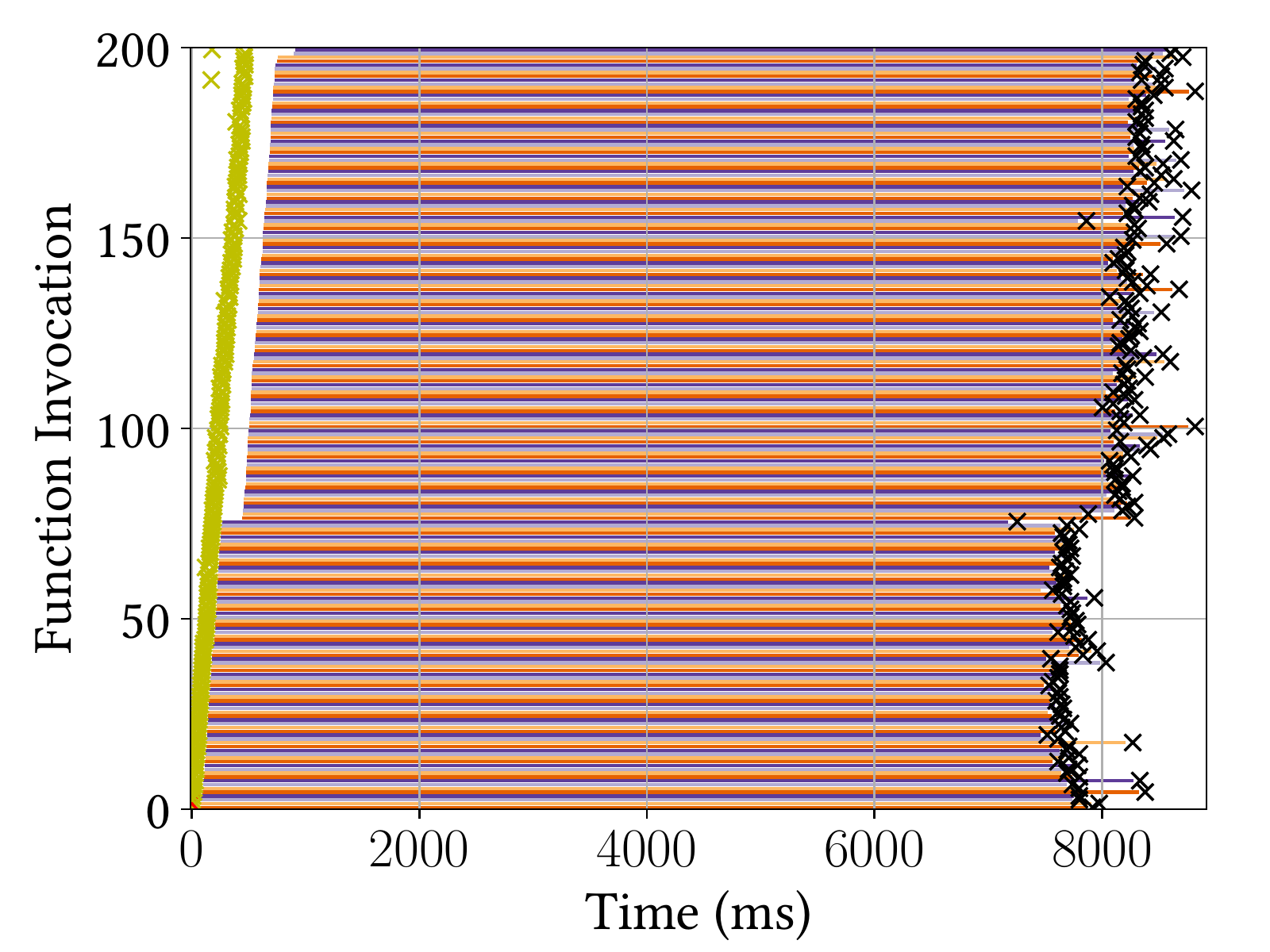}
    \caption{$200/100/C/s$}
    \label{fig:aws:work:200}
  \end{subfigure}

  \begin{subfigure}[b]{0.48\linewidth}
    \includegraphics[width=\linewidth]{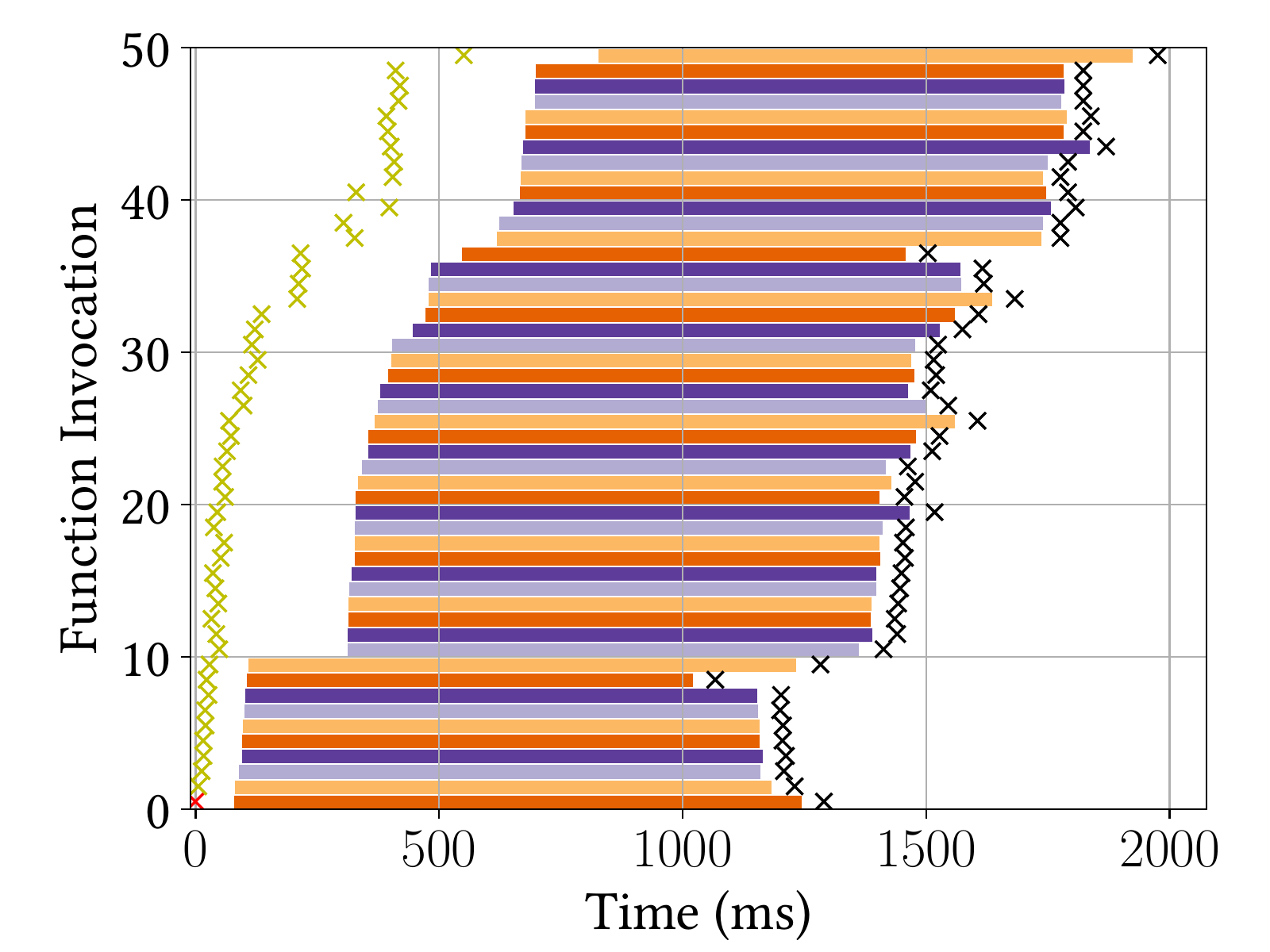}
    \caption{$50/10/C/b$}
    \label{fig:aws:work2g:50}
  \end{subfigure}
  ~
  \begin{subfigure}[b]{0.48\linewidth}
    \includegraphics[width=\linewidth]{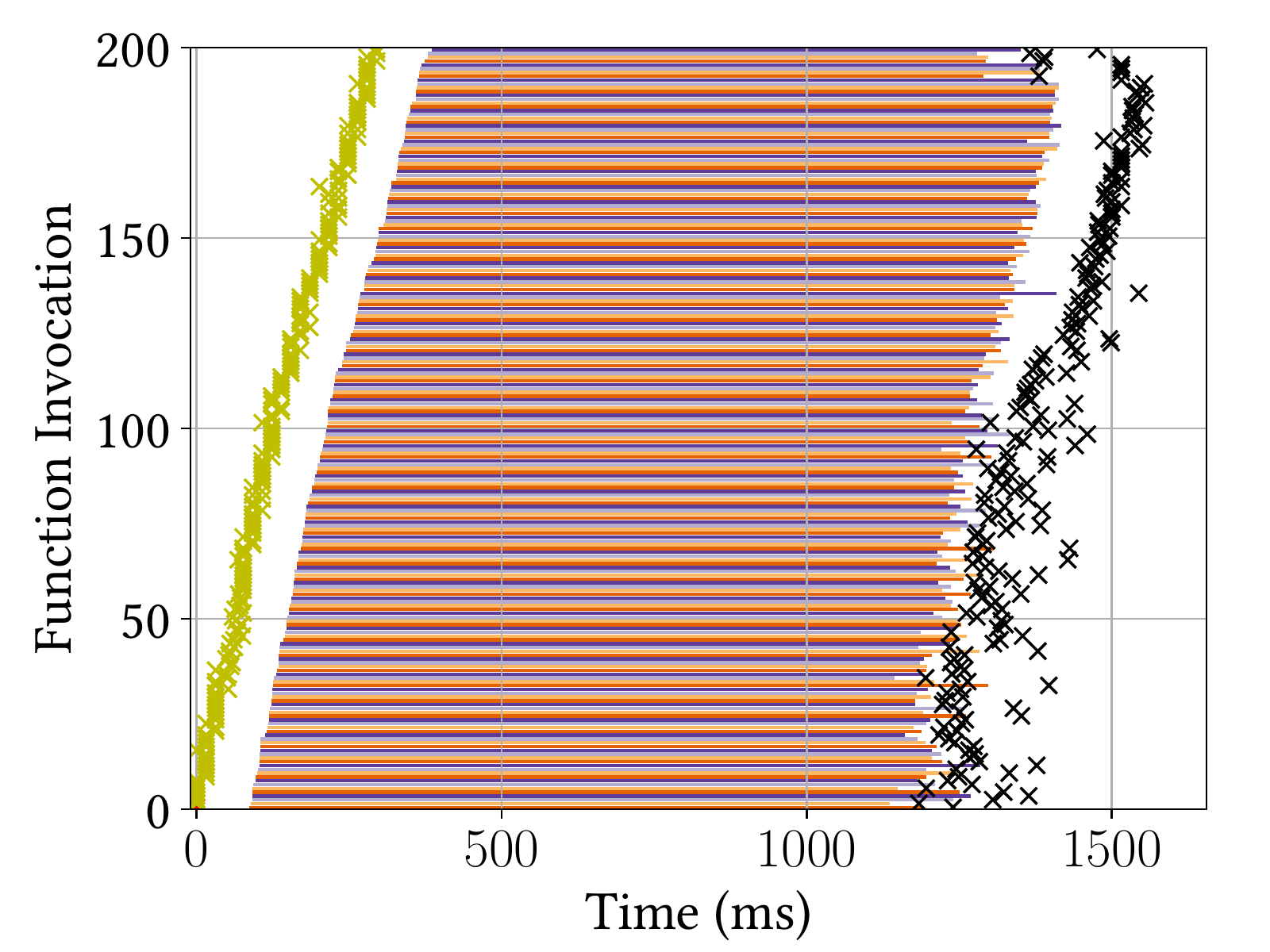}
    \caption{$200/200/C/b$}
    \label{fig:aws:work2g:200}
  \end{subfigure}

  \caption{
    Experiment on AWS.
  }
  \label{fig:aws}
\end{figure}

\section{Experiments on Amazon Web Services}
\label{sec:experiment:aws}
We deploy and update our function with the AWS CLI.
The invocation ID is obtained through the function context object.
The instance ID is the randomly generated identifier present at \texttt{/proc/self/cgroup}, starting with \texttt{sandbox-root}~\cite{wang2018peeking}.

\subsection{Results}

\paragraph{Experiments with sleeping functions}
We start with the \textit{small} ($256$~MiB) functions and the sleeping task.
A first run with $10/0/S/s$ shows how the system creates a different container for each invocation, allowing full parallelism.
A subsequent execution with $50/10/S/s$ results in Figure~\ref{fig:aws:sleep:50}.
Note that the cold start increases invocation latency by $\approx 200$~ms.
Still, the service achieves full parallelism.
Figure~\ref{fig:aws:sleep:500} shows $500/500/S/s$; the service still creates different containers for each invocation.



\paragraph{Experiments with computing functions}
Still with \textit{small} functions, we move to the compute-intensive task.
Running a single invocation, the computation takes $7.7$ seconds average with this configuration.
Figures~\ref{fig:aws:work:50} and~\ref{fig:aws:work:200} show subsequent invocations of this experiment with different parallelism.
The variance of execution time is within one second.

With the \textit{big} functions, which have a full CPU, execution time for the individual run reduces to $1.1$ seconds average.
Figures~\ref{fig:aws:work2g:50} and~\ref{fig:aws:work2g:200} show the experiment with different parallelism.
Execution time is never far from the individual execution with a bit more variance than with the \textit{small} configuration.

\begin{figure}
  \centering
  \begin{subfigure}[b]{0.48\linewidth}
    \includegraphics[width=\linewidth]{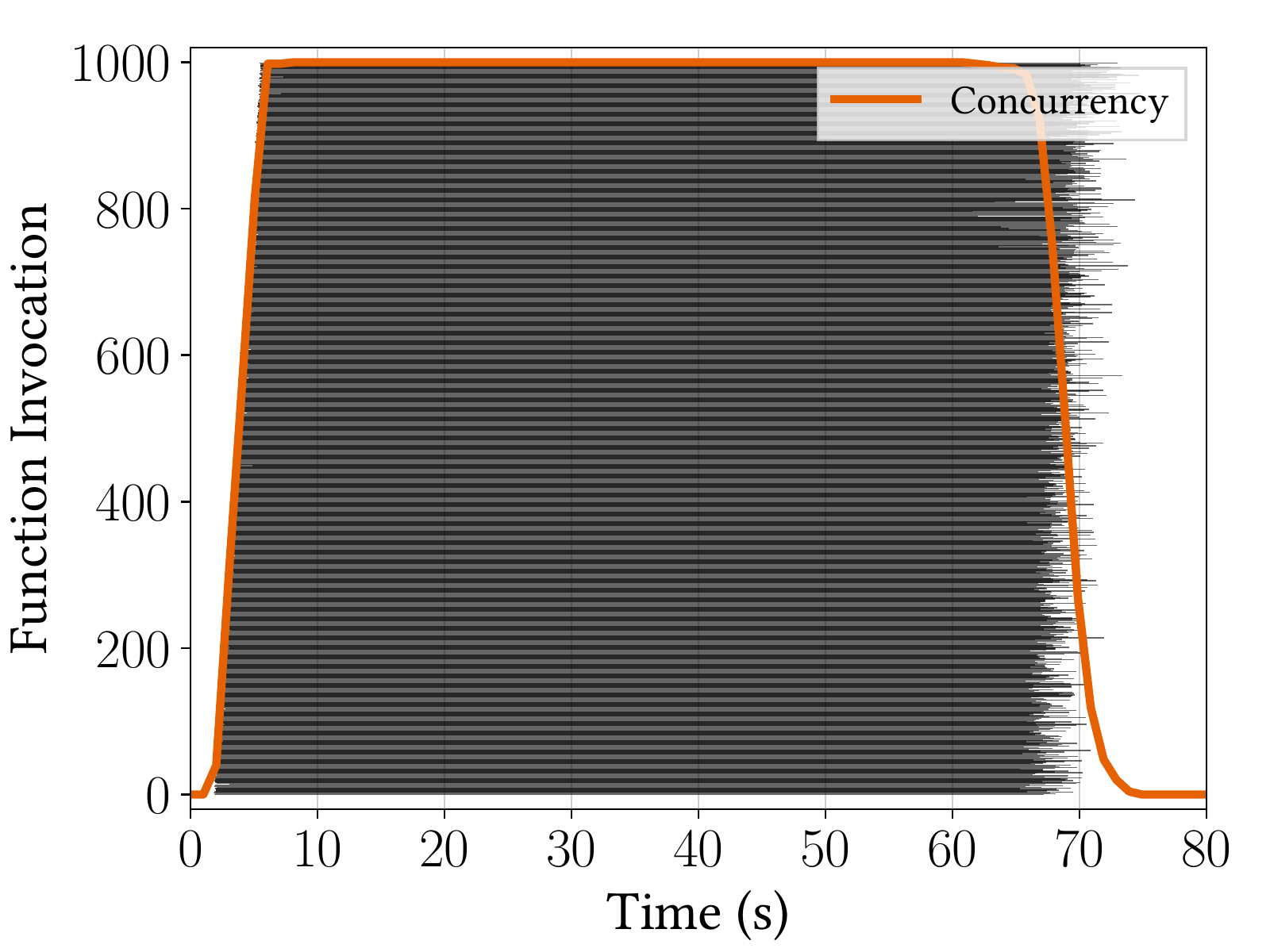}
    \caption{}
    \label{fig:aws:big:time}
  \end{subfigure}
  ~
  \begin{subfigure}[b]{0.48\linewidth}
    \includegraphics[width=\linewidth]{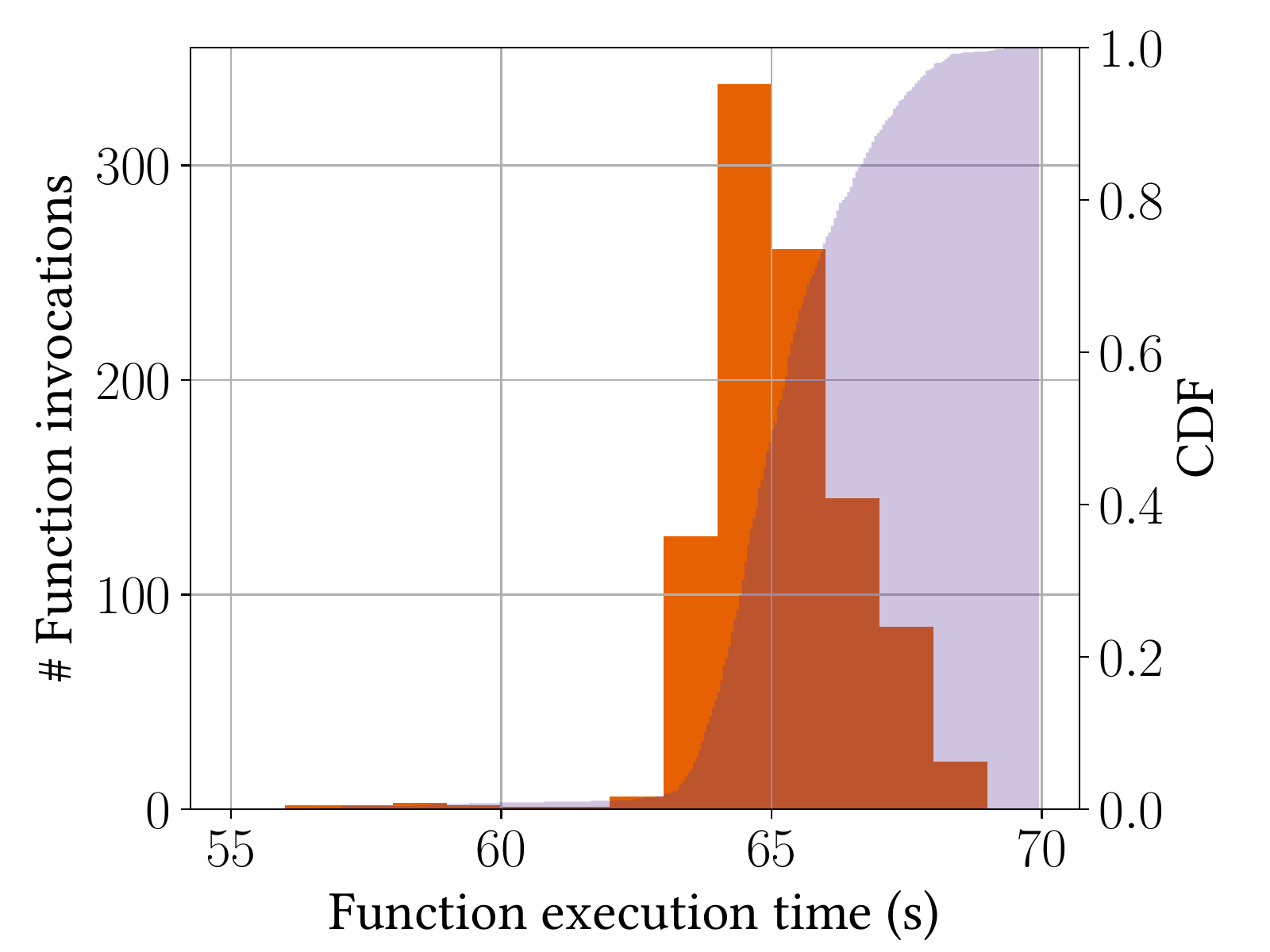}
    \caption{}
    \label{fig:aws:big:hist}
  \end{subfigure}
  \caption{
    Large-scale experiment on AWS.
  }
  \label{fig:aws:big}
\end{figure}

\paragraph{On a bigger scale}
Figure~\ref{fig:aws:big} shows the results of executing the larger configuration with $1000$ parallel requests.
Full parallelism is fulfilled even for big scale executions on AWS Lambda.
The histogram shows that all invocations do not vary much from around $65$~s run time, confirming resource homogeneity.

\subsection{Answers to questions}

\paragraph{Q1}
All experiments show good parallelism, scaling rapidly to the number of requests.
The overhead is small and all invocations run in different instances at the same time.
In particular, the experiment with $500$ concurrent requests shows that the server can keep dealing invocations at the pace the client is able to create.
Our larger configuration confirms that AWS Lambda scales to thousands with asynchronous invocations~\cite{PyWren2017}.

\paragraph{Q2}
CPU resources scale with memory as documented~\cite{lambda:config}.
Function performance is constant with little variance (i.e., there is no interference).
This suggests that provisioning and isolation are strict, not only for memory but also for other resources.
We can clearly see this from the compute-intensive experiments.
Our task takes $\approx 1.1$~s with the full CPU (\textit{big} functions) and $\approx 7.7$~s with the \textit{small} functions.
Since a full CPU is reached at $1792$~MiB, our $256$~MiB functions are $7$ times smaller and should have $1/7$ of CPU.
Accordingly, our \textit{small} functions take $7$ times more than the \textit{big} ones.

\paragraph{Q3}
The experiments also reveal these conclusions:
\begin{inparaenum}[i)]
  \item Scheduling allows generous resource allocation in burst. Containers are immediatly created when none are available.
  \item Instances are set up for processing quite fast, probably a result of using a microVM technology.
  \item Even with cold starts, invocation latency is usually below $300$~ms, including client-cloud latency.
\end{inparaenum}

\section{Experiments on Microsoft Azure}
\label{sec:experiment:azure}
The development and deployment of Function Apps is managed with the Visual Studio Code extensions, as recommended in the documentation.\cite{azure:quickstart}
The invocation ID is obtained through the function context object available as an optional function parameter (inherited from WebJobs).
Fot the instance ID, an environment variable (``\texttt{WEBSITE\_INSTANCE\_ID}") is present from Azure WebJobs and identifies a function instance~\cite{azure:instanceid}.
We also use Live Metrics, an Azure service that shows real time detailed information for a Function App, such as the number of active servers (instances), or CPU and memory usage, among others.

\subsection{Results}

\begin{figure*}
  \centering
  \begin{subfigure}[b]{0.24\linewidth}
    \includegraphics[width=\linewidth]{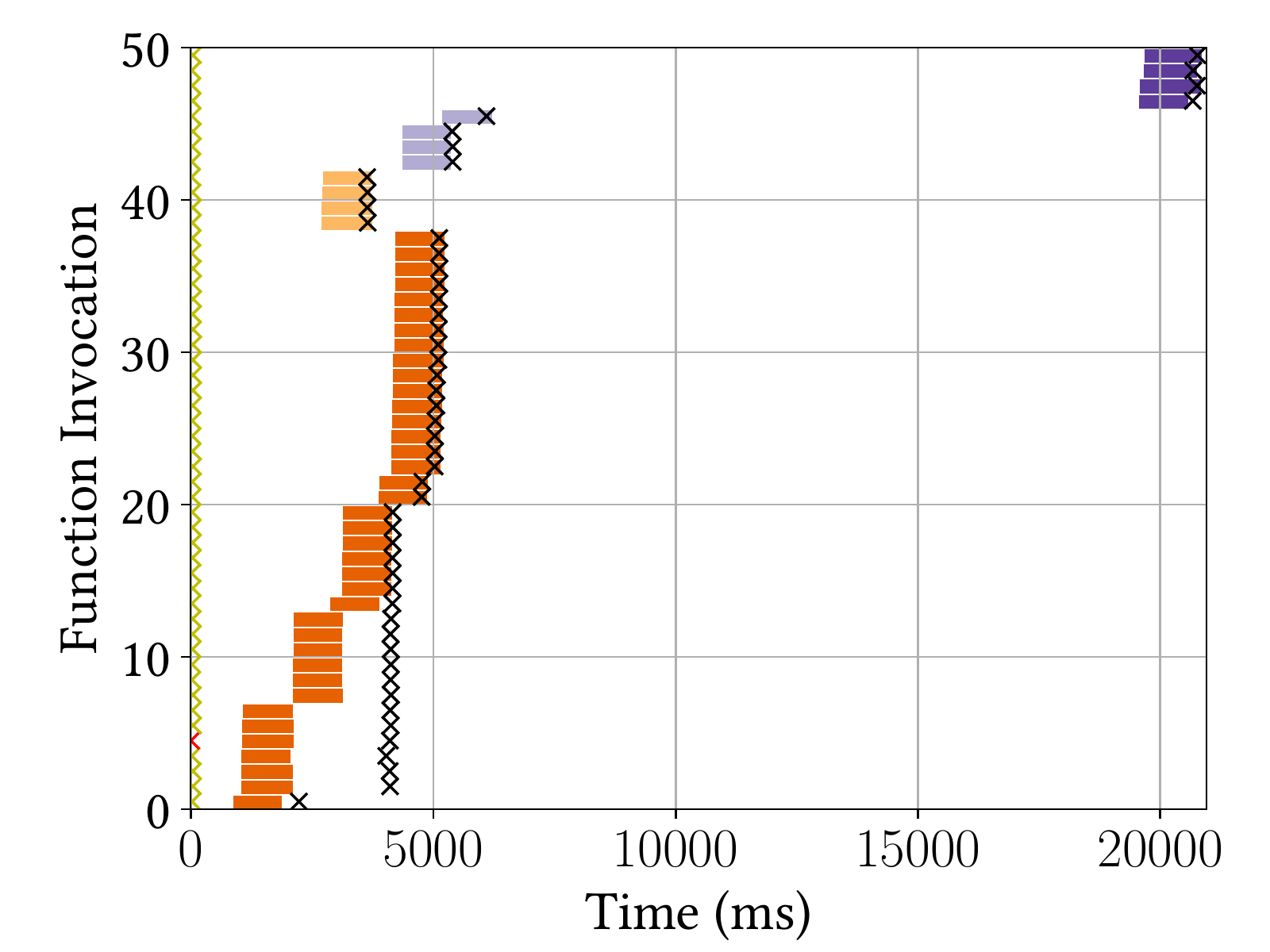}
    \caption{$50/0/S/-$}
    \label{fig:azure:basic:50:cold}
  \end{subfigure}
  ~
  \begin{subfigure}[b]{0.24\linewidth}
    \includegraphics[width=\linewidth]{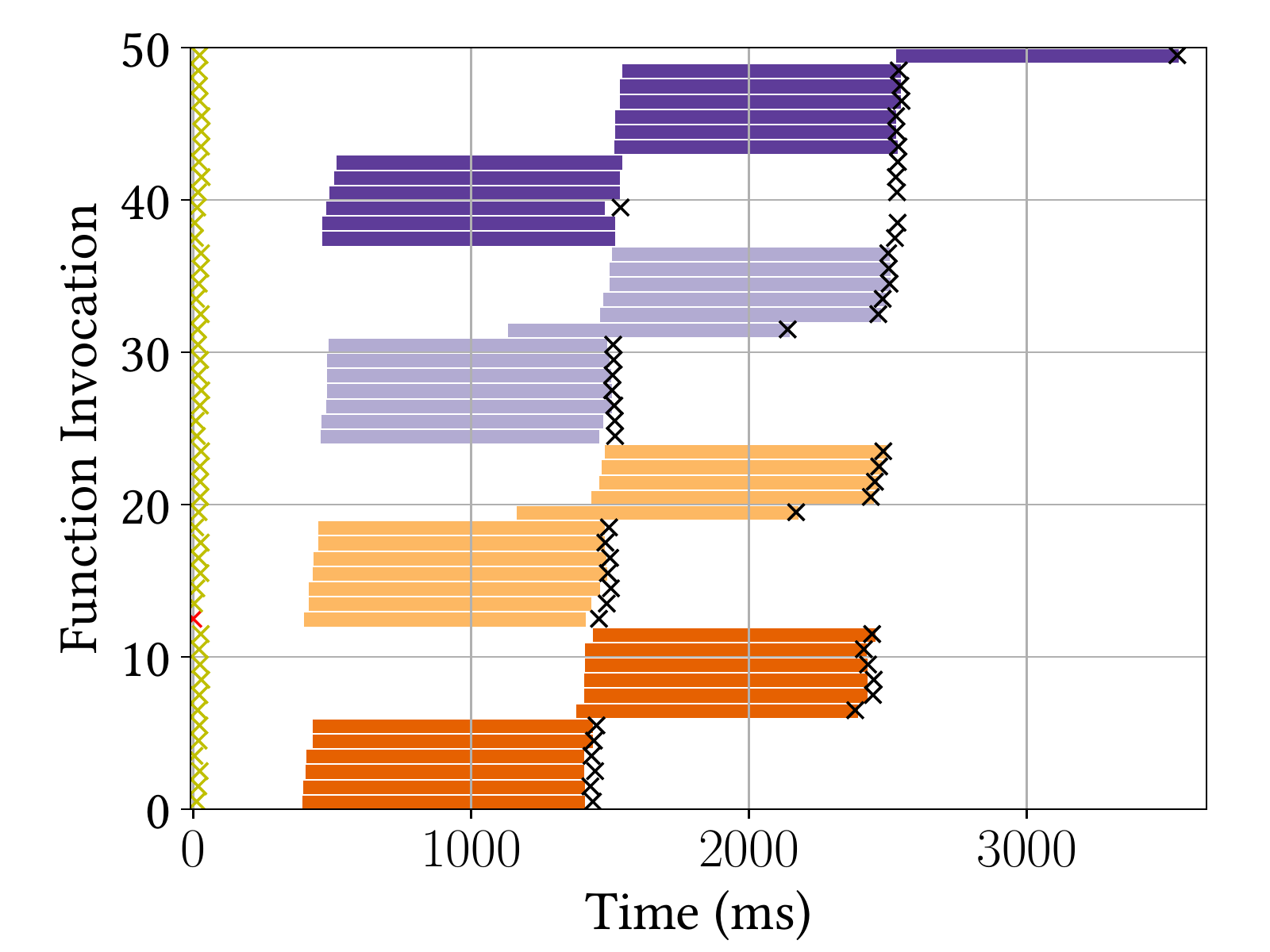}
    \caption{$50/4/S/-$}
    \label{fig:azure:basic:50:warm}
  \end{subfigure}
  ~
  \begin{subfigure}[b]{0.24\linewidth}
    \includegraphics[width=\linewidth]{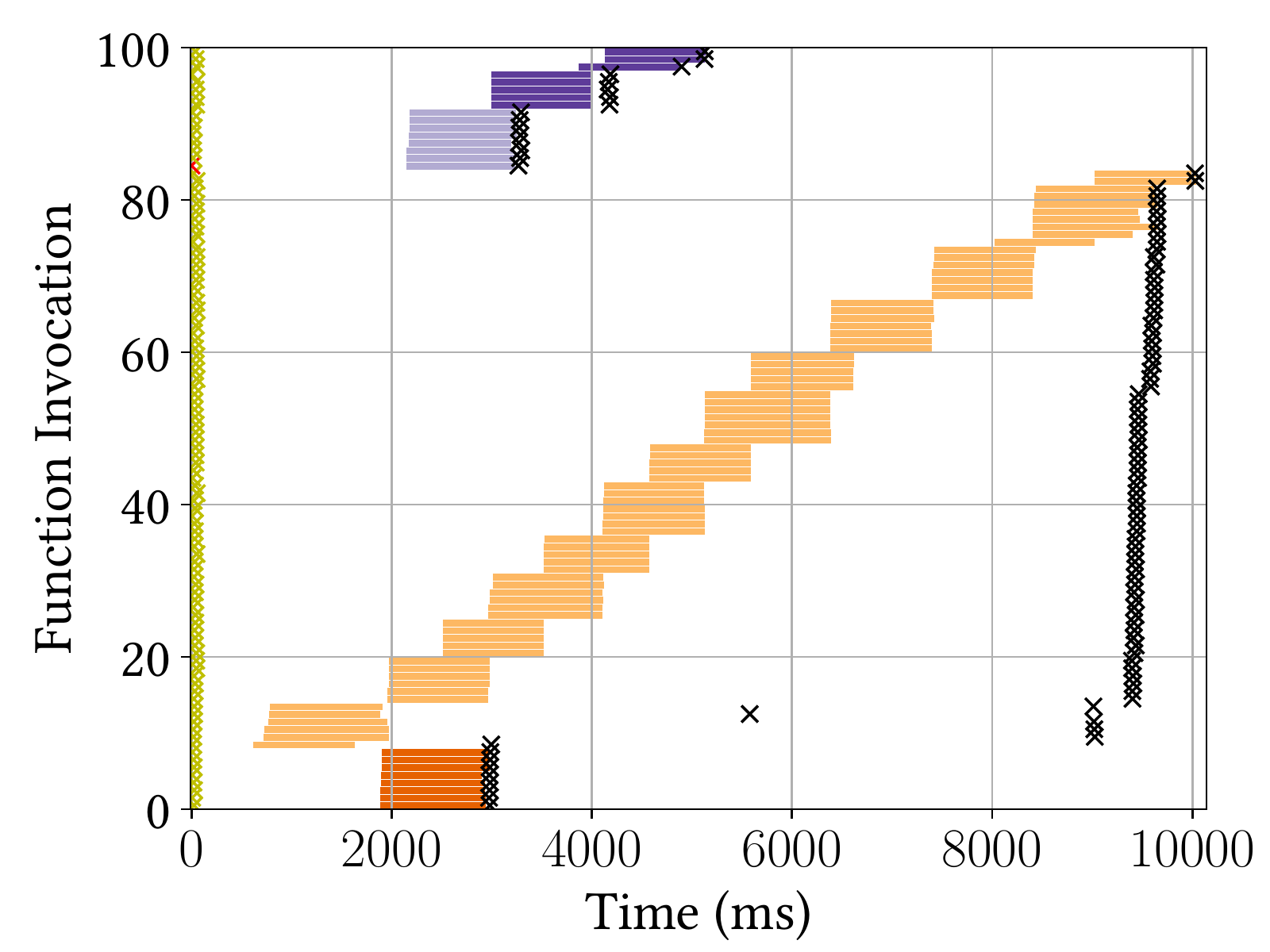}
    \caption{$100/0/S/-$}
    \label{fig:azure:basic:100:1}
  \end{subfigure}
  ~
  \begin{subfigure}[b]{0.24\linewidth}
    \includegraphics[width=\linewidth]{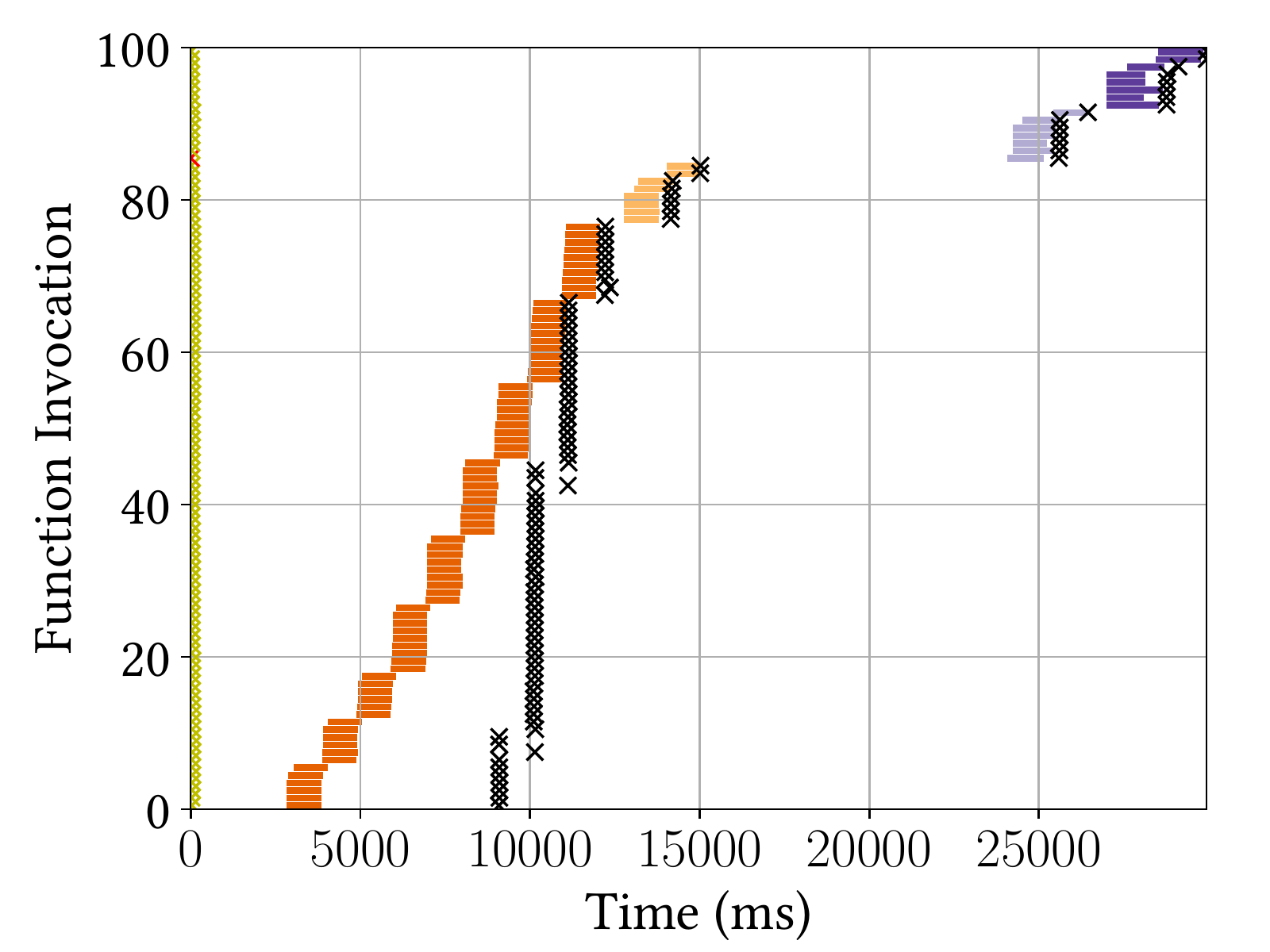}
    \caption{$100/0/S/-$}
    \label{fig:azure:basic:100:2}
  \end{subfigure}


  \begin{subfigure}[b]{0.24\linewidth}
    \includegraphics[width=\linewidth]{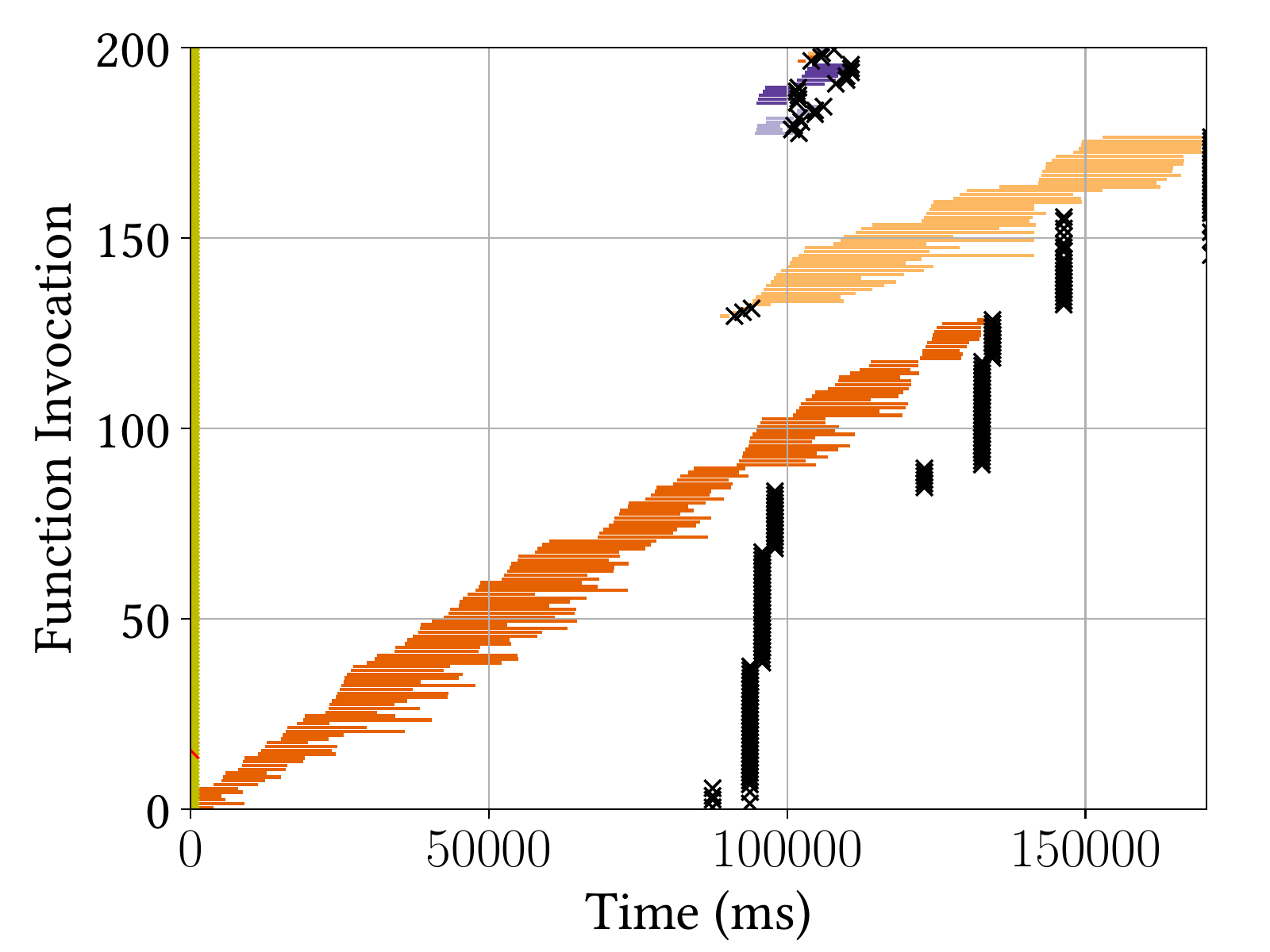}
    \caption{$200/1/C/-$}
    \label{fig:azure:basic:work:200:cold}
  \end{subfigure}
  ~
  \begin{subfigure}[b]{0.24\linewidth}
    \includegraphics[width=\linewidth]{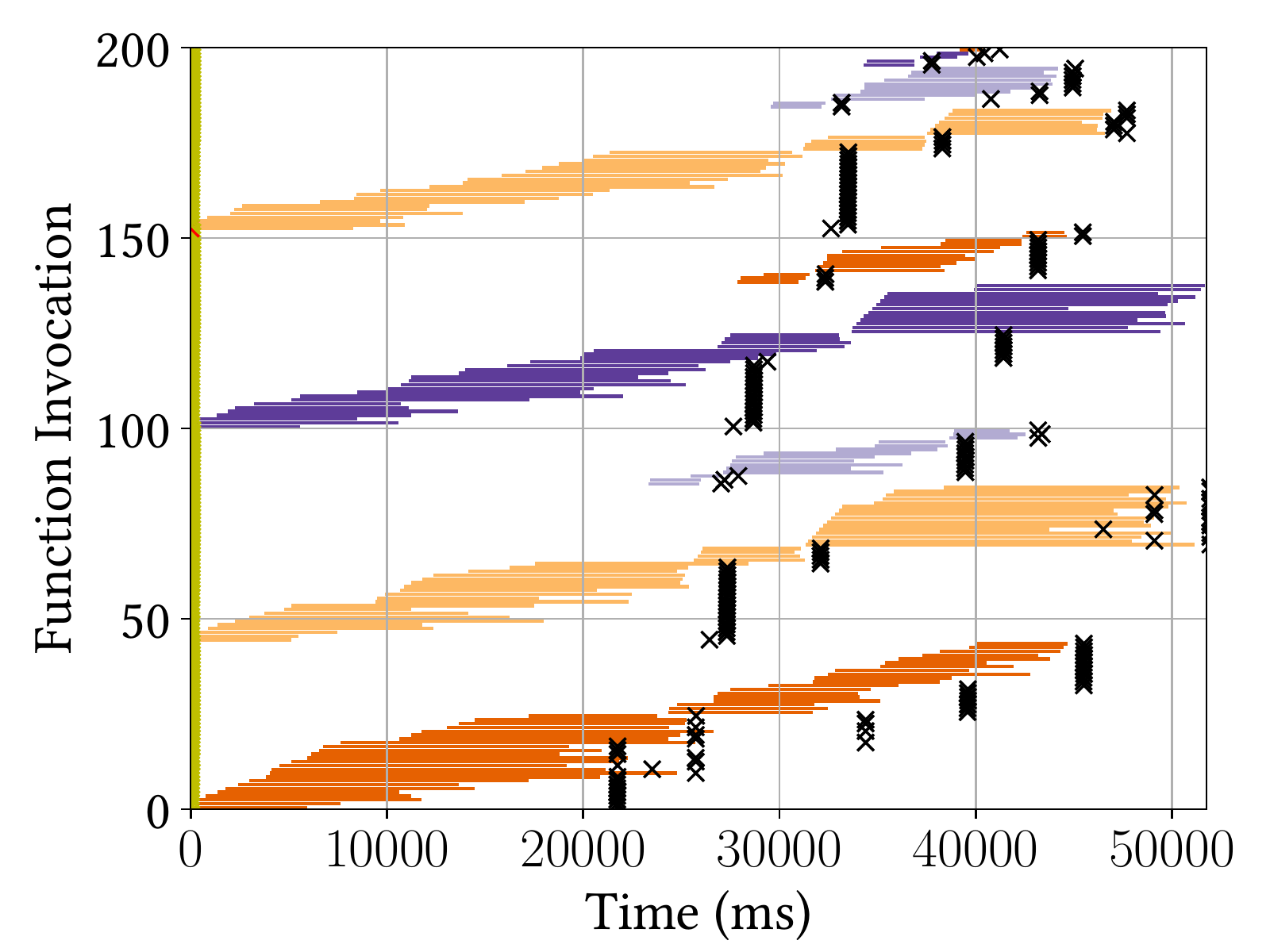}
    \caption{$200/4/C/-$}
    \label{fig:azure:basic:work:200:warm}
  \end{subfigure}
  ~
  \begin{subfigure}[b]{0.24\linewidth}
    \includegraphics[width=\linewidth]{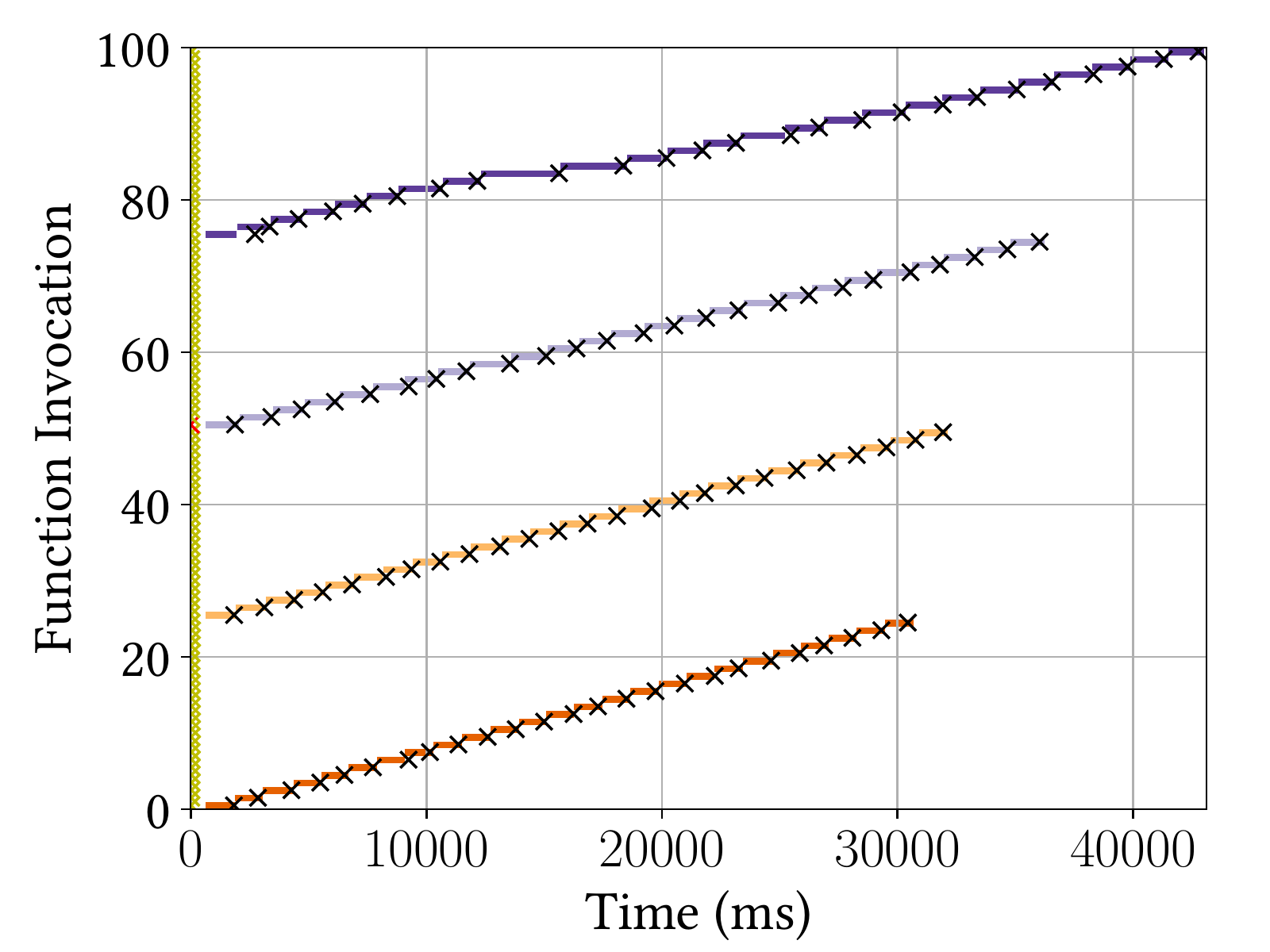}
    \caption{$100/4/C/-$}
    \label{fig:azure:limit1:100}
  \end{subfigure}
  ~
  \begin{subfigure}[b]{0.24\linewidth}
    \includegraphics[width=\linewidth]{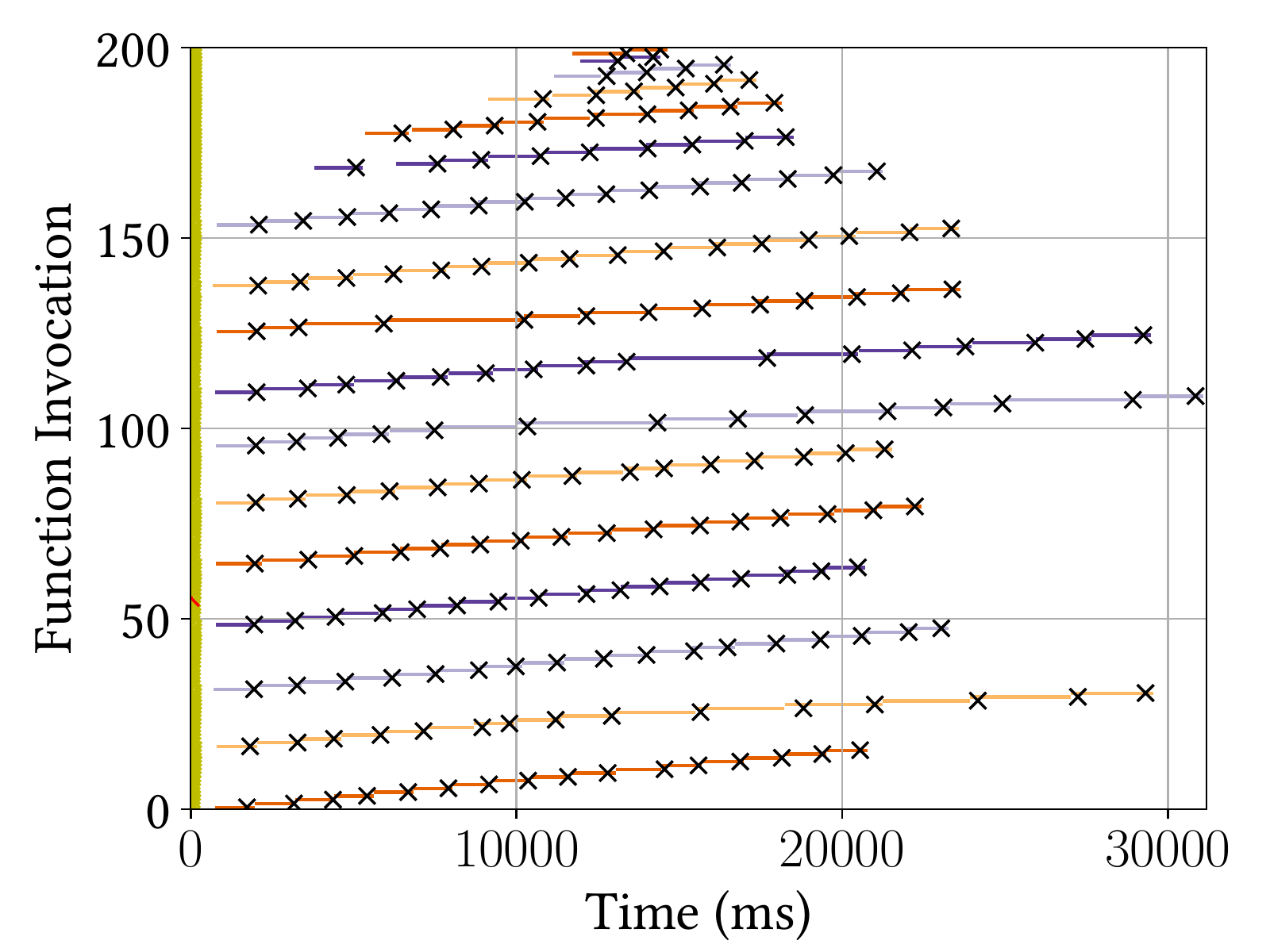}
    \caption{$200/13/C/-$}
    \label{fig:azure:limit1:200}
  \end{subfigure}

  \caption{
    Experiment on Azure.
  }
  \label{fig:azure:basic}
\end{figure*}

\paragraph{Experiments with sleeping functions}
We start with the default configuration and the sleeping task.
A first execution with $50$ parallel requests results in Figure~\ref{fig:azure:basic:50:cold}, which shows a cold start.
With this run, the service ends with $4$ instances.
A subsequent execution of the same experiment results in Figure~\ref{fig:azure:basic:50:warm}.
In this case, the $4$ hosts where already running, and start processing invocations right away.
Figures~\ref{fig:azure:basic:100:1} and \ref{fig:azure:basic:100:2} show the same experiment with $100$ parallel requests, both without previously running instances.
They demonstrate that two executions with the same parameters can be scaled differently in this platform.

\paragraph{Experiments with computing functions}
Now, we switch to the compute-intensive tasks.
Running $50$ or $100$ parallel requests do not get more than a single instance, thus we increase the workload.
Repeating the same experiment ($50$ and $100$ parallel requests) several times in quick succession does not alter results.
We then run $200$ parallel requests, which results in Figure~\ref{fig:azure:basic:work:200:cold}.
The system finally creates new instances (up to $7$ as confirmed with Live Metrics).
Right away, we run the experiment again, which we plot in Figure~\ref{fig:azure:basic:work:200:warm}.
We see that requests only run on $4$ instances at first, but then scale out to $9$.
In this case, Live Metrics tells us that the service created up to $12$ servers, but some of them did not get any work.

\paragraph{Limiting function invocation concurrency per instance}
Since the default configuration is a bad fit for compute-intensive tasks, we run the experiments limiting per-instance invocation concurrency as explained in Section~\ref{sec:arch:azure}.

Due to the CPU-intensive nature of our tasks, our experiment benefits from limiting concurrency to $1$ to avoid resource interference.
Now invocations take the expected time ($\approx 1.2$~s).
In the previous executions, resource sharing was extending execution time by $40$x.
Like before, with larger experiments the system does not create more than $4$ instances (Figure~\ref{fig:azure:limit1:100}) until reaching $200$ concurrent requests.
For instance, Figure~\ref{fig:azure:limit1:200} shows an execution where $13$ instances were already up and ends with $18$ servers processing invocations.
As a note, this last experiment runs $200$ tasks (embarrassingly parallel), each of them with an expected duration of $1.2$ seconds.
Such computation should take $1.2$ seconds plus some system overhead (all tasks are parallel).
However, the whole experiment takes more than $30$ seconds with a maximum parallelism of $18$.

\paragraph{Revisited}
Due to the poor parallelism experienced, we decide to revisit this experiment on March 2021.
The configuration is the same but for the region of deployment.
Since the experiments on the other platforms were performed on US regions, we move to ``Central US''.
This way, we discard the datacenter from causing this problems and avoid peak hours on that region in case heavy traffic of other users may have affected performance.
However, we find the same behavior experienced months before.
Indeed, the low parallelism seems related to the Scale Controller component and its policies for spawning new instances and not to the load in a specific datacenter.

\begin{figure}
  \centering
  \begin{subfigure}[b]{0.48\linewidth}
    \includegraphics[width=\linewidth]{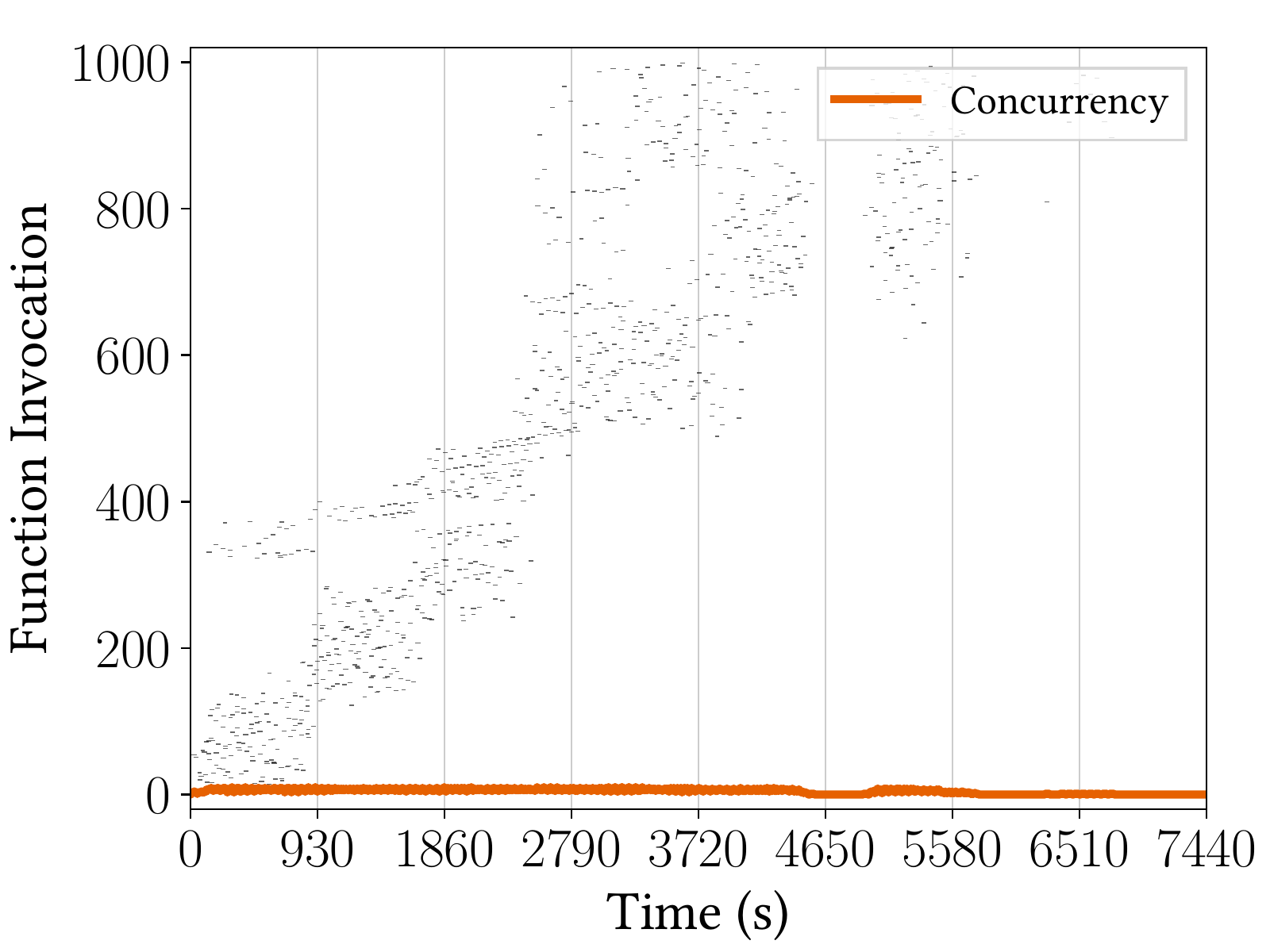}
    \caption{}
    \label{fig:azure:big:time}
  \end{subfigure}
  ~
  \begin{subfigure}[b]{0.48\linewidth}
    \includegraphics[width=\linewidth]{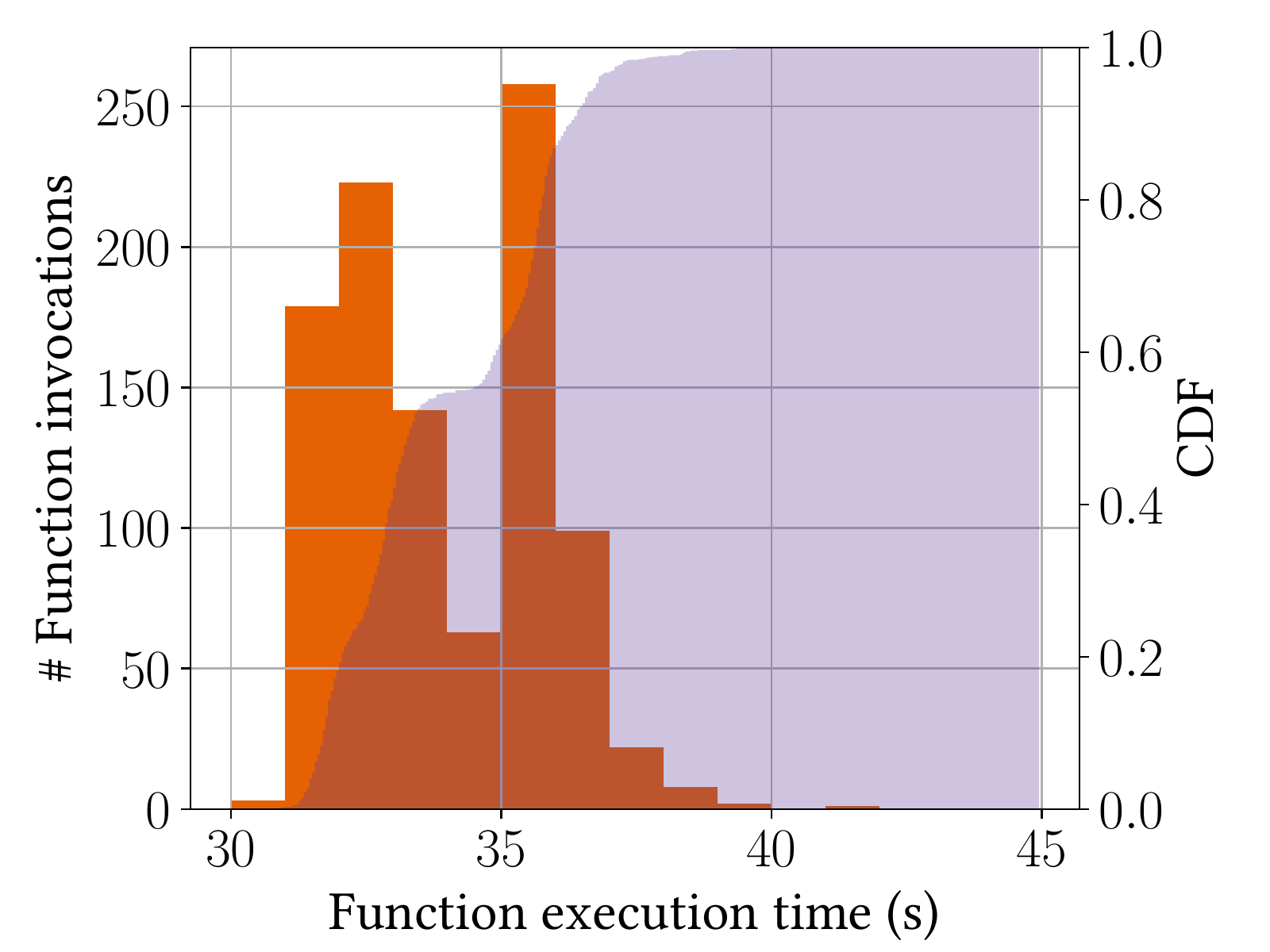}
    \caption{}
    \label{fig:azure:big:hist}
  \end{subfigure}
  \caption{
    Large-scale experiment on Azure.
  }
  \label{fig:azure:big}
\end{figure}

\paragraph{On a bigger scale}
Figure~\ref{fig:azure:big} shows the results of executing the larger configuration with $1000$ parallel requests.
In line with the previous runs, concurrency is very limited with just a few instances, affecting the total execution time.
The histogram shows fairly consistent run times, meaning that, when limiting per instance concurrency, the resources for each invocation are well ensured.
Azure, having always a full vCPU regardless of configuration, has faster execution times than the other platforms (in this experiment the others have less than a vCPU).

\subsection{Answers to questions}

\paragraph{Q1}
Azure Functions is not designed for high parallelism or heavy computation.
Our experiments clearly show that the service is reluctant to scale and function invocations are queued on a few instances.
Also, instances take invocations at irregular intervals, even when processing other invocations.
In general, but most noticeable with computing tasks, the service does not create instances until there is high load, meaning that, in some cases, $100$ requests end up being handled by the same instance.
Changing configuration to limit instance concurrency confirms that the system needs considerable load to spin up new instances.
In particular, only a parallelism of $18$ is achieved when running $200$ concurrent invocations.
We should note that the service does not target this kind of applications, and that their approach is resource-efficient for IO tasks.

\paragraph{Q2}
More than one invocation is assigned to each instance concurrently, producing the stairs-like shape in the plots.
This happens for both sleeping and computing tasks, which unlinks its cause from the resource usage of a task.
The consequence is an important interference that, although sleeping functions obviously do not notice, it heavily affects computing tasks.
Invocations that should take $1.2$ seconds span out to minutes with $200$ concurrent requests (Fig.~\ref{fig:azure:basic:work:200:warm}).
We find a solution for this issue in limiting per-instance concurrency.
Although we still do not reach the desired parallelism for the job, execution time is much better and consistent with this limit.

We also see that responses to the client are throttled when there is high concurrency in an instance, perceived on client times (black \texttt{X}s).
On less busy instances, responses are almost immediate (Figure~\ref{fig:azure:basic:50:warm}).
This hints to more interferences.

\paragraph{Q3}
In the cases that include cold starts, host creations are at least a second apart, in line with the documentation~\cite{azure:scale}.
However, we also see that the delay in host creation can be significant and function requests are assigned to new instances even before they can process them, resulting in important delays.
For example, in Figure~\ref{fig:azure:basic:50:cold} most of the invocations are resolved in the first $6$~s by $3$ fast-spawning hosts, but some of them were assigned to a fourth instance that took almost $20$~s to start, delaying invocations that could have run earlier on the other hosts.

Azure Functions is generally conservative with resources.
For example, we do not see much scale until reaching $200$ parallel requests, and it is restricted by the one ``instance per second" limit.
This prudent scheduling configuration is what mainly differentiates Azure from other providers.
While others create new instances quite eagerly, Azure tends to pack as many invocations as possible to reduce resource consumption.
The approach works really well for the IO-bound tasks the service primarily targets, since it makes better use of resources, reduces costs, and facilitates management.

\section{Experiments on Google Cloud Platform}
\label{sec:experiment:gcp}
We deploy and update our function with the GCP CLI.
The invocation ID is obtained from one of the request headers in the function.
It is also available for the client in the HTTP response.
Differently from other providers, Google erases all information that could identify a container fon the instance ID.
To check if the container is the same, we use global code that generates an identifier during a cold start.
This is reliable since the Python file is only loaded once per container.

\subsection{Results}

\begin{figure}
  \centering

  \begin{subfigure}[b]{0.48\linewidth}
    \includegraphics[width=\linewidth]{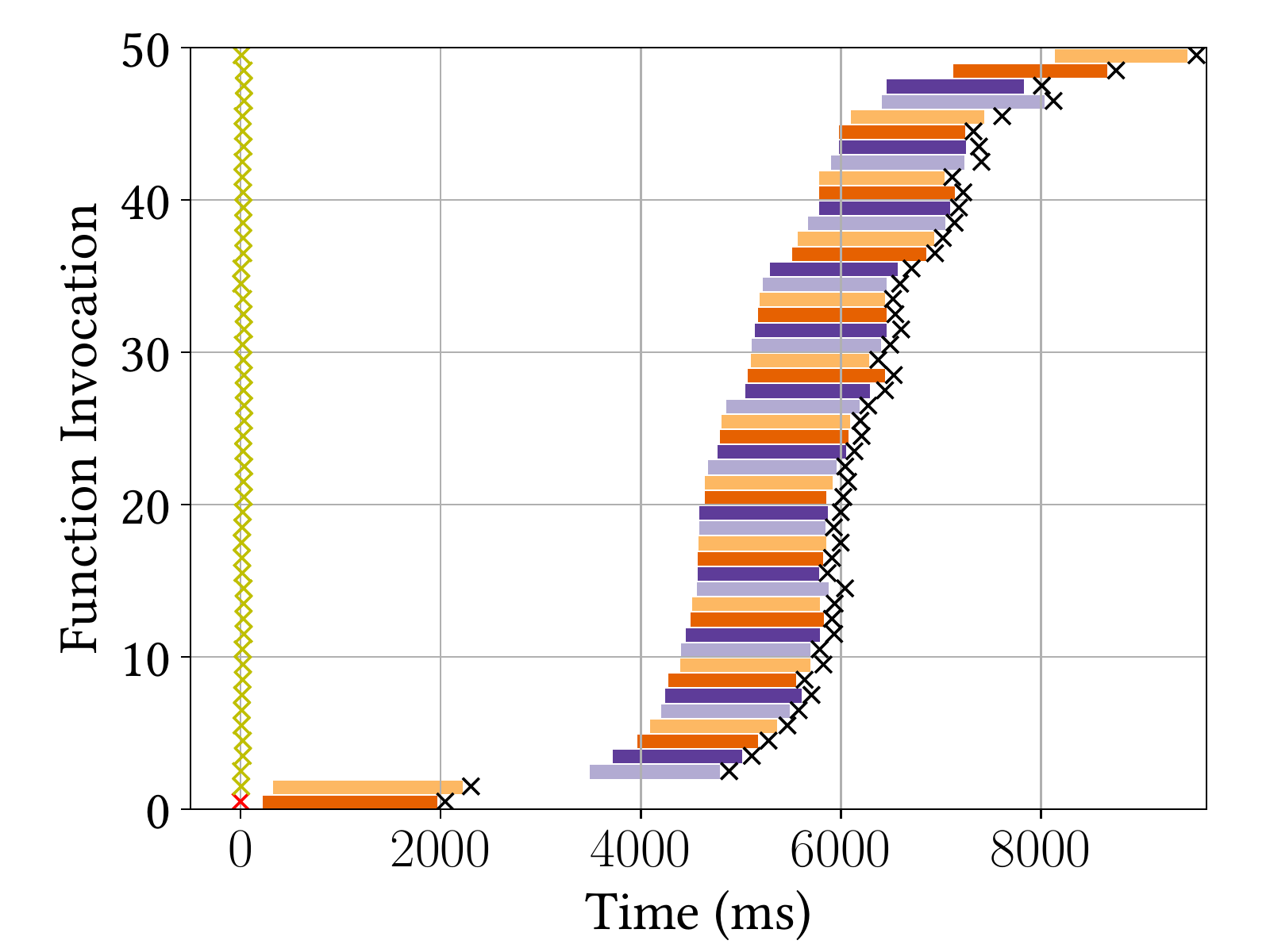}
    \caption{$50/10/S/s$}
    \label{fig:gcp:sleep:50}
  \end{subfigure}
  ~
  \begin{subfigure}[b]{0.48\linewidth}
    \includegraphics[width=\linewidth]{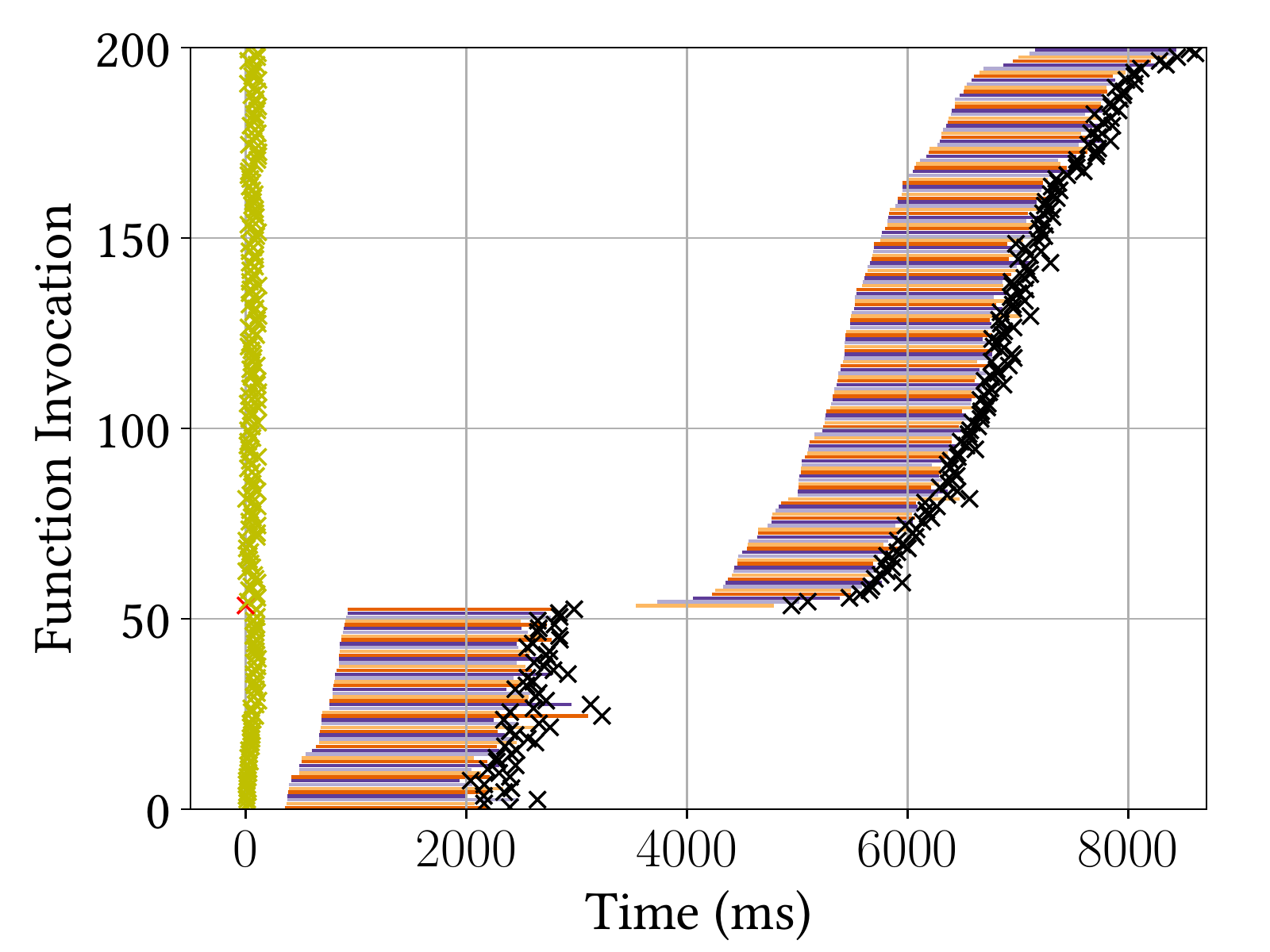}
    \caption{$200/200/S/s$}
    \label{fig:gcp:sleep:200}
  \end{subfigure}

  \begin{subfigure}[b]{0.48\linewidth}
    \includegraphics[width=\linewidth]{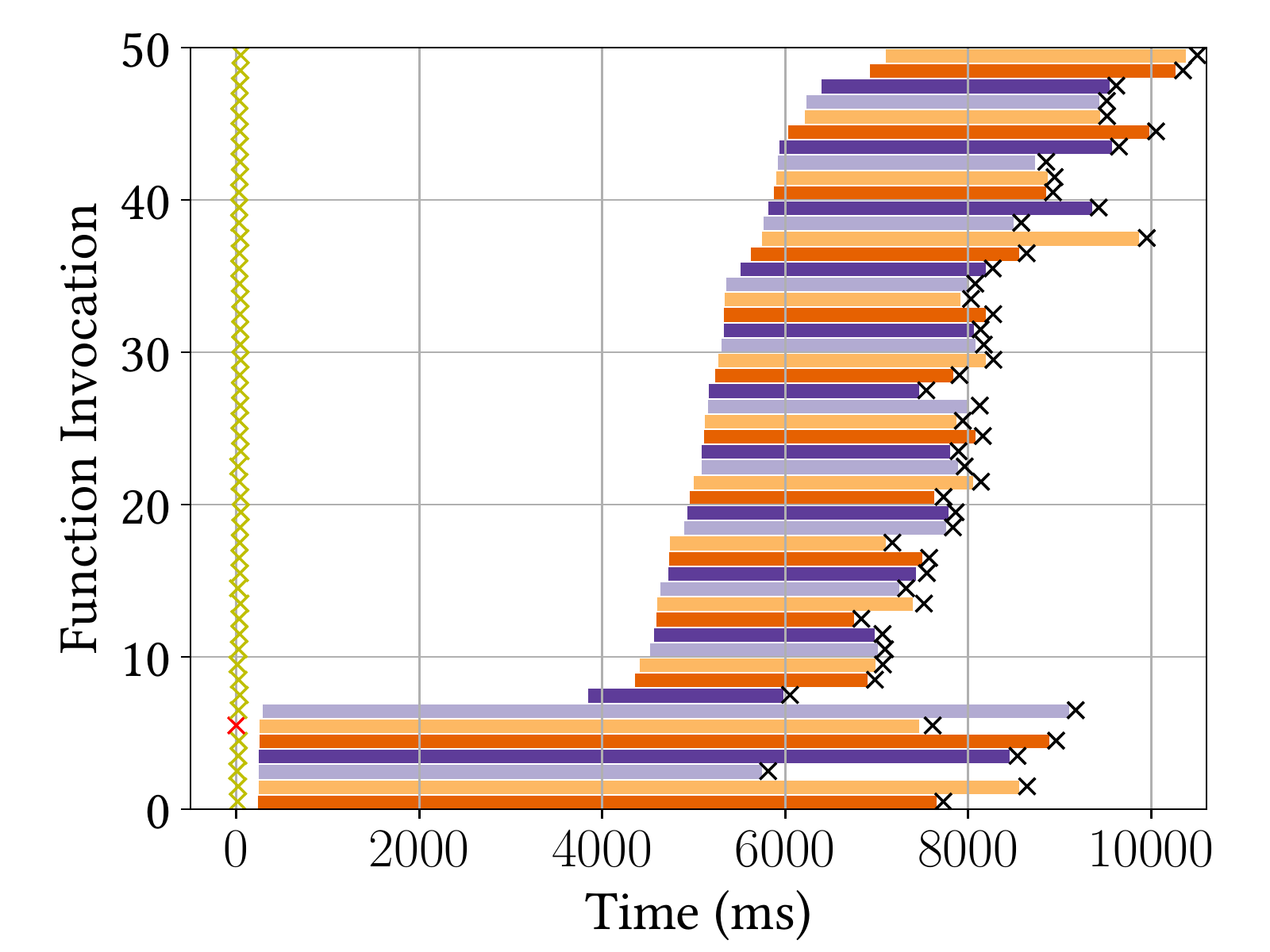}
    \caption{$50/10/C/s$}
    \label{fig:gcp:work:50}
  \end{subfigure}
  ~
  \begin{subfigure}[b]{0.48\linewidth}
    \includegraphics[width=\linewidth]{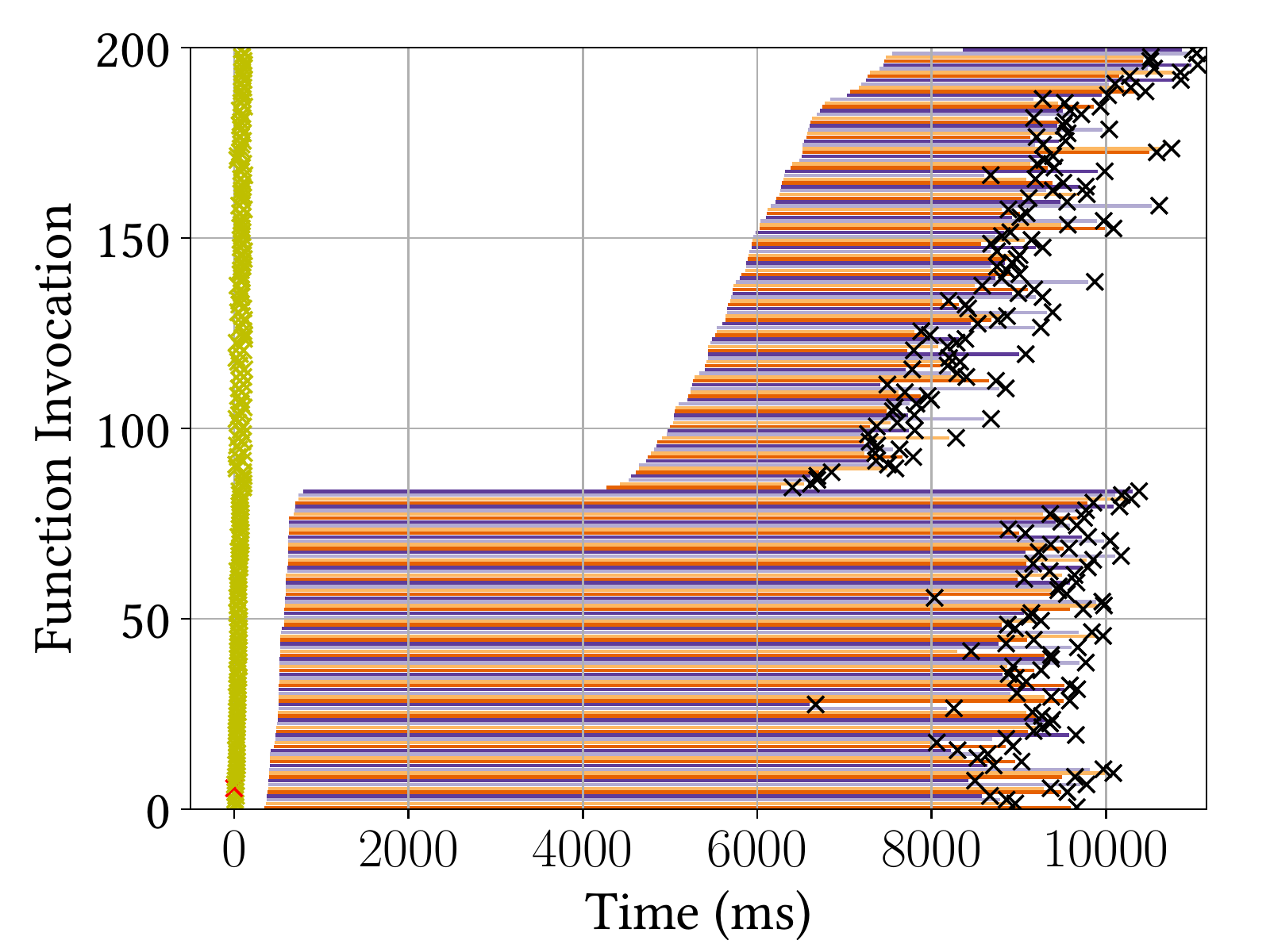}
    \caption{$200/100/C/s$}
    \label{fig:gcp:work:200}
  \end{subfigure}

  \begin{subfigure}[b]{0.48\linewidth}
    \includegraphics[width=\linewidth]{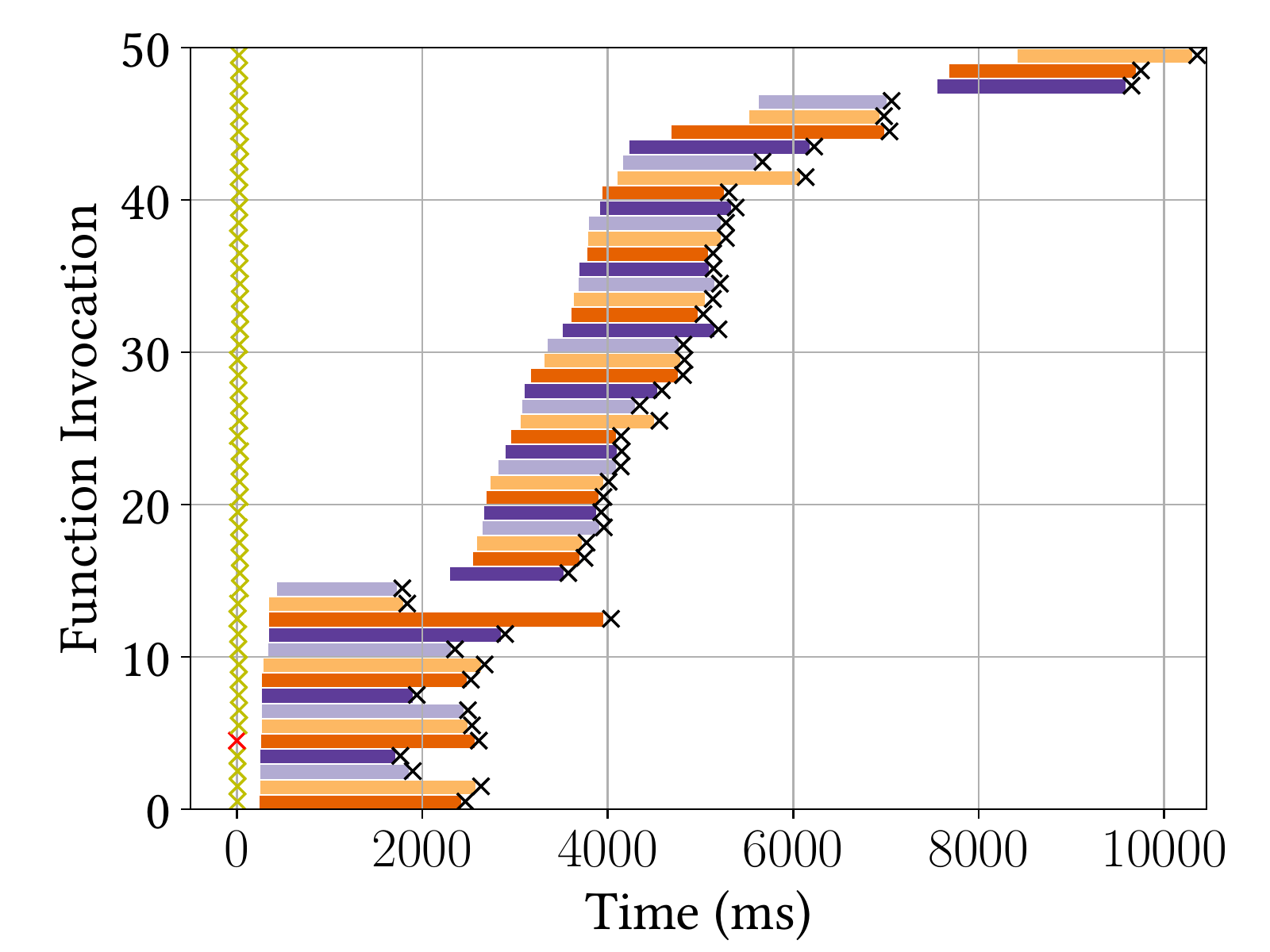}
    \caption{$50/20/C/b$}
    \label{fig:gcp:work2g:50}
  \end{subfigure}
  ~
  \begin{subfigure}[b]{0.48\linewidth}
    \includegraphics[width=\linewidth]{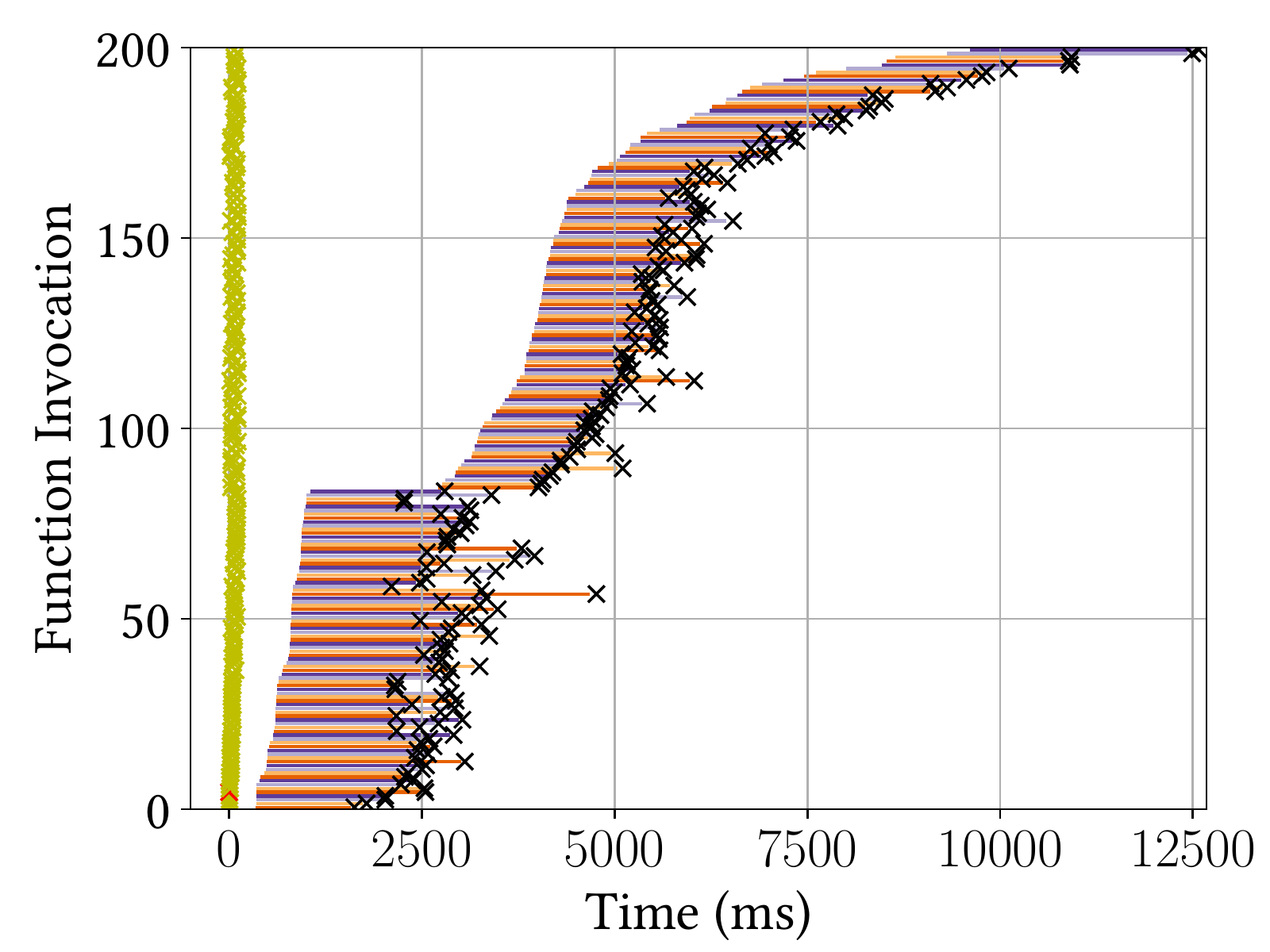}
    \caption{$200/100/C/b$}
    \label{fig:gcp:work2g:200}
  \end{subfigure}

  \caption{
    Experiment on GCP.
  }
  \label{fig:gcp}
\end{figure}

\paragraph{Experiments with sleeping functions}
With the \textit{small} functions ($256$~MiB) and the sleeping task, an execution with $50/10/S/s$ results in Figure~\ref{fig:gcp:sleep:50}.
Each invocation runs on a different container.
Note how the run only keeps $2$ instances warm from a previous execution of $10$.
Figure~\ref{fig:gcp:sleep:200} shows $200/200/S/s$.
It is the second consecutive execution with this configuration, so we expect all instances warm; however, most hit a cold start.
Still, the service runs each request on a different container.

\paragraph{Experiments with computing functions}
Still with \textit{small} functions, we test the compute-intensive tasks.
We start running the function individually, and assess that the computation takes $5.5$~s with this configuration.
Figures~\ref{fig:gcp:work:50} and~\ref{fig:gcp:work:200} show invocations of this experiment with different parallelism, where we clearly see the performance difference between cold and warm containers.
On cold invocations, the computation takes $3.5$~s, while warm executions take up to $10$~s.
Also, warm containers are recycled very quickly.
For instance, the $200$-requests execution, run right after a $100$ one, only finds $84$ warm containers.

With $2$~GiB functions (\textit{big} configuration), the maximum memory configurable for GCP, function time for the individual execution reduces to $1.3$~s.
Figures~\ref{fig:gcp:work2g:50} and~\ref{fig:gcp:work2g:200} show subsequent invocations of this experiment with different parallelism.
Like previously, the system keeps full parallelism.
However, execution time still varies significantly from $1.3$ to $4$~s.

\begin{figure}
  \centering
  \begin{subfigure}[b]{0.48\linewidth}
    \includegraphics[width=\linewidth]{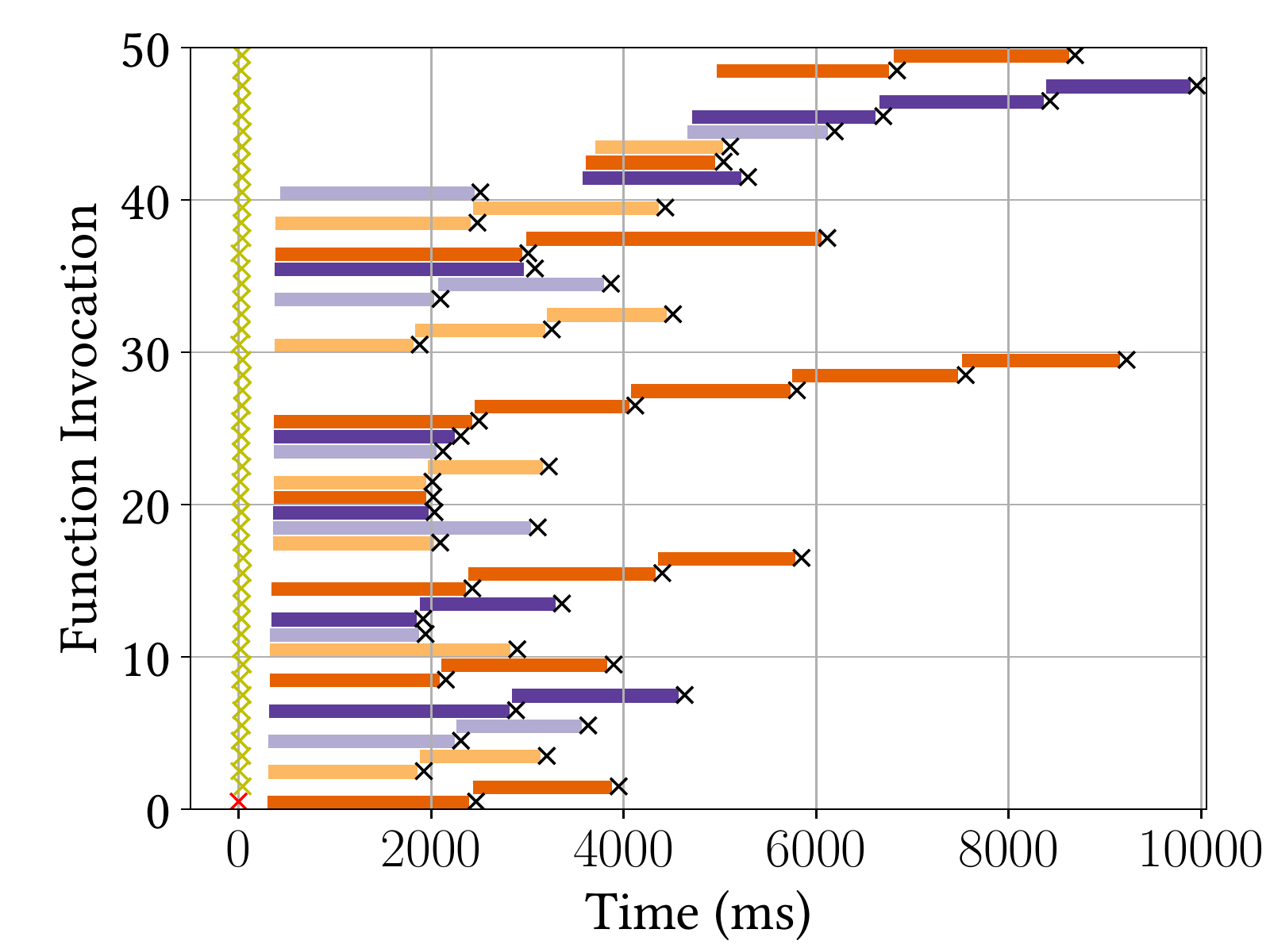}
    \caption{$50/50/C/b$}
    \label{fig:gcp:work2g:bad:50}
  \end{subfigure}
  ~
  \begin{subfigure}[b]{0.48\linewidth}
    \includegraphics[width=\linewidth]{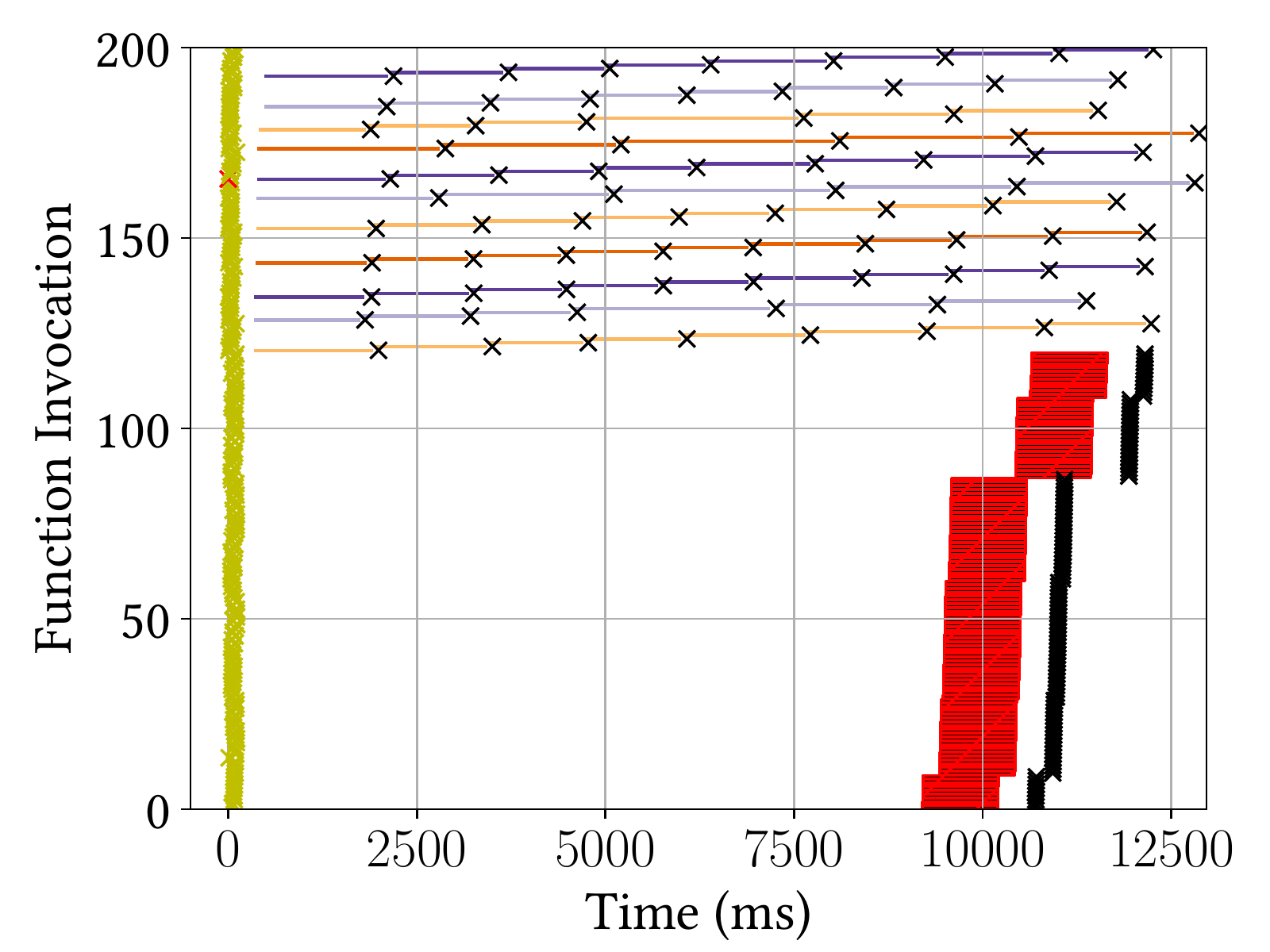}
    \caption{$200/100/C/b$}
    \label{fig:gcp:work2g:bad:200}
  \end{subfigure}

  \caption{
    Server errors on GCP.
    The red bars are rejected requests.
  }
  \label{fig:gcp:work2g:bad}
\end{figure}

These are the best scenarios experienced.
However, the system seems to throttle \textit{big} functions, queueing some invocations and even rejecting them.
Figure~\ref{fig:gcp:work2g:bad} shows samples of such cases, experienced after performing less than $1000$ requests.

\begin{figure}
  \centering\begin{subfigure}[b]{0.48\linewidth}
    \includegraphics[width=\linewidth]{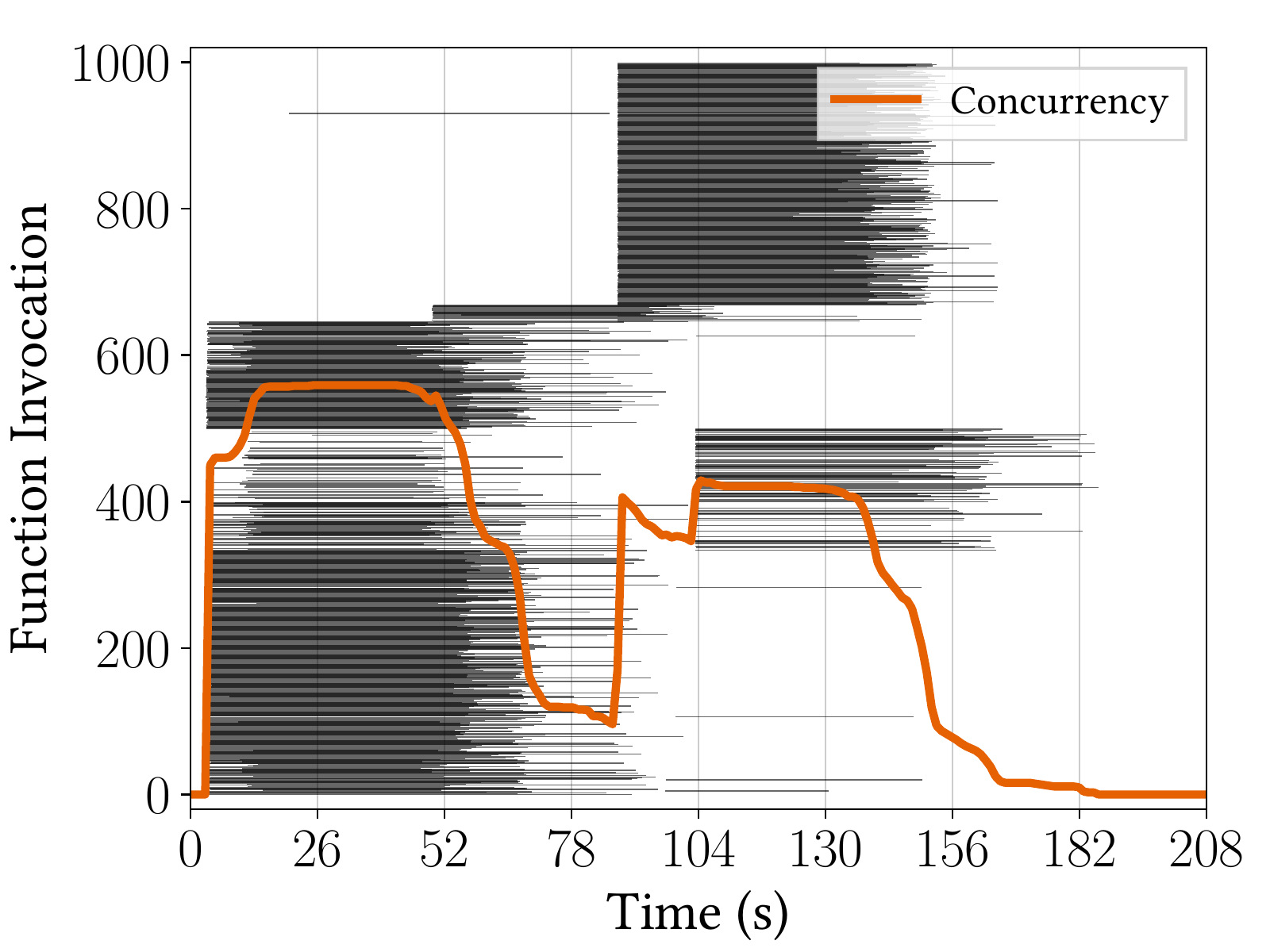}
    \caption{}
    \label{fig:gcp:big:time}
  \end{subfigure}
  ~
  \begin{subfigure}[b]{0.48\linewidth}
    \includegraphics[width=\linewidth]{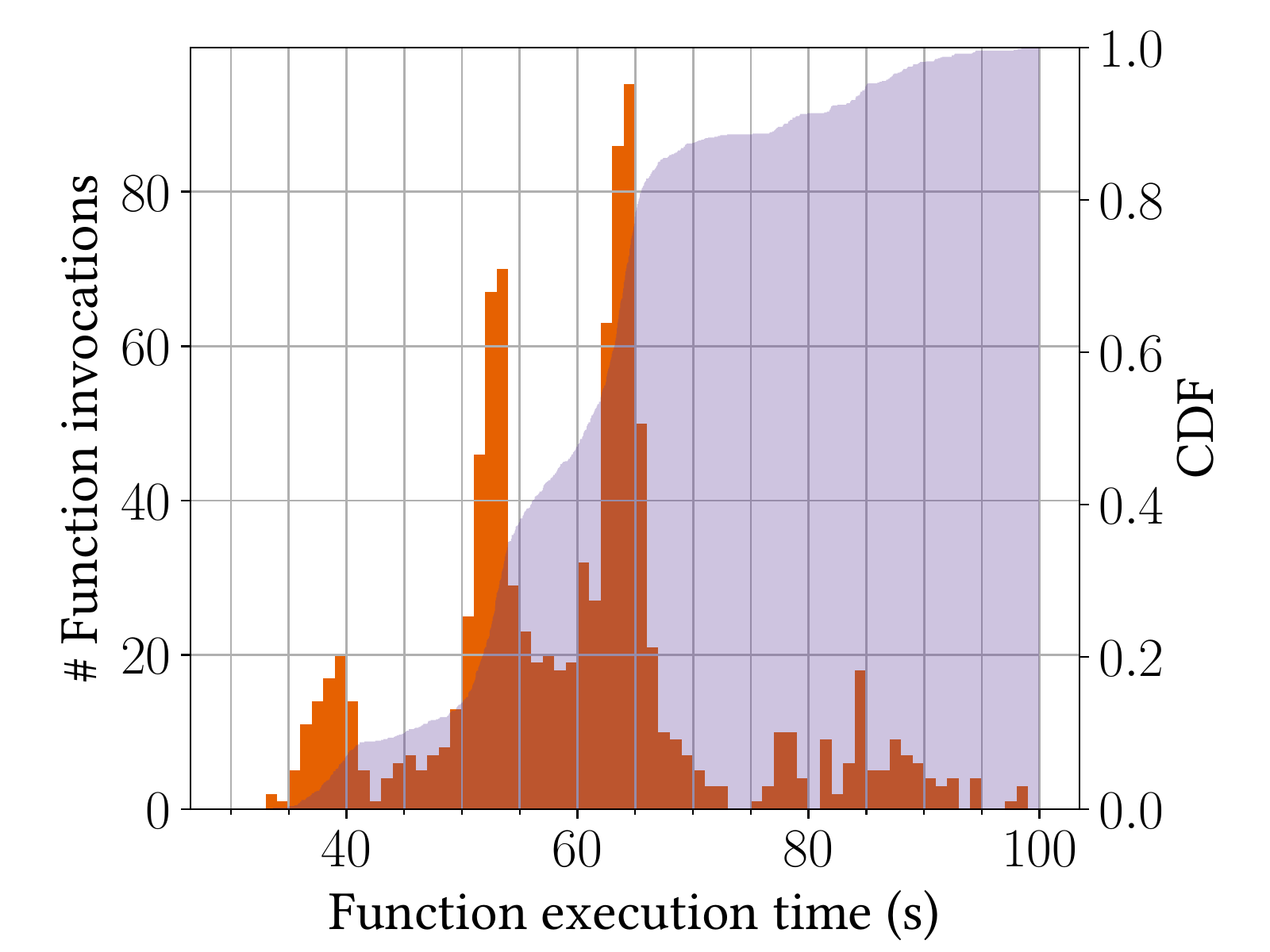}
    \caption{}
    \label{fig:gcp:big:hist}
  \end{subfigure}
  \caption{
    Large-scale experiment on GCP.
  }
  \label{fig:gcp:big}
\end{figure}

\paragraph{On a bigger scale}
Figure~\ref{fig:gcp:big} depicts the results when running the larger configuration with $1000$ asynchronous invocations.
We see that, despite requesting $1000$ invocations at once, only $550$ functions run in parallel at first, and then another batch of $450$ functions are run later.
With the help of the histogram, we also notice a wide variety of function execution times.
This behavior seems to confirm the differences between cold and warm invocations seen before, but also evinces further interferences in resources and/or heterogeneity of resources.

\subsection{Answers to questions}

\paragraph{Q1}
Mostly, all invocations get a new instance, which allows good parallelism.
However, the scheduling looks more complicated than in other platforms and imposes several rate limits.
For instance, functions with more memory are less elastic.
We experienced a lot of throttling with $2$~GiB functions and even failed requests.
Given the size of our experiments, this suggests a more restrictive rate limit than stated in the documentation~\cite{gcp:limits}.
While this does not affect functions at small scale, it is an issue for large-scale embarrassingly parallel tasks.
Also, the service removes idle containers very quickly and subsequent runs of the experiment do not all find warm containers, and there are always cold starts.
This can be an important issue for latency-sensitive applications, and also hinders parallelism.
As an example, although invocations run on different instances, not all of them are running in parallel, simultaneously.
E.g., from $200$ requests less than $100$ run in parallel and the big scale experiment only found a concurrency of $550$.

\paragraph{Q2}
With the information gathered from the environment, we see that all invocations run on a $2$~GiB microVM.
This is different from AWS, where each microVM is configured with its memory corresponding to the function configuration.
The microVMs also have $2$ vCPUs, which in most instances run at $2.7$~GHz, and some at $2.3$~GHz.
Since all functions run on equally-sized microVMs, the different CPU limits in the documentation~\cite{gcp:pricing'n'types} are probably imposed through CPU slices.

However, in experiments with the compute task, execution time is not consistent across invocations, suggesting that the limit is not well ensured.
For instance, the $256$~MiB functions complete in $2$ and up to $10$~s.
Even with $2$~GiB functions (corresponding to a full microVM), performance is inconsistent, ranging from $1.3$ to $4$~s.
The most surprising finding is that there seems to be a significant performance difference between warm and cold invocations, being cold ones much faster.

\paragraph{Q3}
We can add the following conclusions:
\begin{inparaenum}[i)]
  \item Scheduling is based on several parameters (e.g., function size, invocation rate, function run time, etc.), and it affects scalability.
  \item Cold starts usually induce a delay around $3$ seconds, but it increases with parallelism and memory size.
\end{inparaenum}

\section{Experiments on IBM Cloud}
\label{sec:experiment:ibm}
We deploy and update our function with the IBM Cloud CLI on the default package in a simple namespace, by directly uploading the source code.
The invocation ID is at the environment variable ``\texttt{\_\_OW\_ACTIVATION\_ID}''.
The most reliable way to identify a container is through the randomly generated identifier present at \texttt{/proc/self/cgroup}; Docker writes the container name there~\cite{ibm:dockermetrics}.
We obtain the system uptime to identify the VM where each container runs.
Even collected from a container, the uptime corresponds to the container host, which is the Invoker VM.
Although not fully reliable, it can help us guess container co-residency.

\subsection{Results}

\begin{figure}
  \centering
  \begin{subfigure}[b]{0.48\linewidth}
    \includegraphics[width=\linewidth]{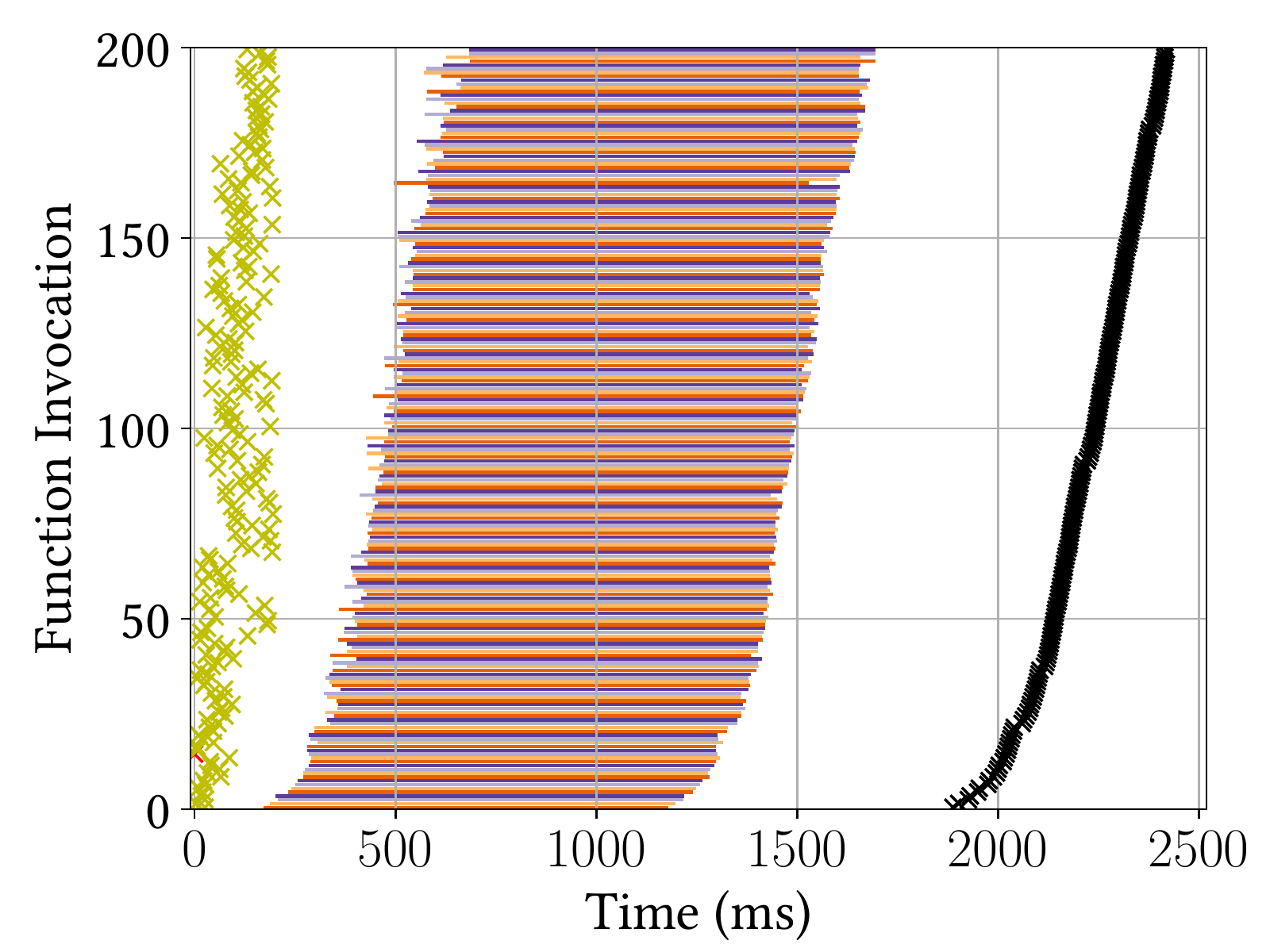}
    \caption{Plot without the VM inference}
    \label{fig:ibm:guessing:pre}
  \end{subfigure}
  ~
  \begin{subfigure}[b]{0.48\linewidth}
    \includegraphics[width=\linewidth]{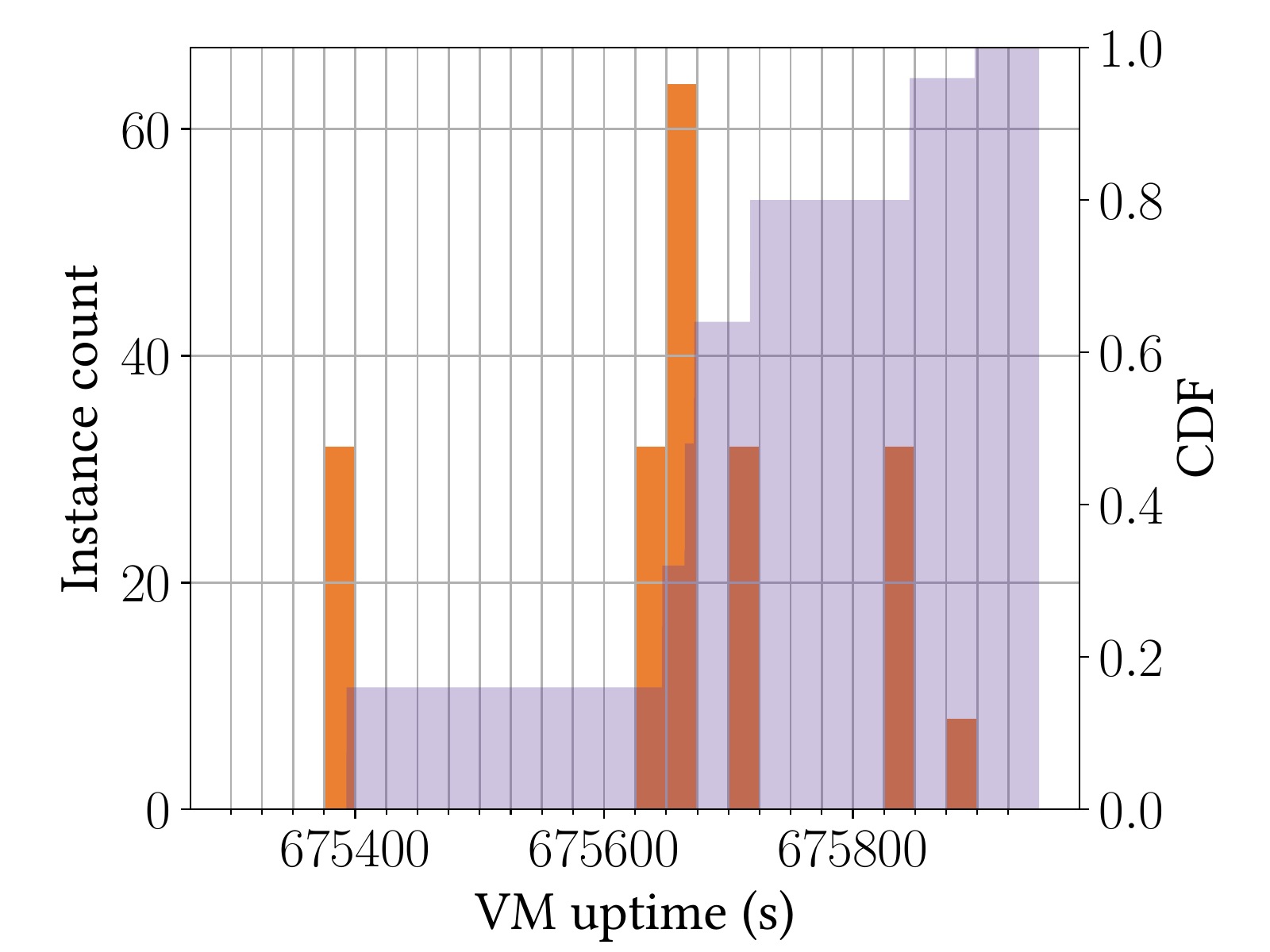}
    \caption{Uptime distribution}
    \label{fig:ibm:guessing:uptime}
  \end{subfigure}
  \caption{
    Guessing the VM on IBM for an execution with $200$ requests.
  }
  \label{fig:ibm:guessing}
\end{figure}

\paragraph{Guessing the VM from the system uptime}
The plot for a $200$-requests execution would look like Figure~\ref{fig:ibm:guessing:pre}.
Each invocation is running on a different container, but some of them could be on the same VM.
By getting the system uptime, we can display container co-residency.%
\footnote{
  A detailed description of a similar method for machine identification was introduced by Lloyd et al.~\cite{lloyd:investigation}.
}
We represent the system uptime gotten at each function instance in Figure~\ref{fig:ibm:guessing:uptime}.
If the uptime gotten by different invocations is similar, they are likely co-residents of the same VM.
Since invocations are not exactly simultaneous (they do not read the uptime at the same instant), never two of them will get the exact same uptime.
However, since the whole experiment lasts $3$ seconds, two co-resident invocations will get an uptime different by at most $3$ seconds (usually in the same second since it is collected near function start).
The CDF gives a very precise view.
Each step in the curve is all the invocations that got a similar uptime, and thus co-residents.
With the information from the histogram we can count how many invocations run on each VM.
If two VM uptimes are too close, the accompanying histogram may pack invocations from different machines in the same bar, but we can still distinguish them with the CDF.
Since we know that all invocations run concurrently, this gives us the VM maximum concurrency.
For instance, we see that most histogram bars count $32$.
One reaches $64$, but we see in the CDF that it comprises two steps, thus being in fact two VMs.
This means that it is very likely that each machine holds a maximum of $32$ containers in this experiment.

\begin{figure}
  \centering
  
  \begin{subfigure}[b]{0.48\linewidth}
    \includegraphics[width=\linewidth]{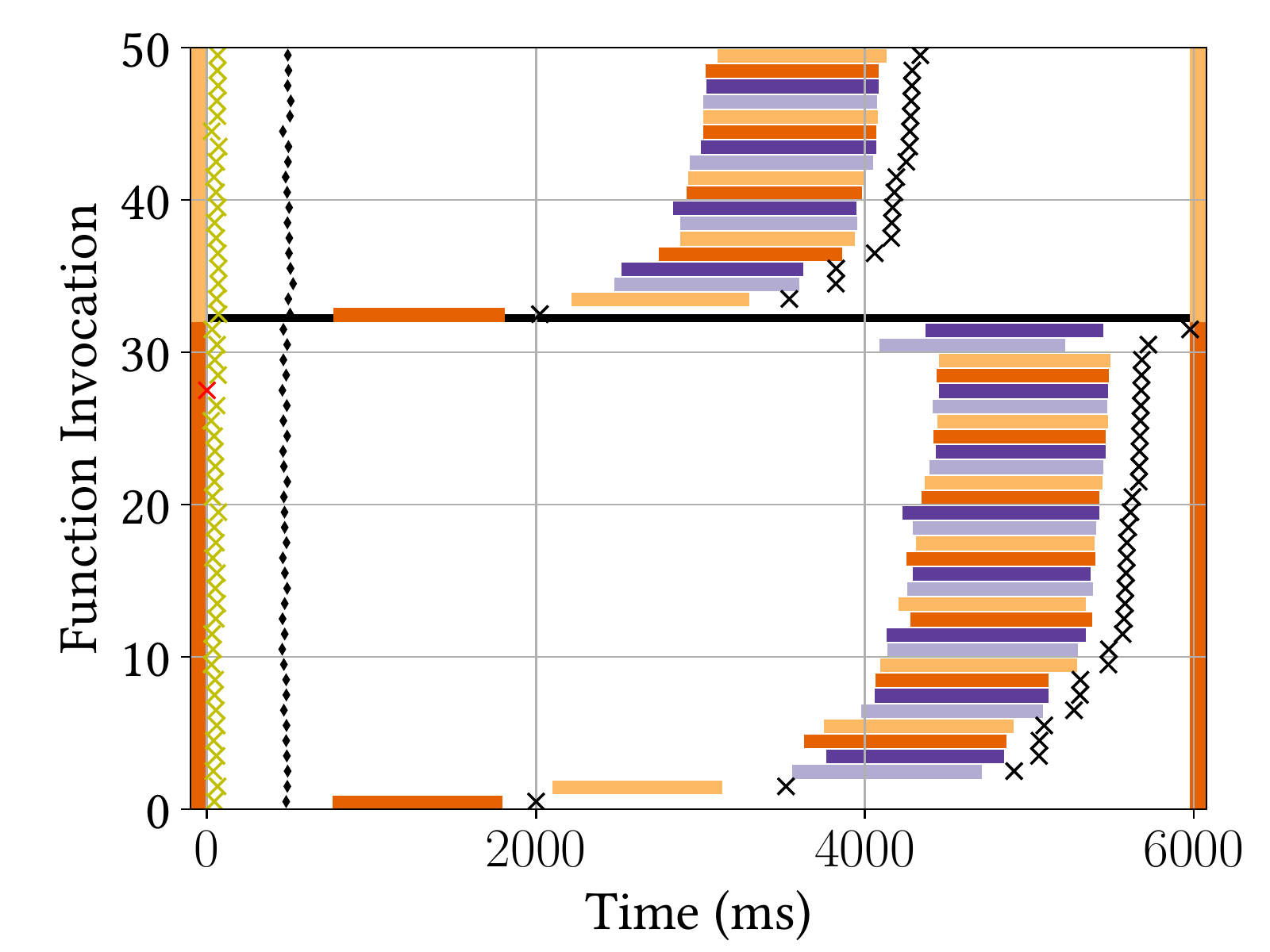}
    \caption{$50/0/S/s$}
    \label{fig:ibm:sleep:50}
  \end{subfigure}
  ~
  \begin{subfigure}[b]{0.48\linewidth}
    \includegraphics[width=\linewidth]{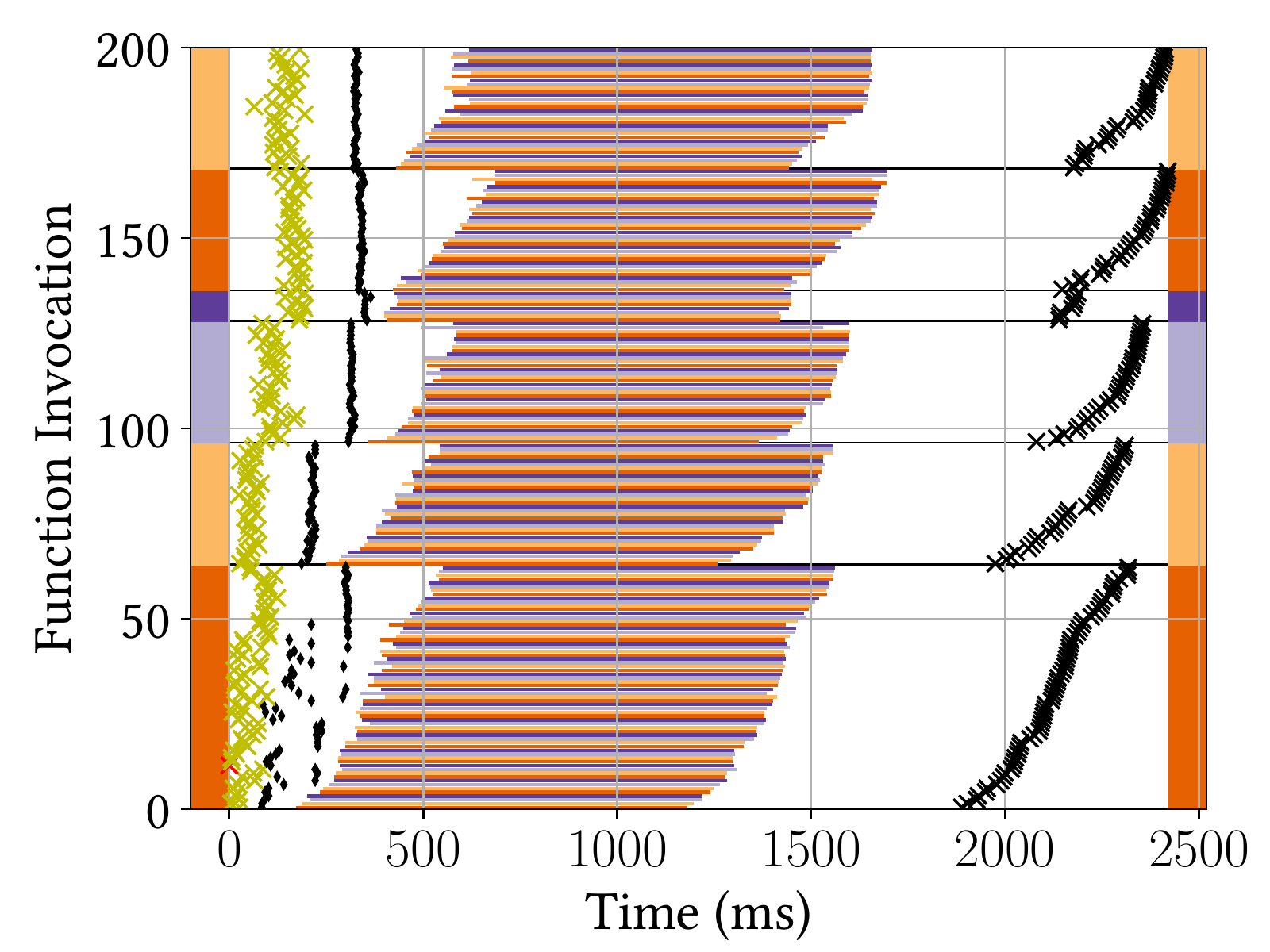}
    \caption{$200/200/S/s$}
    \label{fig:ibm:sleep:200}
  \end{subfigure}

  \begin{subfigure}[b]{0.48\linewidth}
    \includegraphics[width=\linewidth]{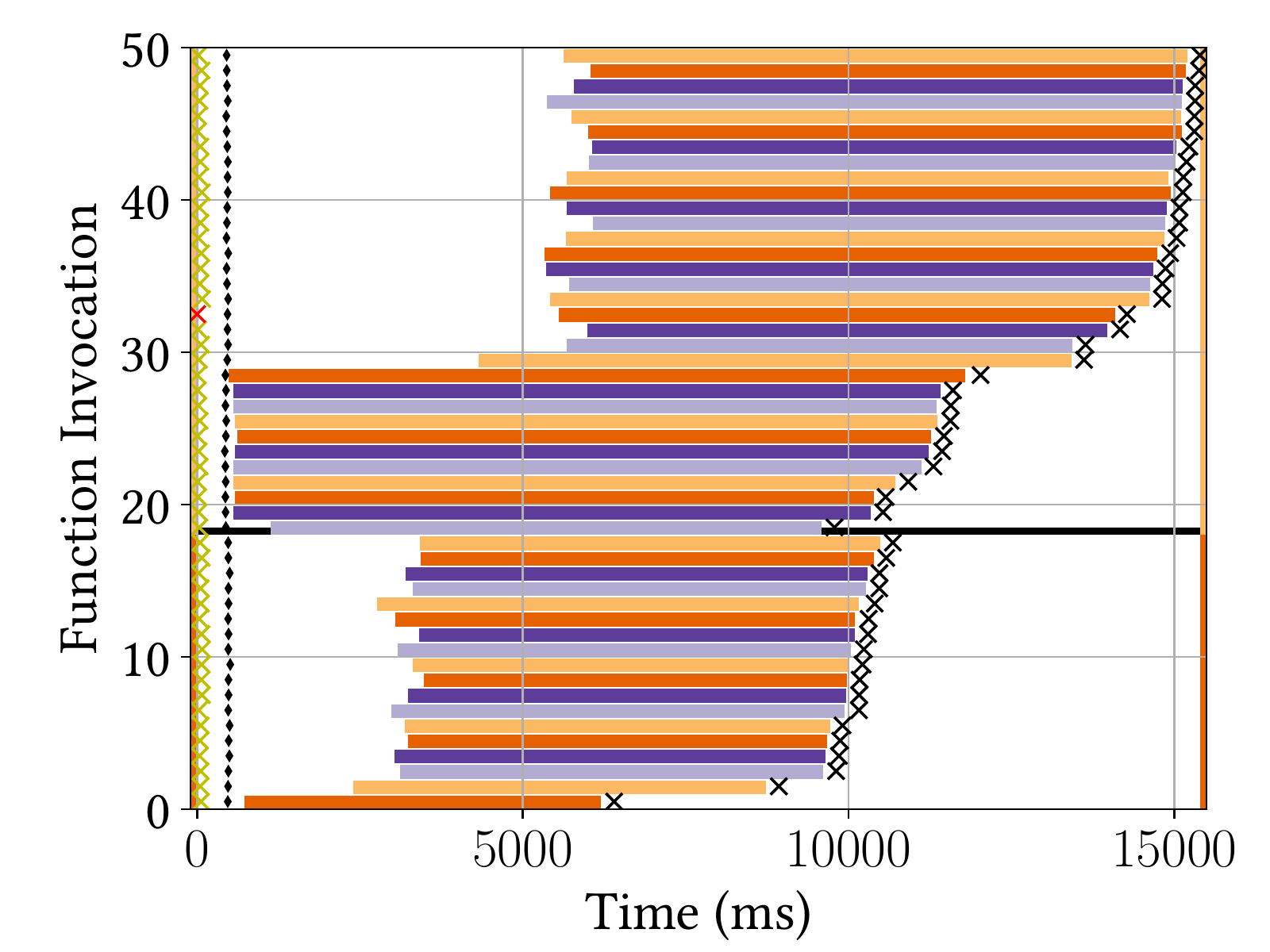}
    \caption{$50/10/C/s$}
    \label{fig:ibm:work:50}
  \end{subfigure}
  ~
  \begin{subfigure}[b]{0.48\linewidth}
    \includegraphics[width=\linewidth]{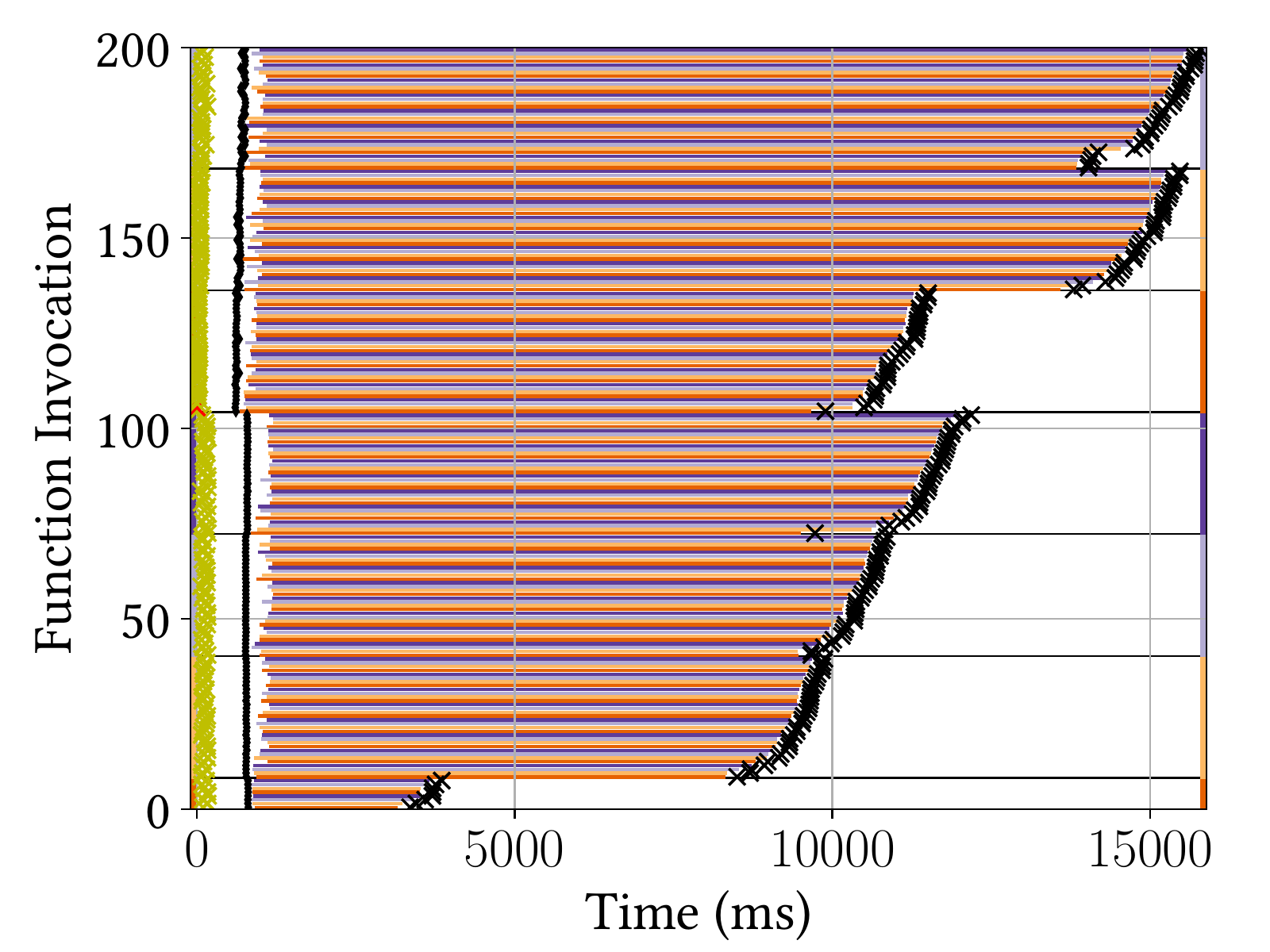}
    \caption{$200/200/C/s$}
    \label{fig:ibm:work:200}
  \end{subfigure}

  \begin{subfigure}[b]{0.48\linewidth}
    \includegraphics[width=\linewidth]{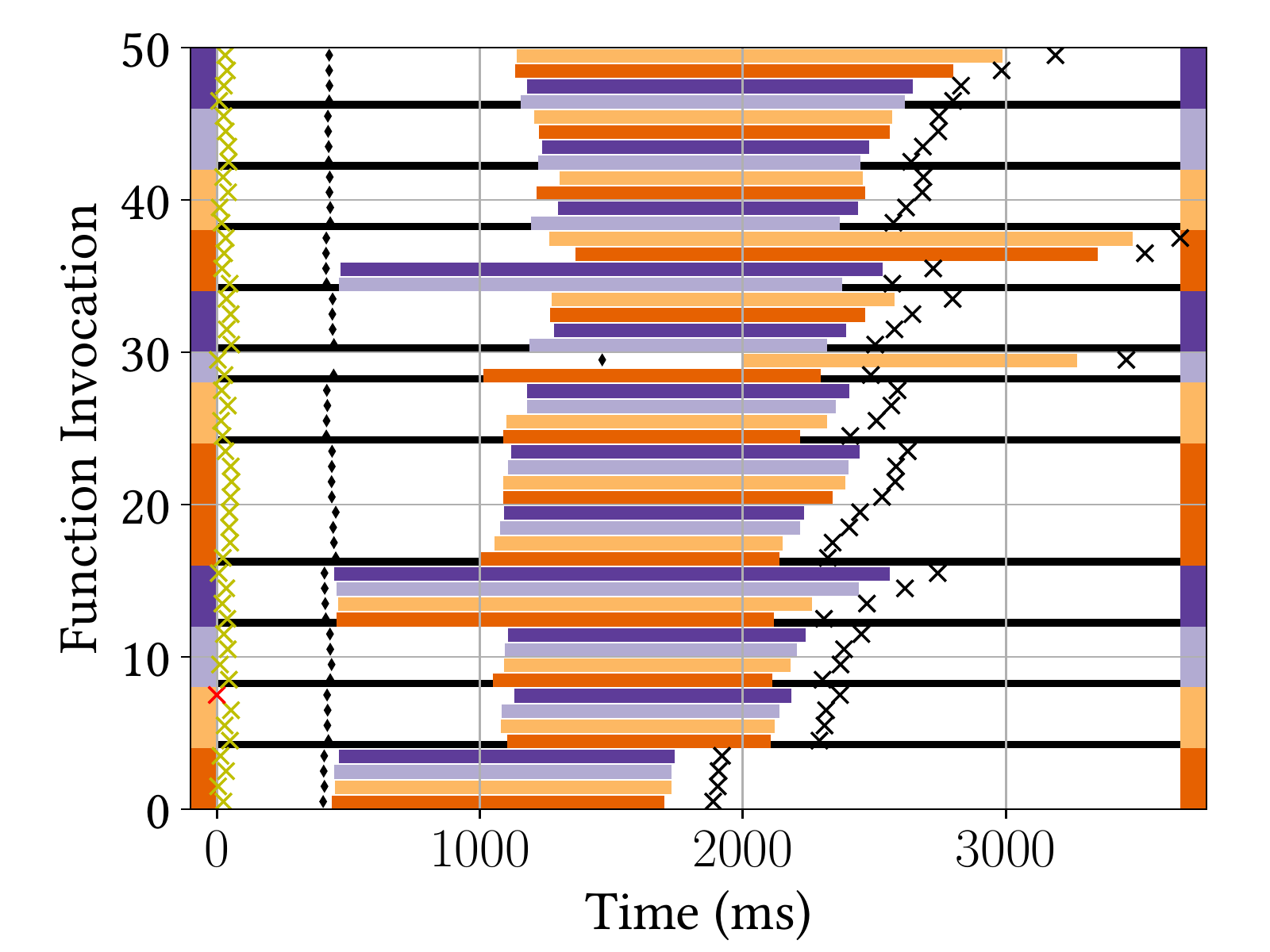}
    \caption{$50/10/C/b$}
    \label{fig:ibm:work2g:50}
  \end{subfigure}
  ~
  \begin{subfigure}[b]{0.48\linewidth}
    \includegraphics[width=\linewidth]{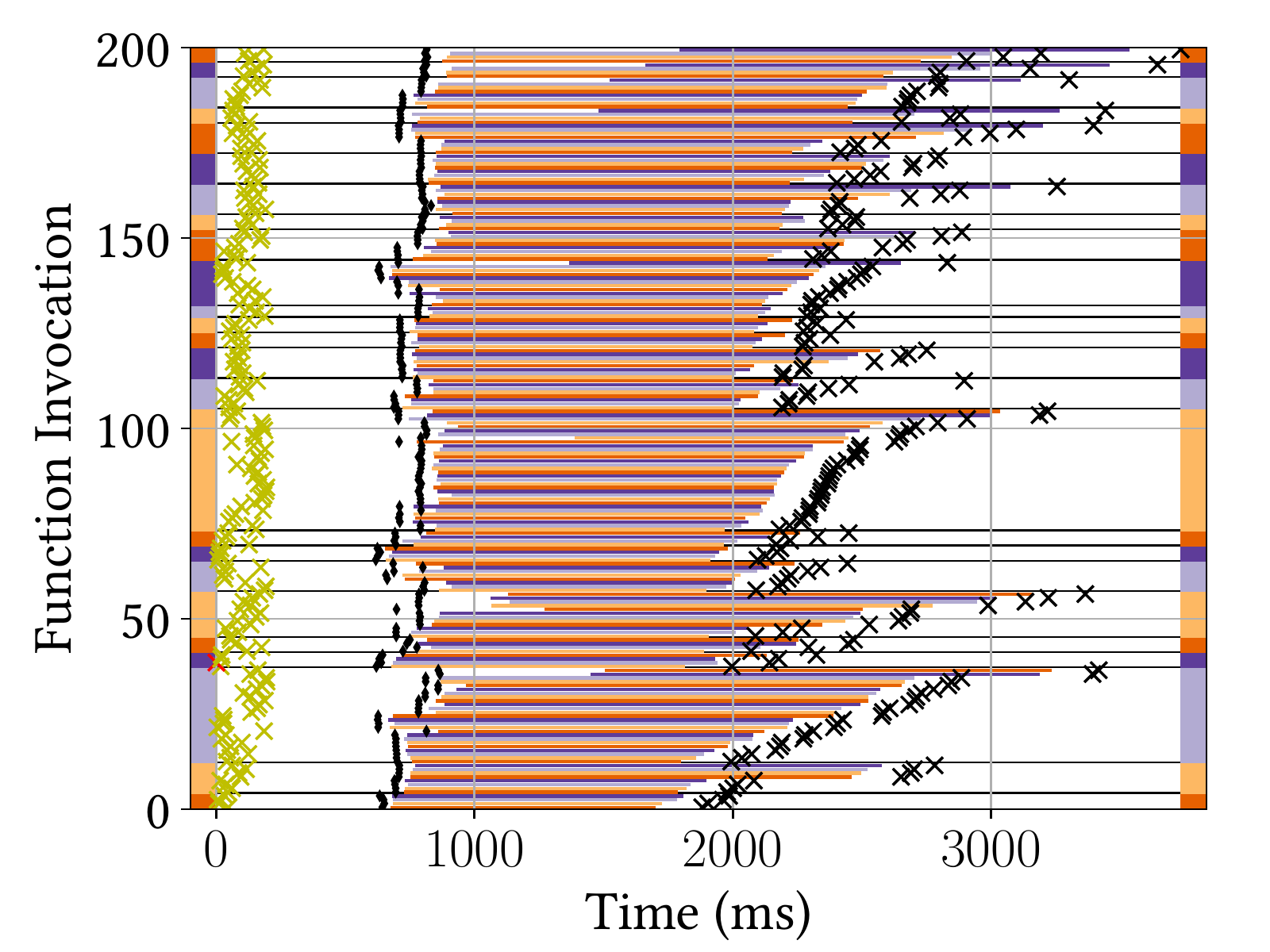}
    \caption{$200/200/C/b$}
    \label{fig:ibm:work2g:200}
  \end{subfigure}

  \caption{
    Experiment on IBM.
  }
  \label{fig:ibm}
\end{figure}

We merge this data into our plot to build Figure~\ref{fig:ibm:sleep:200}.
The color blocks at the sides indicate the guessed VM based on the system uptime.
There is also a black line that separates VMs for clarity.
Additionally, the service collects the time the invocation has been waiting in the system.
We plot it as black diamonds to indicate when the system received the request.

\paragraph{Experiments with sleeping functions}
With $256$~MiB functions and the sleeping task, we first run a cold execution with $10$ parallel requests.
A subsequent execution with $50$ requests results in Figure~\ref{fig:ibm:sleep:50}.
Figure~\ref{fig:ibm:sleep:200} shows $200/200/S/s$.

\begin{figure}
  \centering
  \begin{subfigure}[b]{0.48\linewidth}
    \includegraphics[width=\linewidth]{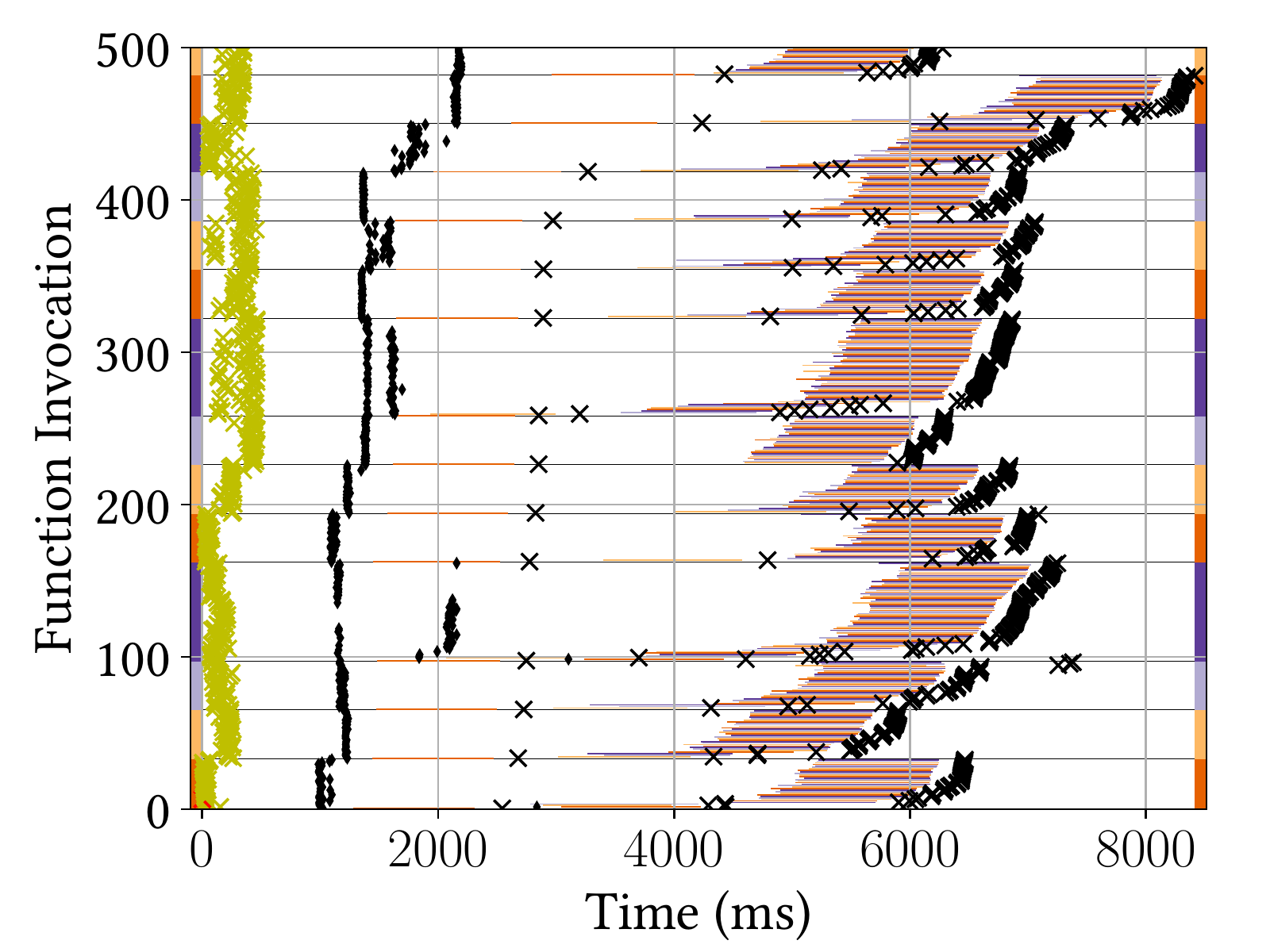}
    \caption{Parallelism}
    \label{fig:ibm:sleep:500}
  \end{subfigure}
  ~
  \begin{subfigure}[b]{0.48\linewidth}
    \includegraphics[width=\linewidth]{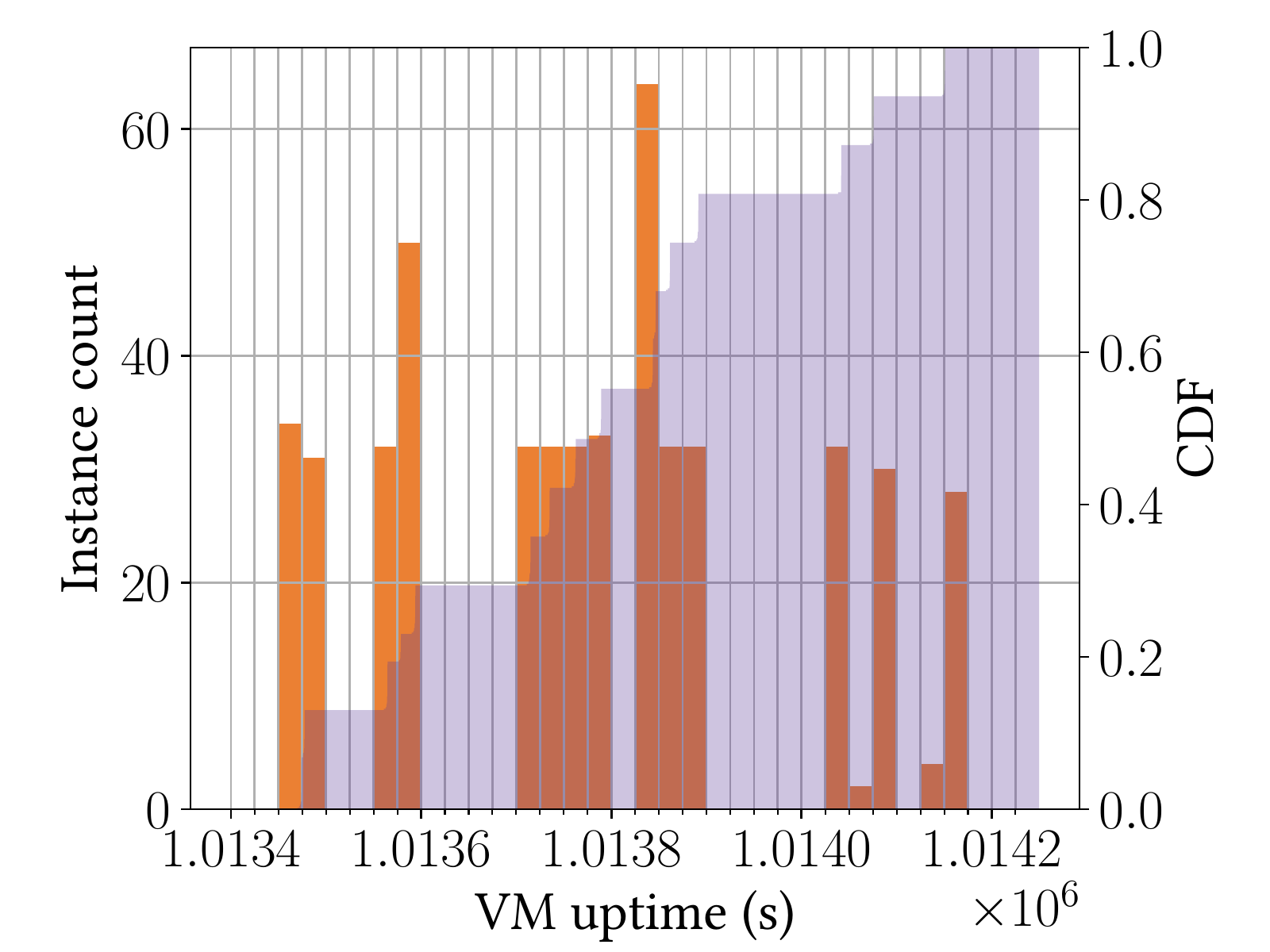}
    \caption{Uptime distribution}
    \label{fig:ibm:sleep:500:uptime}
  \end{subfigure}

  \caption{
    $500/0/S/s$ on IBM.
  }
\end{figure}

Figure~\ref{fig:ibm:sleep:500} shows a cold start for $500$ parallel invocations.
In this case, two side blocks are twice as big as the others.
However, in the CDF (Figure~\ref{fig:ibm:sleep:500:uptime}) it is clear that each step is in fact of $32$ containers.
This case uses more VMs than the previous, and it is easier to find several machines with very similar uptime.
This experiment also shows an interesting behavior of cold starts in OpenWhisk.
Each VM has one invocation that runs almost as in a warm start, while the others take some extra seconds.
This corresponds to the fact that OpenWhisk starts an empty container on each Invoker machine before receiving any request.
It also validates that the bigger side blocks are in fact two VMs since they have two of these early invocations.

\paragraph{Experiments with computing functions}
Still with \emph{small} functions, we switch to the compute-intensive task.
An individual execution assesses that the computation takes $1.3$~s.
Figures~\ref{fig:ibm:work:50} and~\ref{fig:ibm:work:200} show subsequent invocations with different parallelism.
We see clearly that execution time is affected and increases as the VMs fill up with containers.
Our $1.3$~s tasks take from $8$ to $15$~s on machines full with $32$ containers.

With $2$~GiB functions, which should have a full CPU as per our calculations, an individual execution still takes $1.3$~s.
Figures~\ref{fig:ibm:work2g:50} and~\ref{fig:ibm:work2g:200} show subsequent invocations with different parallelism.
In this case, function run time is more consistent and maintains around $1.3$~s.
However, some executions span for up to an additional second, which hints us to other resource interferences.
Here, the $200$ execution requires more VMs than any previous experiment, leading to a similar situation than with the previous $500$ execution (Fig.~\ref{fig:ibm:sleep:500}).
In its uptime distribution (omitted), we identify $50$ steps, which proves that each VM holds four $2$~GiB containers.

\begin{figure}
  \centering
  \begin{subfigure}[b]{0.48\linewidth}
    \includegraphics[width=\linewidth]{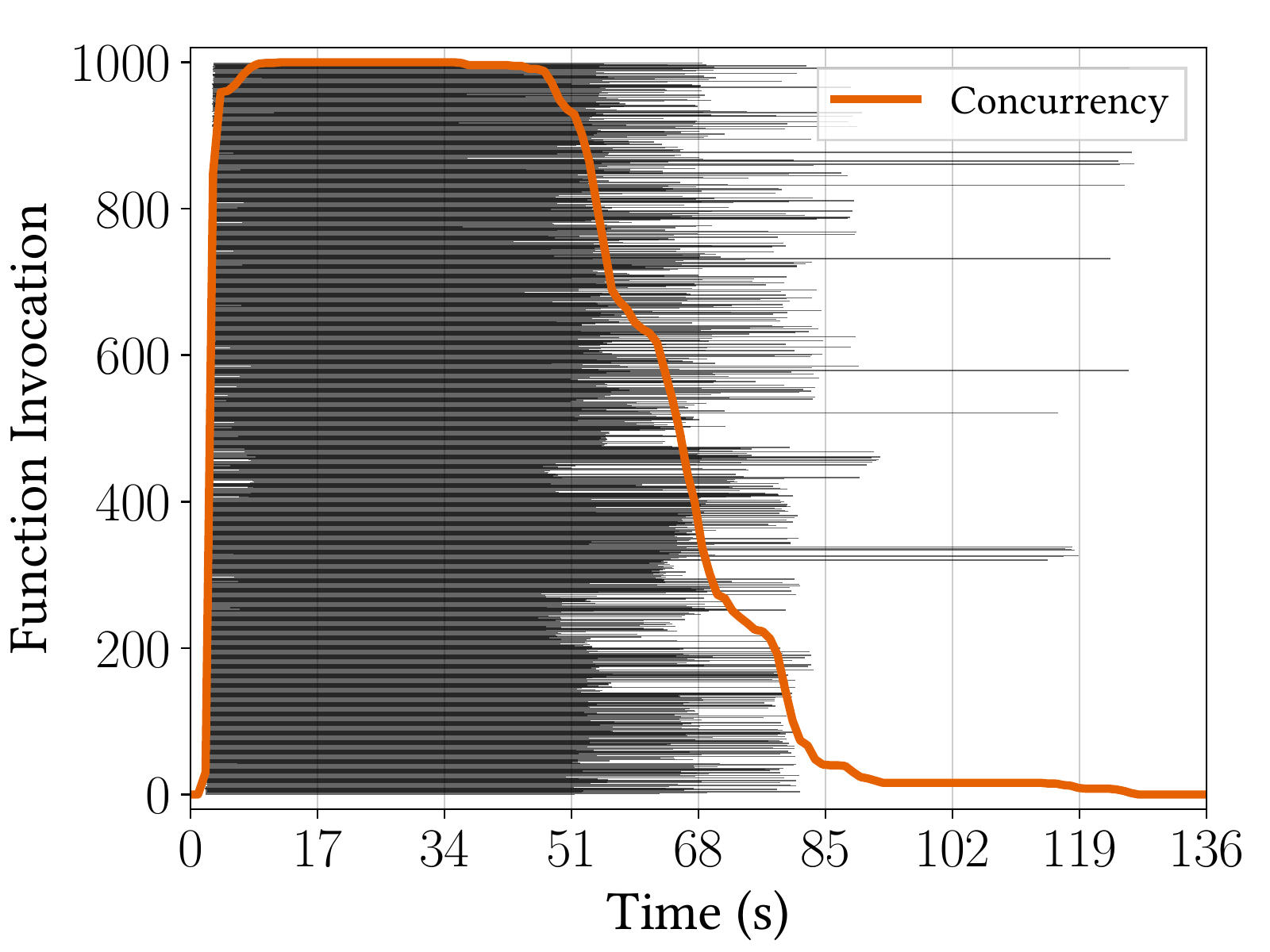}
    \caption{}
    \label{fig:ibm:big:time}
  \end{subfigure}
  ~
  \begin{subfigure}[b]{0.48\linewidth}
    \includegraphics[width=\linewidth]{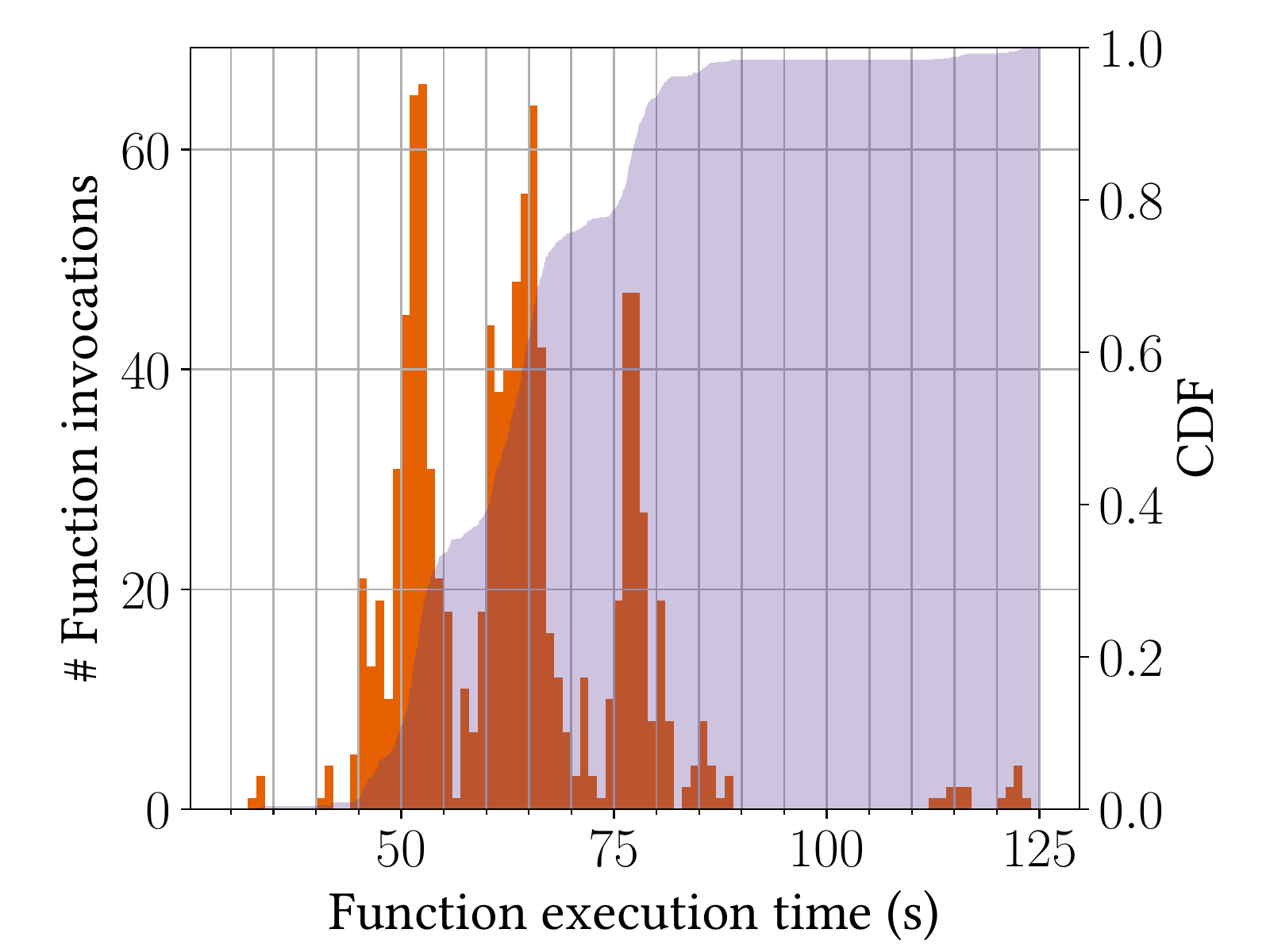}
    \caption{}
    \label{fig:ibm:big:hist}
  \end{subfigure}
  \caption{
    Large-scale experiment on IBM.
  }
  \label{fig:ibm:big}
\end{figure}

\paragraph{On a bigger scale}
The plot for the configuration with $1000$ invocations appears in Figure~\ref{fig:ibm:big}, showing full parallelism from the start.
However, several invocations take significantly longer to finish computation, doubling total completion time.
The histogram shows this wide distribution of function run time.
Resource heterogeneity and the interferences we perceived in previous experiments are possible causes of this variability.

\subsection{Answers to questions}
\label{sec:experiment:ibm:disc}

\paragraph{Q1}
Generally, IBM Cloud Functions shows compelling parallelism with all new invocations starting a new container if there is none immediately available.
This allows high-level parallelism as invocations come and enables full parallelism in all our experiments.
This behavior presents a good fit for parallel tasks.
Nonetheless, we have seen two unusual exceptions were an invocation got delayed in the system and reused a container.

\paragraph{Q2}
We can infer function resource management and VM distribution from the experiments.
Gathering system information, each machine presents $4$ CPUs and $16$~GiB of RAM.
However, only $8$~GiB are assigned to functions on each machine.
We deduce this by seeing that a single VM only allocated $32$ instances of $256$~MiB, or $4$ of $2048$~MiB.

The compute tasks show that CPU is not strictly limited by the system, but the amount of memory given to each container will determine how much interference with others there will be, and thus how much CPU can be guaranteed to each one.
This resolves that each container with $2$~GiB of memory will get a full CPU, but could use up to four if the remaining of the VM is not used.
$256$~MiB containers will get at least $0.125$ CPU in a congested machine, but could also get all $4$ CPUs in a free machine.
It is a generous policy where the provider gives users more resources than requested.

We have seen this resource interference clearly in our experiments.
With \textit{small} functions, an individual invocation takes the same as in a $2$~GiB function: $1.3$ seconds.
However, with parallel requests, the functions run in groups of $32$ per VM and get $0.125$ of CPU each, which means a time increase of $8\times$: $10.4$~s.
We see this behavior in the plots, although with considerable variance ($10$-$15$~s).
In contrast, the invocations that run on less crowded machines run much faster (see Figure~\ref{fig:ibm:work:200})

\paragraph{Q3}
The experiments also sketch that:
\begin{inparaenum}[i)]
  \item Scheduling is straightforward: if upon request arrival there are no containers idle, a new one is created.
  \item Cold starts can be as low as $1$ or $2$~s, but grow with parallelism.
  \item We see that each VM provides a container pre-warmed. Although it can be helpful for certain applications, it is not that important for parallel workloads.
  \item The non-strict resource assignment is a good advantage, but the user should be conscious of it to avoid unexpected behavior.
\end{inparaenum}

\section{Experiment summary}
\label{sec:experiment:summary}

Table~\ref{tab:sum} summarizes the metrics defined in Section~\ref{sec:experiment:metrics} as perceived in Sections~\ref{sec:experiment:aws} through \ref{sec:experiment:ibm}.
We discuss them next:

\begin{table}
  \caption{
    Summary of experiment results. Parallelism metrics from Section~\ref{sec:experiment:metrics}.
  }
  \label{tab:sum}
  \centering
  \begin{tabular}{lllll}
      & \textbf{AWS} & \textbf{Azure} & \textbf{GCP} & \textbf{IBM}  \\
    \toprule
    \textit{Cold start ($\approx$)} & $300$ ms  & $2$-$20$ s & $2$-$6$ s & $1$-$4$ s \\
    \textit{Completion time} & $1.5$~s & $31$~s & $12.5$~s & $3.5$~s \\
    \multirow{2}{*}{\textit{Parallel degree}} & $200$ & $18$ & $<100$ & $200$ \\
     & $100\%$ & $11\%$ & $<50\%$ & $100\%$ \\
    \textit{Failures} & None & None & Rejects & None \\
    \bottomrule
  \end{tabular}
\end{table}

\textbf{AWS Lambda.}
Currently, cold starts tend to stay around $300$~ms~\cite{faasdom,serverlessbench}.
Our experiments (see Section~\ref{sec:experiment:aws}) match this tendency consistently, without substantial changes with increased concurrency.
AWS Lambda completes the $200$ requests in just $1.5$ seconds (Figure~\ref{fig:aws:work2g:200}), which is the fastest with just half a second of overhead.
This is possible because all invocations run on different instances and instantiation is quick, hence the parallel degree of $200$ ($100\%$).
We did not experience any failure.

\textbf{Azure Functions.}
Instances generally start in $2$-$6$~s~\cite{faasdom,serverlessbench}.
However, we find much larger delays (Section~\ref{sec:experiment:azure}), sometimes over $20$~s.
This could be explained by increased delay in finding resources or the scale controller delaying instantiation and not directly by the overhead of creating an instance.
The $200$ requests experiment is completed in about $31$~s (Figure~\ref{fig:azure:limit1:200}).
This is precisely because the service only used a maximum of $18$ instances, which is only an $11\%$ of the total invocations.
However, none of the invocations failed or were rejected.

\textbf{Google Cloud Functions.}
Other benchmarks~\cite{faasdom,serverlessbench} place Google's cold starts around $3$ seconds.
Our experiments (Section~\ref{sec:experiment:gcp}) show a similar trend: \textit{small} instances starting in $4$~s and \textit{big} ones in $2$~s.
However, they experience increasing delay with parallelism; up to $8$~s (Figure~\ref{fig:gcp:work2g:50}).
The completion time for the $200$ requests is at $12.5$~s (Figure~\ref{fig:gcp:work2g:200}).
Although all invocations run in different instances, Google did not keep all of them warm, and only $84$ are available from the start.
The presence of cold starts delays some invocations and expands completion time.
Incidentally, the number of instances used at the same time never reaches $100$, leaving parallelism below $50\%$.
Additionally, many requests are throttled or rejected with the \textit{big} setup, as shown in Figure~\ref{fig:gcp:work2g:bad}.

\textbf{IBM Cloud Functions.}
We typically see cold starts ranging from $1$ to $2$~s, and the experiments (Section~\ref{sec:experiment:ibm}) indicate it increases with scale, reaching up to $5$~s with $500$ requests.
The $200$-request experiment finishes in $3.5$~s.
All invocations run on different instances, achieving the maximum parallelism of $200$ ($100\%$), which leaves all overhead to instance creation delay.
All invocations completed without failures.

\section{Do FaaS platforms fit parallel computation?}
\label{sec:fit}

With the analysis of the architectures in Section~\ref{sec:arch}, the empirical study in Sections~\ref{sec:experiment} to~\ref{sec:experiment:ibm}, and the metrics summary in Section~\ref{sec:experiment:summary}, we can finally answer the main question proposed in this paper: \emph{Do FaaS platforms fit parallel computation?}.
We do so through the discussion of our main conclusions next.

\emph{Not all FaaS platforms follow the same architecture}, which has high impact on parallel performance.
Two aspects directly influence their support for parallel computations:

\paragraph{Virtualization technologies} They establish how secure and isolated are function instances and how much it takes to start them.
As discussed in Section~\ref{sec:arch}, Table~\ref{tab:archs} shows a relation between the technology and the general architecture design, both impacting invocation latency.
Table~\ref{tab:sum} reveals that platforms with lighter technologies generally provide better cold starts.
AWS Lambda shows the best latency with its Firecracker microVMs.

\paragraph{Scheduling approach} It defines resource management and how invocations traverse the system.
We identified two approaches in Section~\ref{sec:arch}.
%
%
The push-based approach is generous with resources since it can rush decisions and immediately spin up instances when none is available.
AWS and IBM clearly show this on Figures~\ref{fig:aws} and \ref{fig:ibm}.
It improves parallelism, but to be efficient for the provider, resources need to be managed at fine granularity and instances spawn very quickly.
%
The pull-based approach utilizes resources more efficiently, packing more invocations on the same instances.
Usefully, it can enhance management for the provider, and reduce costs for the users.
A downside is that its reactive elasticity is slower to adapt to current demand and is very dependent on its tunning.
Azure is fairly restrictive in that way, as experienced in Section~\ref{sec:experiment:azure}.

\medskip
\emph{Azure Functions stands out from the other platforms} when dealing with parallelism.
Its behavior is very different due to its particular scheduling (how invocations are sent to instances) and resource management (how instances are created and removed).
These characteristics, described in Section~\ref{sec:arch:azure} and visualized in Section~\ref{sec:experiment:azure}, explain the poor elasticity experienced by Kuhlenkamp et al.~\cite{benchmarkingBerlin}, and the limited request throughput assessed by Maissen et al.~\cite{faasdom}, among other works~\cite{wang2018peeking,lee2018evaluation}.
The service is tuned for efficiency in cost and resource management.
It packs invocations on a few instances to maximize resource utilization and reduce costs for the users and management for the provider.
This configuration makes sense, since the service is built atop Azure WebJobs, focused on web applications, and it is great for short IO-bound tasks where the high per-instance concurrency is a big ally.
However, it does not work well for parallel, compute-intensive tasks (see, e.g., Figure~\ref{fig:azure:basic:work:200:warm}), since scaling is degraded in favor of instance concurrency.
Even when limiting instance concurrency to enhance compute-bound applications, the service prefers queueing invocations to a few instances before starting new ones, incurring in significant delays (Figure~\ref{fig:azure:limit1:200}).

\emph{Performance for parallel computations changes considerably between platforms}, since none was, at least initially, designed for this kind of applications.
AWS and IBM's services are able to provide full parallelism for parallel workloads, as demonstrated by PyWren~\cite{PyWren2017} and IBM-PyWren~\cite{IBMPyWren}.
Our experiments show in detail how each invocation is dealt by a different instance and invocation latency is kept low, enabling all tasks to run in parallel.
Google's platform also shows similar scaling behavior in our detailed tests.
However, as discussed earlier (Section~\ref{sec:experiment:summary}), we start to see failed invocations with relatively small parallelism (the aforementioned papers run thousands of parallel functions).
Finally, we already discussed above how Azure Functions is not prepared for these tasks (Table~\ref{tab:sum}), and would struggle to support them.

Our conclusions help explain several benchmarking works in the literature~\cite{benchmarkingBerlin,wang2018peeking,faasdom,cloudbuttonbench,lee2018evaluation}.
Indeed, they already point to the good performance of AWS and IBM or the sometimes strange behavior in GCP.
And most importantly, the difference in performance for Azure was already sketched in the literature~\cite{allbutone}.
However, in this paper we analyzed the different platforms from the perspective of parallelism and took a deep look into the different architecture designs, which adds new information and helps to understand the causes of these behaviors.

In sum, \emph{FaaS is not inherently good for parallel computations} and performance strongly depends on the platform design and configuration by the provider.
Consequently, users must be aware of the parallel capabilities of the platform they choose in order to understand how their applications will behave.

\section{Conclusion and future insights}
\label{sec:conclusions}
In this paper, we have analyzed the architectures of four major FaaS platforms: AWS Lambda, Azure Functions, Google Cloud Functions, and IBM Cloud Functions.
Our research focused on the capabilities and limitations the services offer for highly parallel computations.
The design of the platforms revealed two important traits influencing their performance: virtualization technology and scheduling approach.
We further explored the platforms with detailed experiments to plot parallel executions and show task distribution in the platform.
The experiments evidenced that the different approaches to architecture heavily affect how parallelism is achieved.
AWS, IBM, and GCP run different function instances for each function invocation, while Azure packs invocations in a few instances.
In consequence, parallelism is thwarted on the latter (only $18\%$ of invocations run in parallel) and parallel computations suffer big overhead (a $1$~s computation takes $31$~s).
AWS and IBM always achieve good parallelism ($100\%$).
However, although GCP's approach is also prone to parallelism, our experiments show conflicting performance.
The appearance of failed invocations produces stragglers in the computation and increases complexity for the user, who must manage the errors.

In the future, we see FaaS platforms improving on two aspects.
On one hand, \emph{virtualization technologies} are one of the most important factors for parallel computing in the serverless model.
This is because they establish the granularity of resource management, the quickness to create instances, and the complexity of scheduling.
In other words, it takes an important role in invocation latency and overall cost (for the user and the provider).
We already see AWS improving this aspect with Firecracker, and we expect further improvements in this line from the other providers as well.
On the other hand, the \emph{scheduling approach} is also a key component.
Our exploration revealed that a reactive model is too slow to scale, so a proactive push-based architectural approach is more adequate.
Achieving high levels of parallelism requires being able to provide resources rapidly.
Then it is critical to be efficient when dealing with incoming invocations, and proactive approaches are faster than reactive ones.
We expect FaaS platforms to move in this direction and further improve their scheduling mechanisms.

An example of this evolution is granular computing~\cite{granular}, where microsecond-scale tasks come into place.
Such short tasks need even smaller overhead, and hence scheduling time needs to decrease in orders of magnitude.
Likewise, tiny tasks only need few resources, so the system should be able to provide them at finer granularity.
Granular computing is very akin to FaaS parallel computing as it benefits from the same properties.
However, it requires new lightweight virtualization technologies and improved scheduling to appear in the next years.

Hopefully, better virtualization technologies are likely to also improve scheduling.
A faster start up time reduces invocation latency and thus the weight of creation penalty in the scheduler decision-taking.
Finer granularity also allows to securely run invocations of different tenants in fewer machines, increasing resource utilization and reducing cost.

We envision that FaaS platforms will continue to evolve in this direction for the future years, all in all, enhancing performance of FaaS services for parallel workloads, but also enabling new kinds of applications and use cases.

\section*{Acknowledgement}
Thanks to Aitor Arjona, Josep Sampe, and Pol Roca for their contributions on the big scale experiments.
Thanks to Marc S{\'a}nchez-Artigas for his valuable insights and reviews.
We would like to thank our editor and the anonymous reviewers for their insightful comments which helped enhance the quality of this article.
This work has been partially supported by the EU project H2020 ``CloudButton: Serverless Data Analytics Platform'' (825184)
and by the Spanish government (PID2019-106774RB-C22).
Daniel Barcelona-Pons's work is financed by a Mart\'{i} i Franqu\`{e}s programme grant (URV).

\bibliography{references}

\begin{thebibliography}{10}
\expandafter\ifx\csname url\endcsname\relax
  \def\url#1{\texttt{#1}}\fi
\expandafter\ifx\csname urlprefix\endcsname\relax\def\urlprefix{URL }\fi
\expandafter\ifx\csname href\endcsname\relax
  \def\href#1#2{#2} \def\path#1{#1}\fi

\bibitem{PyWren2017}
E.~Jonas, Q.~Pu, S.~Venkataraman, I.~Stoica, B.~Recht, {Occupy the Cloud:
  Distributed Computing for the 99\%}, in: Proceedings of the 2017 Symposium on
  Cloud Computing, SoCC'17, 2017.
\newblock \href {http://dx.doi.org/10.1145/3127479.3128601}
  {\path{doi:10.1145/3127479.3128601}}.

\bibitem{excamera}
S.~Fouladi, R.~S. Wahby, B.~Shacklett, K.~V. Balasubramaniam, W.~Zeng,
  R.~Bhalerao, A.~Sivaraman, G.~Porter, K.~Winstein, {Encoding, Fast and Slow:
  Low-Latency Video Processing Using Thousands of Tiny Threads}, in: 14th
  {USENIX} Symposium on Networked Systems Design and Implementation
  ({NSDI}'17), 2017.

\bibitem{crucial2019}
D.~Barcelona-Pons, M.~S\'{a}nchez-Artigas, G.~Par\'{\i}s, P.~Sutra,
  P.~Garc\'{\i}a-L\'{o}pez, {On the FaaS Track: Building Stateful Distributed
  Applications with Serverless Architectures}, in: Proceedings of the 20th
  International Middleware Conference, Middleware '19, ACM, New York, NY, USA,
  2019, pp. 41--54.
\newblock \href {http://dx.doi.org/10.1145/3361525.3361535}
  {\path{doi:10.1145/3361525.3361535}}.

\bibitem{gg}
S.~Fouladi, F.~Romero, D.~Iter, Q.~Li, S.~Chatterjee, C.~Kozyrakis, M.~Zaharia,
  K.~Winstein, {From Laptop to Lambda: Outsourcing Everyday Jobs to Thousands
  of Transient Functional Containers}, in: 2019 {USENIX} Annual Technical
  Conference ({USENIX} {ATC} 19), {USENIX} Association, Renton, WA, 2019, pp.
  475--488.

\bibitem{numpywren}
V.~Shankar, K.~Krauth, Q.~Pu, E.~Jonas, S.~Venkataraman, I.~Stoica, B.~Recht,
  J.~Ragan{-}Kelley, numpywren: serverless linear algebra, CoRR abs/1810.09679.
\newblock \href {http://arxiv.org/abs/1810.09679} {\path{arXiv:1810.09679}}.

\bibitem{locus}
Q.~Pu, S.~Venkataraman, I.~Stoica, {Shuffling, Fast and Slow: Scalable
  Analytics on Serverless Infrastructure}, in: 16th {USENIX} Symposium on
  Networked Systems Design and Implementation ({NSDI} 19), {USENIX}
  Association, Boston, MA, 2019, pp. 193--206.

\bibitem{GIMENEZALVENTOSA2019259}
V.~Giménez-Alventosa, G.~Moltó, M.~Caballer, A framework and a performance
  assessment for serverless mapreduce on aws lambda, Future Generation Computer
  Systems 97 (2019) 259 -- 274.
\newblock \href {http://dx.doi.org/10.1016/j.future.2019.02.057}
  {\path{doi:10.1016/j.future.2019.02.057}}.

\bibitem{servermix}
P.~García-López, M.~Sánchez-Artigas, S.~Shillaker, P.~Pietzuch,
  D.~Breitgand, G.~Vernik, P.~Sutra, T.~Tarrant, A.~J. Ferrer, Servermix:
  Tradeoffs and challenges of serverless data analytics (2019).
\newblock \href {http://arxiv.org/abs/1907.11465} {\path{arXiv:1907.11465}}.

\bibitem{berkeleyServerless}
E.~Jonas, J.~Schleier-Smith, V.~Sreekanti, C.-C. Tsai, A.~Khandelwal, Q.~Pu,
  V.~Shankar, J.~Menezes~Carreira, K.~Krauth, N.~Yadwadkar, J.~Gonzalez, R.~A.
  Popa, I.~Stoica, D.~A. Patterson, Cloud programming simplified: A berkeley
  view on serverless computing, Tech. Rep. UCB/EECS-2019-3, EECS Department,
  University of California, Berkeley (Feb 2019).

\bibitem{hellersteinServerless}
J.~M. Hellerstein, J.~M. Faleiro, J.~Gonzalez, J.~Schleier{-}Smith,
  V.~Sreekanti, A.~Tumanov, C.~Wu, Serverless computing: One step forward, two
  steps back, in: {CIDR} 2019, 9th Biennial Conference on Innovative Data
  Systems Research, Asilomar, CA, USA, January 13-16, 2019, Online Proceedings,
  2019.

\bibitem{benchmarkingBerlin}
J.~Kuhlenkamp, S.~Werner, M.~C. Borges, D.~Ernst, D.~Wenzel, {Benchmarking
  Elasticity of FaaS Platforms as a Foundation for Objective-Driven Design of
  Serverless Applications}, in: Proceedings of the 35th Annual ACM Symposium on
  Applied Computing, SAC ’20, Association for Computing Machinery, New York,
  NY, USA, 2020, p. 1576–1585.
\newblock \href {http://dx.doi.org/10.1145/3341105.3373948}
  {\path{doi:10.1145/3341105.3373948}}.

\bibitem{cloudflaredef}
Cloudflare, {What Is Function as a Service (FaaS)?},
  \url{https://www.cloudflare.com/learning/serverless/glossary/function-as-a-service-faas/},
  {retrieved Apr. 2020} (2020).

\bibitem{wang2018peeking}
L.~Wang, M.~Li, Y.~Zhang, T.~Ristenpart, M.~Swift, {Peeking Behind the Curtains
  of Serverless Platforms}, in: 2018 {USENIX} Annual Technical Conference
  ({USENIX} {ATC} 18), {USENIX} Association, Boston, MA, 2018, pp. 133--146.

\bibitem{parallelOrchWOSC}
D.~Barcelona-Pons, P.~Garc\'{\i}a-L\'{o}pez, A.~Ruiz, A.~G\'{o}mez-G\'{o}mez,
  G.~Par\'{\i}s, M.~S\'{a}nchez-Artigas, {FaaS Orchestration of Parallel
  Workloads}, in: Proceedings of the 5th International Workshop on Serverless
  Computing, WOSC ’19, Association for Computing Machinery, New York, NY,
  USA, 2019, p. 25–30.
\newblock \href {http://dx.doi.org/10.1145/3366623.3368137}
  {\path{doi:10.1145/3366623.3368137}}.

\bibitem{faasdom}
P.~Maissen, P.~Felber, P.~Kropf, V.~Schiavoni, {FaaSdom: A Benchmark Suite for
  Serverless Computing}, in: Proceedings of the 14th ACM International
  Conference on Distributed and Event-Based Systems, DEBS ’20, Association
  for Computing Machinery, New York, NY, USA, 2020, p. 73–84.
\newblock \href {http://dx.doi.org/10.1145/3401025.3401738}
  {\path{doi:10.1145/3401025.3401738}}.

\bibitem{serverlessbench}
B.~Strehl, {{$\lambda$} Serverless Benchmark},
  \url{https://serverless-benchmark.com/} (2020).

\bibitem{SCHEUNER2020EVAL}
J.~Scheuner, P.~Leitner, Function-as-a-service performance evaluation: A
  multivocal literature review, Journal of Systems and Software (2020)
  110708\href {http://dx.doi.org/https://doi.org/10.1016/j.jss.2020.110708}
  {\path{doi:https://doi.org/10.1016/j.jss.2020.110708}}.

\bibitem{blog:Shilkov}
M.~Shilkov, {From 0 to 1000 Instances: How Serverless Providers Scale Queue
  Processing}, \url{https://blog.binaris.com/from-0-to-1000-instances/},
  {retrieved Apr. 2020} (2018).

\bibitem{pawlik2019performance}
M.~Pawlik, K.~Figiela, M.~Malawski, {Performance considerations on execution of
  large scale workflow applications on cloud functions} (2019).
\newblock \href {http://arxiv.org/abs/1909.03555} {\path{arXiv:1909.03555}}.

\bibitem{cloudbuttonbench}
CloudButton, {CloudButton Serverless Benchmark},
  \url{https://cloudbutton.github.io/benchmarks/} (2020).

\bibitem{sequoia}
A.~Tariq, A.~Pahl, S.~Nimmagadda, E.~Rozner, S.~Lanka, Sequoia: Enabling
  quality-of-service in serverless computing, in: Proceedings of the 11th ACM
  Symposium on Cloud Computing, SoCC '20, Association for Computing Machinery,
  New York, NY, USA, 2020, p. 311–327.
\newblock \href {http://dx.doi.org/10.1145/3419111.3421306}
  {\path{doi:10.1145/3419111.3421306}}.

\bibitem{lee2018evaluation}
H.~Lee, K.~Satyam, G.~Fox, {Evaluation of production serverless computing
  environments}, in: 2018 IEEE 11th International Conference on Cloud Computing
  (CLOUD), IEEE, 2018, pp. 442--450.
\newblock \href {http://dx.doi.org/10.1109/CLOUD.2018.00062}
  {\path{doi:10.1109/CLOUD.2018.00062}}.

\bibitem{allbutone}
J.~Kuhlenkamp, S.~Werner, M.~C. Borges, D.~Ernst, {All but One: FaaS Platform
  Elasticity Revisited}, SIGAPP Appl. Comput. Rev. 20~(3) (2020) 5–19.
\newblock \href {http://dx.doi.org/10.1145/3429204.3429205}
  {\path{doi:10.1145/3429204.3429205}}.

\bibitem{opensource:bench}
S.~K. {Mohanty}, G.~{Premsankar}, M.~{di Francesco}, {An Evaluation of Open
  Source Serverless Computing Frameworks}, in: 2018 IEEE International
  Conference on Cloud Computing Technology and Science (CloudCom), 2018, pp.
  115--120.

\bibitem{shahrad2019architectural}
M.~Shahrad, J.~Balkind, D.~Wentzlaff, {Architectural Implications of
  Function-as-a-Service Computing}, in: Proceedings of the 52nd Annual IEEE/ACM
  International Symposium on Microarchitecture, MICRO ’52, Association for
  Computing Machinery, New York, NY, USA, 2019, p. 1063–1075.
\newblock \href {http://dx.doi.org/10.1145/3352460.3358296}
  {\path{doi:10.1145/3352460.3358296}}.

\bibitem{pocketpre}
A.~Klimovic, Y.~Wang, C.~Kozyrakis, P.~Stuedi, J.~Pfefferle, A.~Trivedi,
  Understanding ephemeral storage for serverless analytics, in: 2018 {USENIX}
  Annual Technical Conference ({USENIX} {ATC} 18), {USENIX} Association,
  Boston, MA, 2018, pp. 789--794.

\bibitem{pocket}
A.~Klimovic, Y.~Wang, P.~Stuedi, A.~Trivedi, J.~Pfefferle, C.~Kozyrakis,
  Pocket: Elastic ephemeral storage for serverless analytics, in: 13th {USENIX}
  Symposium on Operating Systems Design and Implementation ({OSDI} 18),
  {USENIX} Association, Carlsbad, CA, 2018, pp. 427--444.

\bibitem{comparisonWOSC}
P.~Garc{\'\i}a~L{\'o}pez, M.~S{\'a}nchez-Artigas, G.~Par{\'\i}s,
  D.~Barcelona~Pons, {\'A}.~Ruiz~Ollobarren, D.~Arroyo~Pinto, {Comparison of
  FaaS Orchestration Systems}, in: 2018 IEEE/ACM International Conference on
  Utility and Cloud Computing Companion (UCC Companion), IEEE, 2018, pp.
  148--153.

\bibitem{microsoft2020serverless}
M.~Shahrad, R.~Fonseca, Íñigo Goiri, G.~Chaudhry, P.~Batum, J.~Cooke,
  E.~Laureano, C.~Tresness, M.~Russinovich, R.~Bianchini, {Serverless in the
  Wild: Characterizing and Optimizing the Serverless Workload at a Large Cloud
  Provider} (2020).
\newblock \href {http://arxiv.org/abs/2003.03423} {\path{arXiv:2003.03423}}.

\bibitem{openwhisk}
{Apache OpenWhisk}, \url{https://openwhisk.apache.org/}.

\bibitem{refArchFaas}
E.~{van Eyk}, J.~{Grohmann}, S.~{Eismann}, A.~{Bauer}, L.~{Versluis},
  L.~{Toader}, N.~{Schmitt}, N.~{Herbst}, C.~L. {Abad}, A.~{Iosup}, The spec-rg
  reference architecture for faas: From microservices and containers to
  serverless platforms, IEEE Internet Computing 23~(6) (2019) 7--18.
\newblock \href {http://dx.doi.org/10.1109/MIC.2019.2952061}
  {\path{doi:10.1109/MIC.2019.2952061}}.

\bibitem{lambda:config}
AWS, {Configuring Functions in the AWS Lambda Console},
  \url{https://docs.aws.amazon.com/lambda/latest/dg/configuration-console.html}
  (2020).

\bibitem{lambda:limits}
AWS, {AWS Lambda limits},
  \url{https://docs.aws.amazon.com/lambda/latest/dg/gettingstarted-limits.html}
  (2020).

\bibitem{aws:securitypaper}
AWS, {Security Overview of AWS Lambda},
  \url{https://d1.awsstatic.com/whitepapers/Overview-AWS-Lambda-Security.pdf}
  (2019).

\bibitem{aws:firecracker}
A.~Agache, M.~Brooker, A.~Iordache, A.~Liguori, R.~Neugebauer, P.~Piwonka,
  D.-M. Popa, {Firecracker: Lightweight Virtualization for Serverless
  Applications}, in: 17th {USENIX} Symposium on Networked Systems Design and
  Implementation ({NSDI} 20), {USENIX} Association, Santa Clara, CA, 2020, pp.
  419--434.

\bibitem{lambda:scale}
AWS, {AWS Lambda function scaling},
  \url{https://docs.aws.amazon.com/lambda/latest/dg/invocation-scaling.html}
  (2020).

\bibitem{azure:scale}
Microsoft, {Azure Functions scale and hosting},
  \url{https://docs.microsoft.com/en-us/azure/azure-functions/functions-scale}
  (2020).

\bibitem{azure:concurrency}
Microsoft, {Azure Functions HTTP output bindings - Settings},
  \url{https://docs.microsoft.com/en-us/azure/azure-functions/functions-bindings-http-webhook-output#hostjson-settings}
  (2020).

\bibitem{gcp:concepts}
Google, {Cloud Functions Execution Environment},
  \url{https://cloud.google.com/functions/docs/concepts/exec} (2020).

\bibitem{gcp:usegvisor}
Google, {GKE Sandbox: Bring defense in depth to your pods},
  \url{https://cloud.google.com/blog/products/containers-kubernetes/gke-sandbox-bring-defense-in-depth-to-your-pods}
  (2019).

\bibitem{gcp:gvisor}
gVisor Authors, {What is gVisor?}, \url{https://gvisor.dev/docs/} (2020).

\bibitem{gcp:limits}
Google, {Google Cloud Function Quotas},
  \url{https://cloud.google.com/functions/quotas} (2020).

\bibitem{gcp:pricing'n'types}
Google, {Google Cloud Function Pricing},
  \url{https://cloud.google.com/functions/pricing} (2020).

\bibitem{ibm:how-works}
IBM, {How Cloud Functions works},
  \url{https://cloud.ibm.com/docs/openwhisk?topic=openwhisk-about} (2020).

\bibitem{ibm:limits}
IBM, System details and limits,
  \url{https://cloud.ibm.com/docs/openwhisk?topic=openwhisk-limits} (2020).

\bibitem{openwhisk:concurrency}
{OpenWhisk Concurrency},
  \url{https://github.com/apache/openwhisk/blob/master/docs/concurrency.md}.

\bibitem{azure:pricing}
Microsoft, {Estimating Consumption plan costs},
  \url{https://docs.microsoft.com/en-us/azure/azure-functions/functions-consumption-costs}
  (2020).

\bibitem{lloyd:investigation}
W.~{Lloyd}, S.~{Ramesh}, S.~{Chinthalapati}, L.~{Ly}, S.~{Pallickara},
  Serverless computing: An investigation of factors influencing microservice
  performance, in: 2018 IEEE International Conference on Cloud Engineering
  (IC2E), 2018, pp. 159--169.
\newblock \href {http://dx.doi.org/10.1109/IC2E.2018.00039}
  {\path{doi:10.1109/IC2E.2018.00039}}.

\bibitem{azure:quickstart}
Microsoft, {Create a function in Azure using Visual Studio Code},
  \url{https://docs.microsoft.com/en-us/azure/azure-functions/functions-create-first-function-vs-code}
  (2020).

\bibitem{azure:instanceid}
{Azure runtime environment},
  \url{https://github.com/projectkudu/kudu/wiki/Azure-runtime-environment}
  (2019).

\bibitem{ibm:dockermetrics}
Docker, Runtime metrics,
  \url{https://docs.docker.com/config/containers/runmetrics/} (2020).

\bibitem{IBMPyWren}
J.~Samp{\'e}, G.~Vernik, M.~S\'{a}nchez-Artigas, P.~Garc\'{\i}a-L\'{o}pez,
  {Serverless Data Analytics in the IBM Cloud}, in: Proceedings of the 19th
  International Middleware Conference Industry, Middleware '18, ACM, New York,
  NY, USA, 2018, pp. 1--8.
\newblock \href {http://dx.doi.org/10.1145/3284028.3284029}
  {\path{doi:10.1145/3284028.3284029}}.

\bibitem{granular}
C.~Lee, J.~Ousterhout, Granular computing, in: Proceedings of the Workshop on
  Hot Topics in Operating Systems, HotOS ’19, Association for Computing
  Machinery, New York, NY, USA, 2019, p. 149–154.
\newblock \href {http://dx.doi.org/10.1145/3317550.3321447}
  {\path{doi:10.1145/3317550.3321447}}.

\end{thebibliography}

\end{document}